\documentclass[preprint,aps,floatfix,nofootinbib,a4paper, superscriptaddress]{revtex4-1}

\usepackage{amsmath, amssymb, amsfonts, amsthm, latexsym, epsfig, mathrsfs, xcolor, bbm, slashed, float, thmtools}

\usepackage[all]{xy}

\usepackage[inline]{enumitem}

\usepackage{setspace}
\usepackage[marginal, multiple]{footmisc}

\usepackage[utf8]{inputenc}
\usepackage[T1]{fontenc}
\usepackage{lmodern}

\usepackage{microtype}

\usepackage[colorlinks, allcolors=blue!70!black, linktocpage]{hyperref}

\numberwithin{equation}{section}

\usepackage{cleveref}

\let\OLDthebibliography\thebibliography
\renewcommand\thebibliography[1]{%
	\setstretch{1.079} 
	\OLDthebibliography{#1}%
	\small %
	\setlength{\itemsep}{0.2\baselineskip} 
}

\let\OLDfootnote\footnote
\renewcommand\footnote[1]{%
    \count\footins = 1000%
	\setlength{\footnotesep}{0.75\baselineskip}%
	{\footnotesize \OLDfootnote{#1}}%
}

\setlist[enumerate]{noitemsep, label=(\arabic*), ref=(\arabic*)}

\renewcommand\thesection{\arabic{section}}
\renewcommand\thesubsection{\arabic{subsection}}

\makeatletter
\def\p@subsection{\thesection.}
\def\p@subsubsection{\thesection.\thesubsection.}
\makeatother 

\theoremstyle{plain}

\newtheorem{lemma}{Lemma}[section]
\newtheorem{prop}{Proposition}[section]
\newtheorem{corollary}{Corollary}[section]

\theoremstyle{definition}
\newtheorem{definition}{Definition}[section]

\declaretheorem[style=remark,qed=$\blacksquare$,numberwithin=section]{remark} 

\crefname{equation}{Eq.}{Eqs.}
\creflabelformat{equation}{#2#1#3}

\crefname{section}{Sec.}{Secs.}
\crefname{appendix}{Appendix}{Appendices}
\crefname{figure}{Fig.}{Figs.}

\crefname{definition}{Def.}{Defs.}
\crefname{prop}{Prop.}{Props.}
\crefname{lemma}{Lemma}{Lemmas}
\crefname{corollary}{Cor.}{Cors.}
\crefname{thm}{Theorem}{Theorems}
\crefname{remark}{Remark}{Remarks}

\crefname{ass}{Assumptions}{Assumptions}
\crefname{property}{Properties}{Properties}

\newcommand{\be}{\begin{equation}\begin{aligned}}
\newcommand{\ee}{\end{aligned}\end{equation}}

\newcommand{\lb}{\left}
\newcommand{\rb}{\right}

\newcommand{\pd}[2][]{\frac{\partial #1}{\partial #2}} 

\let\oldint\int
\renewcommand{\int}{\oldint\limits}

\newcommand{\mc}{\mathcal}

\newcommand{\ms}{\mathscr}
\newcommand{\mf}{\mathfrak}
\newcommand{\bb}{\mathbb}

\newcommand{\nfrac}[2]{{{}^#1\!\!/\!_#2}}

\newcommand{\half}{\nfrac{1}{2}}

\newcommand{\eqsp}{\, ,\quad} 

\newcommand{\pb}[1]{\underleftarrow{#1}}

\newcommand{\hateq}{\mathrel{\mathop {\widehat=} }} 

\newcommand{\Lie}{\pounds} 

\newcommand{\defn}{\mathrel{\mathop:}=} 

\newcommand{\df}[1]{\boldsymbol{#1}} 

\renewcommand{\Re}{{\rm Re}}
\renewcommand{\Im}{{\rm Im}}

\DeclareMathOperator{\thorn}{\text{\rm \th}}
\let\eth\relax
\DeclareMathOperator{\eth}{\text{\rm \dh}}

\newcommand{\wt}{\circeq}

\renewcommand{\bar}{\overline}

\newcommand{\scri}{\ms I}

\DeclareMathOperator{\STF}{STF}

\begin{document}

\setstretch{1.2}


\title{The Wald-Zoupas prescription for asymptotic charges at null infinity in general relativity}


\author{Alexander M. Grant}
\email{alex.grant@virginia.edu}
\affiliation{Department of Physics, Cornell University, Ithaca, NY 14853, USA}
\affiliation{Department of Physics, University of Virginia, Charlottesville, VA 22904, USA}

\author{Kartik Prabhu}
\email{kartikprabhu@ucsb.edu}
\affiliation{Department of Physics, University of California, Santa Barbara, CA 93106, USA}

\author{Ibrahim Shehzad}
\email{is354@cornell.edu}
\affiliation{Department of Physics, Cornell University, Ithaca, NY 14853, USA}


\begin{abstract}
We use the formalism developed by Wald and Zoupas to derive explicit covariant expressions for the charges and fluxes associated with the Bondi-Metzner-Sachs symmetries at null infinity in asymptotically flat spacetimes in vacuum general relativity. Our expressions hold in non-stationary regions of null infinity, are local and covariant, conformally-invariant, and are independent of the choice of foliation of null infinity and of the chosen extension of the symmetries away from null infinity. While similar expressions have appeared previously in the literature in Bondi-Sachs coordinates (to which we compare our own), such a choice of coordinates obscures these properties. Our covariant expressions can be used to obtain charge formulae in any choice of coordinates at null infinity. We also include detailed comparisons with other expressions for the charges and fluxes that have appeared in the literature: the Ashtekar-Streubel flux formula, the Komar formulae, and the linkage and twistor charge formulae. Such comparisons are easier to perform using our explicit expressions, instead of those which appear in the original work by Wald and Zoupas.
\end{abstract}

\maketitle
\tableofcontents

\section{Introduction}
\label{sec:intro}
There has been a considerable revival of interest in asymptotic symmetries and their associated charges in general relativity in recent years. At null infinity in asymptotically flat spacetimes, these symmetries are elements of an infinite-dimensional symmetry group called the Bondi-Metzner-Sachs (BMS) group \cite{BBM, Sachs1}. A major reason for this revival of interest has been the conjecture by Strominger that these symmetries give rise to an infinite number of conservation laws which constrain classical gravitational scattering in asymptotically flat spacetimes.\footnote{This conjecture has been recently proven under certain regularity conditions on the spacetimes near spatial infinity, see \cite{Tro, KP-GR-match, Prabhu:2021cgk} for details.} In a series of papers, conservation laws associated with these symmetries have also been shown to be related to soft graviton theorems \cite{He:2014laa,Strominger:2014pwa,Kapec:2015vwa,Avery:2015gxa,Campiglia:2015kxa,Campoleoni:2017mbt} and gravitational memory effects \cite{He:2014laa,Strominger:2014pwa,Pasterski:2015tva,HIW,Mao:2017wvx,Pate:2017fgt,Chatterjee:2017zeb} and have even been speculated to have a possible bearing on the black hole information paradox \cite{Hawking:2016msc,Strominger:2017aeh,Hawking:2016sgy,Flanagan:2021ojq}.\\

In this paper we consider the problem of deriving explicit general expressions for the charges associated with asymptotic symmetries at null infinity in vacuum general relativity. We list a set of criteria that these should satisfy to be physically reasonable and, using the prescription provided by Wald and Zoupas in \cite{WZ}, provide an explicit computation of the general expressions for these quantities. While such charge expressions have previously appeared in the literature \cite{Flanagan:2015pxa}, they have been obtained entirely in Bondi-Sachs coordinates. Using a fixed coordinate system from the outset makes the covariance and conformal-invariance of the resulting expressions difficult to verify. In this paper we obtain explicit expressions for these charges in terms of quantities defined on null infinity in a completely covariant manner. These covariant expressions can be used to obtain formulae in \emph{any} choice of coordinates one wishes to choose (see \cref{sec:coord}).\\

We will consider \(4\)-dimensional asymptotically flat spacetimes defined in a coordinate independent manner through Penrose's conformal completion (see \cref{def:asym-flat}). In this picture, the null infinity of the original (physical) spacetime is represented by a smooth null surface \(\scri\) in the conformally-completed (unphysical) spacetime. This null surface has the topology of \(\scri \cong \bb R \times \bb S^2\) where \(\bb R\) represents the null directions and \(\bb S^2\) the angular directions at infinity. It then follows from the definition of asymptotic flatness that there is a certain \emph{universal} structure associated with \(\scri\) (see \cref{sec:univ-str}). This structure is universal in the sense that it is independent of which asymptotically flat physical spacetime is being considered and thus provides a ``fixed background'' at infinity which is common to all physical spacetimes. The infinitesimal diffeomorphisms (i.e. vector fields) which preserve this universal structure form the asymptotic symmetry algebra given by the BMS algebra (\cref{sec:bms}).\\

Then, one wishes to define certain ``conserved quantities'' (similar to mass, energy or angular momentum) associated with any given physical spacetime and an asymptotic symmetry. In general dynamical spacetimes such quantities will not be conserved due to the presence of gravitational radiation. However, one can proceed as follows: let \(S \cong \bb S^2\) be some cross-section (or ``cut'') of null infinity \(\scri\); the choice of \(S\) represents an ``instant of time'' on \(\scri\). Then, for a vector field \(\xi^a\) representing an asymptotic symmetry and to every cross-section \(S\) we define a \emph{charge} \(\mc Q[\xi;S]\), which is to represent the ``not really conserved quantity'' at that ``instant of time''. The change in this charge with ``time'', i.e. between two cross-sections, is then a \emph{flux} \(\mc F[\xi;\Delta\scri]\) where \(\Delta\scri\) is the region of null infinity between those two cross-sections. However, without any further physical criteria one could make up arbitrary expressions for such a charge and flux formula. Even if one requires that any notion of charge associated with time translations and rotations coincide with the mass and angular momentum usually defined in Kerr spacetimes, it is not apparent how to generalize the charge expressions to all BMS symmetries and to non-stationary spacetimes.\\

Next we list some criteria which any physically reasonable notion of asymptotic charges and fluxes should satisfy: 
\begin{enumerate*}
\item Any charge or flux quantities one defines must be independent of any choice of coordinates; this issue is already resolved since we work directly with the covariant formulation of null infinity mentioned above, instead of fixing some coordinates like Bondi-Sachs coordinates.
\item We want to associate the charges and fluxes with the physical spacetime and not with the choice of conformal-completion used to obtain the unphysical conformally-completed spacetime, or with any additional structure used to compute these quantities. Thus, we require that the charges and fluxes be independent of the conformal factor used to obtain the Penrose conformal-completion. Below we will also use a foliation of \(\scri\) to simplify our computations, and so we will also demand that the charges and fluxes be independent of this additional choice of foliation.
\item We also want to associate the charges and fluxes with the BMS symmetries at \(\scri\) and so, we demand that they be independent of any arbitrary extension of these asymptotic symmetries into the spacetime (these extensions are considered ``pure gauge'' and are discussed in \cref{sec:bms-ext}).
\item The charges and fluxes are local and covariant in the following sense: the value of the charge \(\mc Q[\xi;S]\) is obtained as an integral over \(S\) of a \(2\)-form which is constructed from the available fields and the BMS symmetry, and finitely-many of their derivatives at \(S\). The flux \(\mc F[\xi;\Delta\scri]\) is the integral over \(\Delta\scri\) of a \(3\)-form constructed in a similar way from the fields and symmetry in the region \(\Delta\scri\).
\item Finally, since we want the fluxes \(\mc F[\xi;\Delta\scri]\) to characterize physical dynamical processes, like gravitational radiation, we also require that the flux associated with \emph{any} BMS symmetry between \emph{any} two cross-sections vanishes when the physical spacetime is stationary. 
\end{enumerate*}\\

A prescription for obtaining such asymptotic charges and fluxes in any local and covariant Lagrangian theory of gravity was given by Wald and Zoupas \cite{WZ}. We will detail this procedure in \cref{sec:WZ} for our case of interest, namely asymptotically flat spacetimes in vacuum general relativity. This case was also considered in detail by Wald and Zoupas \cite{WZ}, where it was shown, rather indirectly, that the resulting charge expression matches the one given by Dray and Streubel \cite{DS}, which was obtained by generalizing Penrose's formula motivated by twistor theory (see \cref{sec:twistor}). Explicit expressions for these charges in terms of quantities on null infinity were obtained by Flanagan and Nichols \cite{Flanagan:2015pxa}, but their analysis was done in Bondi coordinates (where the covariance and conformal-invariance of the expressions is not manifest) and also in stationary regions of null infinity.\footnote{The method used in \cite{BT} only gives a ``perturbed charge'' expression which is not integrable in phase space; see \cref{eq:charge-bondi-2} and the subsequent discussion. Charge expressions similar to those in \cite{Flanagan:2015pxa} were also obtained in \cite{CJK}, but their charge expression differs from the one obtained using the Wald-Zoupas method, except in the case where the asymptotic symmetry is a pure supertranslation; see also footnote 11 of \cite{Flanagan:2015pxa}.}

 In this paper we compute the formulae for the charges and fluxes, using the Wald-Zoupas prescription, written explicitly in terms of fields defined on null infinity in full generality. We call these the ``Wald-Zoupas (WZ) charge'' and ``Wald-Zoupas (WZ) flux'' respectively. To simplify our computations we will choose an arbitrary but fixed foliation of null infinity, but show that our formulae are independent of this choice of foliation. This computation can be done easily for the flux, since the WZ flux expression is written entirely in terms of quantities at null infinity; our expression for the WZ flux in terms of a choice of foliation is given in \cref{eq:flux-formula}. The computation of the charges, however, is more complicated: unlike the flux, a portion of the expression for the WZ charge involves the limit of the integral of a quantity that is defined in a neighborhood of null infinity. To compute this limit, we make use of Bondi coordinates, much like \cite{Flanagan:2015pxa}; however, we show that the value of this limit is independent of the choices that are made in using Bondi coordinates, such as the choice of foliation, conformal factor, and extension of the BMS vector field off of null infinity. As such, we can convert the value of the limit back into a covariant form, allowing us to write an expression for the WZ charge (see \cref{eq:Q-GR-defn}) in terms of quantities defined on null infinity.
Since this procedure is somewhat involved, we then check explicitly that the change of the charge is consistent with the flux formula in \cref{eq:flux-formula}. These covariant expressions are equal to those given by \cite{WZ}, and are therefore independent of foliation; however, picking a choice of foliation allows us to write the charge in terms of quantities defined entirely at null infinity. These explicit formulae can then be easily compared to the expressions one would obtain in \emph{any} coordinate system of ones choosing, see e.g. \cref{sec:coord}. These expressions can also be compared to other formulations of BMS charges, some of which we detail in \cref{sec:other-charges}.\\

The rest of this paper is organized as follows. We start by reviewing the definition of asymptotic-flatness at null infinity in \cref{sec:AF}, and summarize the various asymptotic fields arising at null infinity that we will use in subsequent computations. In \cref{sec:univ-str}, we discuss the universal structure at null infinity (that is, the structure common to all asymptotically-flat spacetimes at null infinity) and the restrictions on metric perturbations which follow from the definition of asymptotic flatness. In \cref{sec:bms}, we review the BMS algebra and its relevant properties. In \cref{sec:WZ}, we describe the Wald-Zoupas prescription for obtaining the charges and fluxes associated with BMS symmetries and obtain manifestly covariant expressions for these quantities which we call the ``Wald-Zoupas (WZ) charge'' and the ``Wald-Zoupas (WZ) flux.'' Finally, we discuss the construction of Bondi-Sachs and conformal Gau{\ss}ian null coordinate frames near null infinity in \cref{sec:coord} and express the BMS symmetries and the Wald-Zoupas charge in these coordinates. We end with a short discussion of our results in \cref{sec:disc}. While in the body of the paper, we work in a conformal frame where the Bondi condition (see \cref{eq:Bondi-cond}) holds, we include a summary of our main results in more general conformal frames in \cref{sec:non-Bondi}. In \cref{sec:GHP} we rewrite the WZ charge and flux expressions in the Geroch-Held-Penrose (GHP) formalism at null infinity. In \cref{sec:other-charges}, we compare the WZ charge and flux formulae to some of the other charge and flux formulae that have been proposed for BMS symmetries, namely, the Ashtekar-Streubel flux formula, the Komar charge formulae and their linkage versions and Penrose's twistor charge formula. \cref{sec:2d-identities} collects some useful results on symmetric tracefree tensors on a \(2\)-sphere.\\

Our notation and conventions are as follows.
We follow the conventions of Wald \cite{Wald-book} for the metric signature, Riemann tensor, and differential forms, and use abstract index notation with Latin indices \(a,b,c,\ldots\) for tensor fields.
Quantities defined on the physical spacetime are denoted by a ``hat'', while the ones on the conformally-completed, unphysical spacetime are denoted without a ``hat'' (e.g., \(\hat g_{ab}\) is the physical metric while \(g_{ab}\) is the unphysical metric).
In addition, we  use ``$\hateq$'' to denote equality at null infinity and $\underleftarrow{}$ to denote the pullback to null infinity.
We also use both the indexed and index-free notations for differential forms, and in the index-free notation we denote differential forms in bold.
We use ``$\equiv$'' to translate between indexed and index-free notation, writing, e.g., $\df \varepsilon_4 \equiv \varepsilon_{abcd}$.
Since ``$\equiv$'' is used for this translation, we use ``$\defn$'' for definitions.

\section{Asymptotic-flatness at null infinity}
\label{sec:AF}

In this section, we recall the covariant definition of asymptotically-flat spacetimes and define the asymptotic fields and their equations at null infinity that will appear in our later analysis.

\begin{definition}[Asymptotic flatness] \label{def:asym-flat}
A \emph{physical} spacetime $(\hat{M},\hat{g}_{ab})$, which satisfies
the vacuum Einstein equation \(\hat G_{ab} = 0
\), is asymptotically-flat at null infinity if there exists an \emph{unphysical} spacetime $(M,g_{ab})$ with a boundary $\scri = \partial M$ and an embedding of \(\hat M\) into \(M\) (we use this embedding to identify \(\hat M\) as a submanifold of \(M\)),
such that
\begin{enumerate}
\item There exists  a smooth function $\Omega$ (the \emph{conformal factor}) on $M$ satisfying \(\Omega\hateq 0\) and \(\nabla_a\Omega \not\hateq 0\) such that \(g_{ab} = \Omega^2 \hat g_{ab} \) is smooth on \(M\) including at \(\scri\).
\item $\scri$ is topologically $\bb R \times \bb S^2$.
\item Defining $n_a := \nabla_a \Omega$,
 the vector field \(\omega^{-1} n^a\) is complete on \(\scri\)
 for any smooth function \(\omega\) on \(M\) such that \(\omega > 0\)
 on \(M\) and \(\nabla_a(\omega^4 n^a) \hateq 0\).
\end{enumerate}
\end{definition}
Detailed expositions on the motivations for this definition may be found in \cite{Geroch-asymp,Wald-book}. The differentiability conditions on the unphysical spacetime can be significantly weakened, but we restrict to the smooth case for simplicity.\\

Using the conformal transformation relating the unphysical Ricci tensor \(R_{ab}\) to the physical Ricci tensor \(\hat R_{ab}\) (see, e.g,
Appendix~D of \cite{Wald-book}), the vacuum Einstein equation can be written as
\be\label{eq:S-ee}
    S_{ab} = - 2 \Omega^{-1} \nabla_{(a} n_{b)} + \Omega^{-2} n^c n_c
    g_{ab}\,,
\ee
where \(S_{ab}\) is given by
\be\label{shoten}
    S_{ab} = R_{ab} - \tfrac{1}{6} R g_{ab}.
\ee
It follows from \cref{eq:S-ee} and the smoothness of \(\Omega\) and the unphysical metric \(g_{ab}\) at \(\scri\) that
\(n_a n^a \hateq 0\). This implies that \(\scri\) is a smooth null hypersurface in \(M\) with normal \(n_a = \nabla_a \Omega\) and that the vector field \(n^a = g^{ab} n_b\) is a null geodesic generator of \(\scri\).

Further, the Bianchi identity \(\nabla_{[a} R_{bc]de} = 0\) on the unphysical Riemann tensor along with \cref{eq:S-ee} gives the following equations for the unphysical Weyl tensor \(C_{abcd}\) (see \cite{Geroch-asymp} for details)
\begin{subequations}\label{eq:Bianchi-unphys}\begin{align}
    \nabla_{[a} (\Omega^{-1} C_{bc]de}) = 0 \label{eq:curl-weyl}\,.\\
    \nabla^d C_{abcd} = - \nabla_{[a} S_{b]c} \label{eq:Weyl-S}\,.
\end{align}\end{subequations}

Given a physical spacetime \((\hat M, \hat g_{ab})\) there is a lot of freedom in constructing its conformal-completion to obtain the unphysical spacetime \((M, g_{ab})\). Firstly, there is the choice of the embedding of \(\hat M\) into \(M\), but since this is simply used to identify points of \(\hat M\) with those of \(M\) this freedom is innocuous and one can always fix a choice of this embedding map. Another freedom is the choice of the conformal factor \(\Omega\). Note that for the same physical spacetime another choice of the conformal factor \(\Omega' = \omega \Omega\), with \(\omega\vert_\scri > 0\) also satisfies \cref{def:asym-flat} with a different unphysical metric \(g'_{ab} = \Omega'^2 \hat g_{ab} = \omega^2 g_{ab}\). Since we are interested in studying the asymptotic properties of the physical spacetime --- the conformal-completion is used only to bring the asymptotic boundary ``at infinity'' of the physical spacetime to a finite boundary \(\scri\) in the unphysical spacetime --- all physical quantities (such as symmetries, charges and fluxes) must be independent of the choice of conformal factor made.

All of our computations which follow can be done using an arbitrary conformal factor; however it is convenient to fix some of the conformal freedom by imposing the \emph{Bondi condition} as follows. On \(\scri\), let \(\Phi\) be defined by
\be\label{eq:Phi-defn}
    \Phi \defn \tfrac{1}{4} \nabla_a n^a\big\vert_\scri \,.
\ee
Under a change of conformal factor \(\Omega \mapsto \omega \Omega\) and \(g_{ab} \mapsto \omega^2 g_{ab}\) we have
\be\label{eq:Phi-conf-trans}
    \Phi \mapsto \omega^{-1} (\Phi + \Lie_n \ln \omega)\,.
\ee
Now, without loss of generality, we can choose $\omega$ to be a solution of \(\Phi + \Lie_n \ln \omega \hateq 0\) to set $\Phi \hateq 0$ \cite{Geroch-asymp,Wald-book}.\footnote{Note that while this condition can always be imposed on $\scri$, it cannot be imposed at spatial infinity where solutions, $\omega$, to the equation \(\Phi + \Lie_n \ln \omega \hateq 0\) diverge \cite{Wald-book}.} In this choice of the conformal factor, \cref{eq:S-ee} implies the
\emph{Bondi condition}
\be\label{eq:Bondi-cond}
	 \nabla_{a} n_{b}\hateq0,
\ee
and
\be \label{eq:nn-cond}
    n_{a}n^{a} = O(\Omega^{2}).
\ee
We will henceforth work in a conformal frame where the Bondi condition holds. Having imposed the Bondi condition, the remaining freedom in the conformal factor is of the form
\(\Omega \mapsto \omega \Omega\) where
\be\label{eq:conf-freedom}
    \omega\vert_\scri > 0 \eqsp \Lie_n \omega\hateq 0.
\ee
 We reiterate that the choice to work in a conformal frame where the Bondi condition holds is made purely for convenience and is not essential for the calculations in this paper. The main results of this paper, expressed in general conformal frames where the Bondi condition is not imposed, are included in \cref{sec:non-Bondi}. In the body of the paper, our statements about conformal invariance will pertain to conformal transformations that satisfy \cref{eq:conf-freedom}. More general conformal transformations for which $\Lie_{n} \omega \not\hateq 0$ will only be considered in \cref{sec:non-Bondi,sec:GHP}. 

Finally, let \(q_{ab}\) denote the pullback of \(g_{ab}\) to \(\scri\). This defines a degenerate metric on \(\scri\) such that
\be\label{eq:q-cond}
    q_{ab} n^b \hateq 0 \eqsp \Lie_n q_{ab} \hateq 0,
\ee
where the second condition follows from \cref{eq:Bondi-cond}. Thus, \(q_{ab}\) defines a Riemannian metric on the space of generators of \(\scri\) which is diffeomorphic to \(\bb S^2\).

\subsection{Auxiliary foliation and null normal at \(\scri\)}
\label{sec:aux}

For carrying out explicit computations on \(\scri\), it is convenient to pick an additional ``auxiliary'' structure on \(\scri\) given by a choice of foliation of \(\scri\) by a one-parameter family of \emph{cross-sections} which are diffeomorphic to \(\bb S^2\). Note that such a foliation always exists since \(\scri \cong \bb R \times \bb S^2\), though there is no unique choice for this foliation. The results in this paper can be obtained without reference to any choice of foliation, but it is far simpler to use some choice of foliation and then verify that the results are independent of the choice of foliation made.

For any choice of foliation, we obtain a \(1\)-form \(l_a\) on \(\scri\) which is normal to each cross-section of the foliation. Then there exists a \emph{unique} vector field \(l^a\) defined at \(\scri\) such that \(l_a \hateq g_{ab} l^b\) is the normal to the chosen foliation and
\begin{equation}
  l^a l_a \hateq 0 \eqsp l^a n_a \hateq -1 .
\end{equation}
We will call such a vector field associated with the chosen foliation the \emph{auxiliary normal}. Note that the vector field \(l^a\) can be extended arbitrarily away from \(\scri\), and our computations will be independent of which extension is chosen.

Using the auxiliary null normal $l^a$, we now define a tensor $Q_{ab}$ at \(\scri\) by
\begin{equation}\label{eq:Q-defn}
  Q_{ab} \defn g_{ab} + 2 l_{(a} n_{b)}.
\end{equation}
Note that $q_{ab} = \underleftarrow{Q}{}_{ab}$, and so $Q_{ab}$ is a choice of pushforward of the intrinsic (degenerate) metric on \(\scri\).
By the definition of \(Q_{ab}\), we have
\begin{equation}
  Q_{ab} l^b \hateq 0, \qquad Q_{ab} n^b \hateq 0.
\end{equation}
We also define the symmetric trace-free part of a tensor with respect to $Q_{ab}$ by:
\begin{equation}\label{eq:STF-defn}
  \STF A_{ab} \defn \left(Q_{(a}{}^c Q_{b)}{}^d - \tfrac{1}{2} Q_{ab} Q^{cd}\right) A_{cd}.
\end{equation}

Next, we define a volume form on $\scri$ and on the cross-sections of the chosen foliation by
\be \label{eqn:epsilon_32}
  \varepsilon_{abc} \defn l^d \varepsilon_{dabc} \eqsp \varepsilon_{ab} \defn \varepsilon_{abcd} l^c n^d \hateq \varepsilon_{abc} n^c.
\ee

\begin{remark}[Orientation conventions]
\label{rem:orientation}

The orientation of these volume forms are such that $\scri$ is outward-facing as a manifold in $M$, and such that the cross-sections have a future orientation within $\scri$.
Note that this convention for the orientations for $\df \varepsilon_2$ and $\df \varepsilon_3$ is somewhat unexpected.
In Minkowski spacetime, setting $l_a = -\nabla_a u$ and $n_a = \nabla_a \Omega$ (with $\Omega = 1/r$ and $u = t - r$), we have that $\df \varepsilon_3 = - \sin \theta\, d u \wedge d \theta \wedge d \phi$ and $\df \varepsilon_2 = - \sin \theta\, d \theta \wedge d \phi$, which is the opposite of the ``expected'' sign for $\df \varepsilon_2$.
Despite this sign difference, we still have that
\begin{equation}
  \int_S \df \varepsilon_2 > 0;
\end{equation}
this, in fact, is the \emph{definition} of an orientation.
In Minkowski, this can be seen by noting that this means that the bounds of the integrals are in the opposite order (say, the $\theta$ integral is from $\pi$ to $0$, instead of $0$ to $\pi$).

Finally, note that, in our choice of orientation, if we consider a cross-section $S$ as a limit of spheres $S'$ within a spacelike hypersurface $\Sigma$ in the unphysical spacetime, then the orientation for $S'$ that is compatible with that of $S$ is the \emph{inward}-facing one within $\Sigma$.
\end{remark}

The \emph{shear} of the auxiliary normal on \(\scri\) is defined by
\be\label{eq:shear-defn}
    \sigma_{ab} \hateq \STF \nabla_a l_b \hateq \tfrac{1}{2} \STF \Lie_l g_{ab} \, .
\ee
The twist \(\varepsilon^{ab}\nabla_a l_b\) vanishes on \(\scri\) since \(l_a\) is normal to cross-sections of a foliation of \(\scri\). We will mostly not need the expansion of \(l_a\) in our analysis (it is introduced in \cref{sec:CGNC} where needed).

The change in \(l_a\) along the null generators of \(\scri\) is given by
\begin{equation} \label{eq:tau-defn}
    \tau_a \hateq Q_a{}^b n^c \nabla_c l_b.
\end{equation}
By the normalization conditions for $l^a$ and $n^a$ and the Bondi condition we have
\begin{equation} \label{eqn:lie_n_l}
  \tau_a \hateq n^b \nabla_b l_a \hateq \Lie_n l_a,
\end{equation}

We also can define a ``sphere derivative'' along the cross-sections for any tensor field that is orthogonal to $l^a$ and $n^a$ on all indices (at $\scri$):
\begin{equation}
  \mathscr{D}_a T^{b_1 \cdots b_r}{}_{c_1 \cdots c_s} \defn Q^d{}_a Q^{b_1}{}_{e_1} \cdots Q^{b_r}{}_{e_r} Q^{f_1}{}_{c_1} \cdots Q^{f_s}{}_{c_s} \nabla_d T^{e_1 \cdots e_r}{}_{f_1 \cdots f_s}.
\end{equation}
One can easily show that this derivative operator is compatible with both $Q_{ab}$ and $\varepsilon_{ab}$ at $\scri$:
\begin{equation}
  \mathscr{D}_a Q_{bc} \hateq 0, \qquad \mathscr{D}_a \varepsilon_{bc} \hateq 0.
\end{equation}

With our choice of $\df \varepsilon_3$ and $\df \varepsilon_2$, we can derive several useful formulas from Stokes' theorem.
Consider some portion $\Delta \scri$ of $\scri$ whose boundary is given by two cross-sections $S_1$ and $S_2$ in our foliation, with $S_2$ to the future of $S_1$.
Then, for any scalar $\alpha$ we have
\be\label{eq:IBP-Lie}
    \int_{\Delta \scri} \df \varepsilon_3\; \Lie_n \alpha &= \int_{S_2} \df \varepsilon_2\; \alpha - \int_{S_1} \df \varepsilon_2~ \alpha \,.
\ee
Also, for any vector $v^a$ which is orthogonal to both \(n_a\) and \(l_a\) we have
\be\label{eq:IBP-D3}
    \int_{\Delta \scri} \df \varepsilon_3~ \nabla_a v^a \hateq \int_{\Delta \scri} \df \varepsilon_3~ (\ms D_a + \tau_a) v^a &\hateq 0,
\ee
where the second expression is obtained using the Bondi condition along with \cref{eq:Q-defn,eqn:lie_n_l}, and the vanishing of the expression follows from the fact that \(\df\varepsilon_3 \nabla_a v^a\) is an exact \(3\)-form and \(l_a v^a \hateq 0\). Similarly, for any cross-section \(S\) of \(\scri\) we also have
\be\label{eq:IBP-D2}
  \int_S \df \varepsilon_2\; \ms D_a v^a &\hateq 0 \,. 
\ee
We note that \cref{eq:IBP-D3} was erroneously given in Eq.~2.13 of \cite{KP-GR-match} as an integration-by-parts identity on the sphere, instead of $\Delta \scri$ (all instances when \cref{eq:IBP-D3} was used in \cite{KP-GR-match}, however, were correct, as they were applied to $\Delta \scri$).\\

The final quantities that we will need are components of the Weyl tensor at \(\scri\). By the peeling theorem, we have \(C_{abcd} \hateq 0\) at \(\scri\), and thus \(\Omega^{-1}C_{abcd}\) admits a limit to \(\scri\) (see Theorem~11 of \cite{Geroch-asymp}). In the choice of a foliation of \(\scri\) we define the Weyl tensor fields
\begin{subequations} \label{eq:weyl-defn}
  \begin{align}
    \mc R_{ab} &\defn (\Omega^{-1} C_{cdef}) Q_a{}^c n^d Q_b{}^e n^f, &\mc S_a &\defn (\Omega^{-1} C_{cdef}) l^c n^d Q_a{}^e n^f, \\
    \mc P &\defn (\Omega^{-1} C_{cdef}) l^c n^d l^e n^f, &\mc P^* &\defn \tfrac{1}{2} (\Omega^{-1}C_{cdef}) l^c n^d \varepsilon^{ef}, \\
    \mc J_a &\defn (\Omega^{-1}C_{cdef}) n^c l^d Q_a{}^e l^f, &\mc I_{ab} &\defn (\Omega^{-1} C_{cdef}) Q_a{}^c l^d Q_b{}^e l^f. \label{eq:non-peeled}
  \end{align}
\end{subequations}
Note that, due to the symmetries of the Weyl tensor, $\mc R_{ab}$ and $\mc I_{ab}$ are symmetric and traceless.
The relation of these tensors to the Weyl scalar components in the Newman-Penrose notation is given in \cref{eq:weyl-GHP}.

For the fields defined in \cref{eq:weyl-defn}, \cref{eq:curl-weyl} implies the following evolution equations along $\scri$, which can be verified to be conformally-invariant (\cref{eq:weyl-evol-GHP} in the GHP formalism):
\begin{subequations} \label{eq:weyl-evol}
  \begin{align}
    \Lie_n \mc S_a &= (\ms D^b + \tau^b) \mc R_{ab}, \\
    \Lie_n \mc P~ &= (\ms D^a + 2 \tau^a) \mc S_a - \sigma^{ab} \mc R_{ab}, \label{eq:lie-P} \\
     \Lie_n \mc P^* &= -\varepsilon^{ab} (\ms D_a + 2 \tau_a) \mc S_b + \varepsilon_b{}^c \sigma^{ab} \mc R_{ac}, \\
    \Lie_n \mc J_a & = \tfrac{1}{2} (\ms D_b + 3\tau_b ) (Q_a{}^b \mc P - \varepsilon_a{}^b \mc P^*) - 2 \sigma_a{}^b \mc S_b, \label{eq:lie-J} \\
    \Lie_n \mc I_{ab} & =   \STF  (\ms D_a + 4 \tau_a) \mc J_b - \tfrac{3}{2} \sigma_{ac} (Q_b{}^c \mc P - \varepsilon_b{}^c \mc P^*).
  \end{align}
\end{subequations}

Next, we list the conformal transformation weights of these various quantities. Any quantity $\alpha$ is said to have \emph{conformal weight} \(w\) if, under $\Omega \mapsto \omega \Omega$, it transforms as $\alpha \mapsto \omega^w \alpha$.
First, we clearly have that
\begin{equation}
  g_{ab} : w = 2, \qquad \varepsilon_{abcd} : w = 4.
\end{equation}
Next, while in the spacetime
\begin{equation} \label{eqn:n_conf}
  n_a \mapsto \omega n_a + \Omega \nabla_a \omega,
\end{equation}
we therefore have that $n_a : w = 1$ on $\scri$ itself, and so
\begin{equation} \label{eq:fol-conf-wts}
  l_a : w = 1, \qquad (Q_{ab}, \varepsilon_{ab}) : w = 2, \qquad \varepsilon_{abc} : w = 3.
\end{equation}
Moreover, we have that
\begin{equation} \label{eq:shear-conf-wt}
  \sigma_{ab} : w = 1,
\end{equation}
while, under a conformal transformation, $\tau_a$ transforms as
\begin{equation} \label{eq:tau-tr}
  \tau_a \mapsto \tau_a + \mathscr{D}_a \ln \omega.
\end{equation}
Finally, the fields constructed from the Weyl tensor have the following conformal weights:
\begin{equation} \label{eq:c-weights}
  (\mc R_{ab}, \mc I_{ab}) : w = -1 \eqsp (\mc S_a, \mc J_a) : w = -2 \eqsp (\mc P, \mc P^*) : w = -3.
\end{equation}

Finally, we remark that there is a particularly convenient choice of conformal factor and foliation, namely one where $q_{ab}$ is given by the unit 2-sphere metric, and $\tau^a \hateq 0$.
This is the \emph{Bondi frame}, and in particular it is a restriction of the remaining conformal freedom $\Omega \mapsto \omega \Omega$ that is more restrictive than the Bondi condition. Even though Bondi frame can always be achieved by a conformal transformation and a change of foliation, we do not enforce Bondi frame, since we want to consider charges on arbitrary cross-sections of $\scri$ (and fluxes between these cross-sections) and want our expressions to be manifestly conformally invariant (apart from the choice of Bondi condition).

\subsection{News tensor}
\label{sec:News}

In this section, we define the \emph{News tensor}, which characterizes the radiative degrees of freedom of the gravitational field at null infinity.
This tensor can be defined in terms of a foliation, or invariantly from the universal structure that exists on $\scri$; we review both definitions and show that they are equal.
Moreover, we review how the News tensor vanishes in asymptotically stationary regions of $\scri$, which motivates its usage as a characterization of radiation.

We define the News tensor as the (projected) Lie derivative along \(n^a\) of the shear $\sigma_{ab}$:
\begin{equation} \label{eqn:news_l_def}
  N_{ab} \defn 2 Q_a{}^c Q_b{}^d \Lie_n \sigma_{cd} = 2 \STF \Lie_n \sigma_{ab } \, .
\end{equation}
Note that it follows from the above definition that
\be\label{eq:News-ortho-STF}
    N_{ab}n^b \hateq g^{ab}N_{ab} \hateq Q^{ab} N_{ab} \hateq 0\,.
\ee
It is straightforward to verify that the News is conformally invariant.

We can also write the News $N_{ab}$ in terms of $S_{ab}$ as follows.
Consider the quantity $\STF \Lie_n \Lie_l g_{ab}$. Despite involving covariant derivatives of $l_a$, when evaluated at $\scri$, this quantity is independent of the choice of \(l_a\) away from \(\scri\).  By choosing a particular extension of $l_a$, one can show from \cref{eq:shear-defn,eqn:lie_n_l} that
\begin{equation} \label{eqn:lie_nl_g}
  \STF \Lie_n \Lie_l g_{ab} \hateq 2 \STF (\Lie_n \sigma_{ab} - \tau_a \tau_b) \hateq N_{ab} - 2 \STF (\tau_a \tau_b),
\end{equation}
where the second equality follows from \cref{eqn:news_l_def}.
Using $[\Lie_n, \Lie_l] = \Lie_{[n, l]}$, together with \cref{eqn:lie_n_l}, we therefore find that
\begin{equation} \label{eq:news_lie_ln_g}
  N_{ab} \hateq \STF [\Lie_l \Lie_n g_{ab} + 2 (\mathscr{D}_a + \tau_a) \tau_b].
\end{equation}
Using \cref{eq:S-ee}, we therefore find that
\begin{equation} \label{eq:news-Schouten}
  N_{ab} \hateq \STF [S_{ab} + 2 (\mathscr{D}_a + \tau_a) \tau_b].
\end{equation}

Further, from \cref{eq:Weyl-S,eq:news-Schouten}, the News is related to the Weyl tensor components (defined in \cref{eq:weyl-defn}) by:
\begin{equation} \label{eq:weyl-News}
  Q_a{}^c Q_b{}^d \Lie_n N_{cd} \hateq 2 \mc R_{ab} \eqsp \ms D^b N_{ab} \hateq 2 \mc S_a \,,
\end{equation}
and
\begin{equation} \label{eq:Pstar-identity}
  \mc P^* = \varepsilon^{ab} \lb[ \ms D_a (\ms D_c - \tau_c) \sigma_b{}^c - \tfrac{1}{2} N_{ac} \sigma_b{}^c \rb]\,.
\end{equation}
In the GHP notation these relations can be found in \cref{eq:weyl-News-conf,eq:Pstar-identity-GHP}.\\

As defined by \cref{eqn:news_l_def}, it is not obvious that the News tensor is independent of our choice of foliation. However, we show below that this definition of the News coincides with the covariant definition given by Geroch \cite{Geroch-asymp} which makes no reference to any foliation of \(\scri\). While most of the literature exclusively uses only one of these two definitions, we will find it convenient to use either interchangeably, and (since there does not appear to be any proof that we could find), we will now show that these two definitions yield the same tensor.

Consider the \emph{Geroch News tensor} defined in \cite{Geroch-asymp} as
\begin{equation} \label{eqn:news_geroch_def}
  N_{ab} \defn \pb{S}_{ab} - \rho_{ab},
\end{equation}
where $\pb{S}_{ab}$ denotes the pullback of \(S_{ab}\) to $\scri$ and $\rho_{ab}$ is the \emph{unique} symmetric tensor field on $\scri$ constructed from the intrinsic universal structure on \(\scri\) defined in Theorem~5 of \cite{Geroch-asymp}.

First note that the Geroch News tensor is conformally invariant, and satisfies the conditions \cref{eq:News-ortho-STF} (see \cite{Geroch-asymp}). Moreover, projections of Eq.~68 of \cite{Geroch-asymp} show that \cref{eq:weyl-News} also holds for the Geroch News tensor. Thus, if \(\lambda_{ab}\) is the difference of the Geroch News tensor and the one defined in \cref{eqn:news_l_def}, then \(\lambda_{ab}\) is a symmetric tensor field on \(\scri\) which satisfies
\begin{equation}
    \lambda_{ab} n^b \hateq Q^{ab} \lambda_{ab} \hateq Q_a{}^c Q_b{}^d \Lie_n \lambda_{cd} \hateq \ms D^b \lambda_{ab} \hateq 0 \,.
\end{equation}
Thus, \(\lambda_{ab}\) is a tensor field on $S$ that is symmetric, traceless and divergence-free, and therefore vanishes by \cref{prop:div_STF}. Consequently, the News tensor defined in \cref{eqn:news_l_def} is equivalent to the covariant definition by Geroch.

Finally, we review the key property of the News tensor, namely that it characterizes the presence of gravitational radiation in a spacetime.
This can be seen by the following result, which is due to Geroch (see pp. 53-54 of \cite{Geroch-asymp}): consider an asymptotically flat spacetime $(\hat M, \hat g_{ab})$, and some portion $\Delta \scri$ of null infinity.
If $\Delta \scri$ is asymptotically stationary, in the sense that there exists a vector field $t^a$ in a neighborhood of $\Delta \scri$ that is a timelike Killing vector with respect to $\hat g_{ab}$, then $N_{ab} \hateq 0$ on $\Delta \scri$.
Since any notion of gravitational radiation should vanish in asymptotically stationary regions of null infinity, this motivates the News tensor as indicating the presence of radiation.
It should be noted, however, that it is not known whether the converse of this statement is true, namely that all regions $\Delta \scri$ where the News tensor vanishes are asymptotically stationary.

\section{Universal structure and metric perturbations}
\label{sec:univ-str}

In this section, we summarize the \textit{universal structure} at null infinity. This is the structure that is common to the conformal completion of \emph{all} spacetimes that satisfy the definition of asymptotic flatness given in \cref{def:asym-flat} and is thus independent of the specific physical spacetime under consideration.\\

If \((M, g_{ab}, \Omega)\) and \((M', g'_{ab}, \Omega')\) are the unphysical spacetimes corresponding to \emph{any} two asymptotically-flat physical spacetimes then, a priori, \(M\) and \(M'\) are distinct manifolds each with their own boundary \(\scri\) and \(\scri'\). However, we argue below that there exists a smooth diffeomorphism from a neighbourhood of \(\scri\) in \(M\) to a neighbourhood of \(\scri'\) in \(M'\) which can be used to identify these unphysical spacetimes (in this neighbourhood) and which maps \(\scri\) to \(\scri'\). Since we are only interested in the the asymptotic properties near null infinity, we can work with just one manifold \(M\) and one null boundary \(\scri\) to represent null infinity for any two (and thus, all) asymptotically-flat spacetimes. Further, \emph{without any loss of generality}, this diffeomorphism can also be chosen so that \(\Omega' = \Omega\) in a neighborhood of \(\scri\) and \(g'_{ab} \hateq g_{ab}\).\\

 This can be achieved by setting up a suitable, geometrically-defined coordinate system in a neighborhood of \(\scri\) and identifying the two unphysical spacetimes in these coordinates, as we now explain (see also the argument on p.~22 in \cite{Geroch-asymp}). On the null infinity $\scri$ of any asymptotically-flat spacetime, we define a parameter $u$ along the null generators of $\scri$ such that \(n^a\nabla_a u \hateq 1\). We then pick some cross-section \(S_0 \cong \mathbb{S}^{2}\) with constant \(u = u_0\). On \(S_0\) define a coordinate system \(x^A\) (with \(A = 1,2\)), and extend these coordinates to all of \(\scri\) by parallel transport along \(n^a\):
\be \label{eq:x-ext-eq}
 n^a \nabla_a x^A \hateq 0\,.
 \ee
This gives us a coordinate system \( (u, x^A)\) on \(\scri\).
Next, since $\Omega\hateq 0$ and \(n_a \hateq \nabla_a \Omega \not\hateq 0\), we can use \(\Omega\) as a coordinate transverse to \(\scri\). As discussed above, there is considerable freedom in the choice of the conformal factor at \(\scri\) which we need to fix to specify the choice of coordinate used. First, as before, we pick the conformal factor so that the Bondi condition (\cref{eq:Bondi-cond}) is satisfied, which leaves us the freedom to change the conformal factor on the cross-sections of \(\scri\). To fix this freedom we proceed as follows. Consider the induced metric \(q_{ab}\) on the cross-section \(S_0\) chosen above. It follows from the \emph{uniformization theorem} (for instance see Ch.~8 of \cite{Bieri-CK-ext0})\footnote{The uniformization theorem is a \emph{global} result depending on the topology of the \(2\)-dimensional space. Locally, all metrics of a particular signature on a \(2\)-surface are conformally-equivalent, Problem~2, Ch.~3 of \cite{Wald-book}.} that any metric on \(S_0\) is conformal to the unit round metric on \(\bb S^2\) (that is, the metric with constant Ricci scalar equaling \(2\)). Thus, we can always choose the conformal factor so that the metric \(q_{ab}\) on this cross-section \(S_0\) is also the unit round metric of \(\bb S^2\) and from \cref{eq:q-cond}, this holds on any cross-section. Thus, we choose as our transverse coordinate the choice of conformal factor \(\Omega\) which satisfies the Bondi condition and makes the metric on the cross-sections of \(\scri\) to be the unit round metric on \(\bb S^2\). This gives us a coordinate system \((\Omega, u, x^A)\) at \(\scri\) for \emph{any} asymptotically-flat spacetime. Since this construction can be done for any asymptotically-flat spacetime, we can, without any loss of generality, identify the null infinities of all asymptotically-flat spacetimes by identifying their points in these coordinates. This shows that there exists a diffeomorphism between neighbourhoods of null infinities of different spacetimes such that we can identify their boundaries with one ``abstract'' manifold \(\scri\) and also that we can choose the same conformal factor \(\Omega\) in this neighbourhood.

Next, we show that the unphysical metric at \(\scri\) can be chosen to be the same for the conformal completion of any physical spacetime. Consider a foliation of \(\scri\) by cross-sections \(S_u\) of constant \(u\). Then, by \cref{eq:x-ext-eq} the null generator \(n^a\), in these coordinates, can be written as \(n^a \hateq \partial/\partial u\). The \(1\)-form on \(\scri\) normal to this foliation is given by $l_{a} \hateq -\nabla_{a} u$. Using $n_{a} = \nabla_{a} \Omega$, $l_{a} \hateq -\nabla_{a} u$ and  $n^{a} l_{a} \hateq - 1$ , we obtain the following expression for the line element of the unphysical metric on $\scri$ in the coordinate system \((\Omega,u,x^A)\)
\be\label{eq:scri-metric}
    ds^2 \hateq 2 d\Omega du + s_{AB} dx^A dx^B \,,
\ee
where \(s_{AB}\) is the unit round metric on \(\bb S^2\) in the chosen coordinates \(x^A\).\footnote{The precise choice of coordinates \(x^A\) on the cross-sections is irrelevant; one could pick polar coordinates \(x^A = (\theta,\phi)\) to put the unit round sphere metric in the standard form \(s_{AB} dx^{A} dx^{B} = d\theta^{2} + \sin^{2} \theta d\phi^{2}\), but any other coordinate system is just as good.} Note that the form of the unphysical metric \cref{eq:scri-metric} is completely independent of which physical spacetime is under consideration, that is, \emph{all} asymptotically-flat spacetimes have the same universal unphysical metric at null infinity. Different choices of the physical spacetime are only reflected in the unphysical metric \emph{away} from \(\scri\). Note that the definition of asymptotic-flatness (\cref{def:asym-flat}) includes an embedding map from the physical spacetime manifold \(\hat M\) into the unphysical manifold \(M\). The existence of the universal structure at \(\scri\) described above implies that one can embed \emph{any} physical spacetime into an unphysical spacetime by identifying some physical spacetime coordinates with the coordinates \((\Omega,u, x^A)\) constructed above so that the unphysical metric \(g_{ab}\) takes the form \cref{eq:scri-metric}. In the conformal-completion formalism this is the version of the statement that ``asymptotically-flat spacetimes behave like the Minkowski spacetime to leading order at infinity'' and the difference between two physical spacetimes only shows up at ``sub-leading order''.\\

Note that since the manifold \(\scri\) is universal, the choice of the foliation can also be made independently of the physical spacetime. It follows that the auxiliary normal $l_{a} \hateq -\nabla_{a} u$ is also universal, and (from \cref{eq:scri-metric}) we also have \(l^{a} \hateq - \partial/\partial \Omega\) which is the auxiliary null normal and is also universal.

Many of the choices made in constructing the coordinates at \(\scri\) are irrelevant to this argument and are made just for convenience. For instance, the choice of the unit-metric \(s_{AB}\) is irrelevant. In \emph{any} asymptotically-flat spacetime, we can instead choose the freedom in the conformal factor at \(\scri\) so that the induced metric on the cross-sections \(q_{ab} = q^{(0)}_{ab}\) where \(q^{(0)}_{ab}\) is \emph{any fixed} metric on \(\bb S^2\). Similarly, one could have chosen a different foliation of \(\scri\) if one wishes. These choices simply correspond to the freedom of choosing the embedding map from the physical spacetime into the unphysical spacetime. Then the rest of the construction proceeds as before and \(g_{ab}\vert_\scri\) is universal.

\subsection{Metric perturbations near \(\scri\)}
\label{sec:metric-pert}

Next we consider linearized perturbations of the physical metric and derive the conditions on the perturbations arising from requiring asymptotic flatness. We consider a one-parameter family of physical metrics $\hat{g}_{ab} (\lambda)$ with \(\hat g_{ab} = \hat g_{ab}(\lambda=0)\) being any chosen background spacetime metric, and define the physical metric perturbation by
\be
    \delta \hat g_{ab} \defn \lb. \frac{d}{d\lambda} \hat g_{ab}(\lambda) \rb\vert_{\lambda = 0}\,.
\ee
We also use the notation \(\delta\) to denote perturbations of other quantities defined in a similar way as above. We emphasize that the \(\delta\) denotes changes of quantities when the physical metric is varied; the quantities appearing in the universal structure described above including the unphysical metric at \(\scri\), the conformal factor and the choice of foliation do not vary under \(\delta\). Note, this does not mean that the conformal factor and foliation cannot be changed --- our expressions for the charges and fluxes are independent of the conformal factor and foliation --- it means that these quantities can always be held fixed when the physical metric is varied.

Taking the one-parameter family of physical metrics \(\hat g_{ab}(\lambda)\) to all be asymptotically-flat, we want to consider the behaviour of the metric perturbations at null infinity. Let \(g_{ab}(\lambda)\) be the one-parameter family of unphysical metrics obtained by the conformal completion of \(\hat g_{ab}(\lambda)\). As discussed above, without loss of generality we can take all the unphysical metrics \(g_{ab}(\lambda)\) to be defined on the same manifold \(M\), with a common boundary \(\scri\) describing null infinity. Further, the conformal factor \(\Omega\) can also be chosen to be independent of the parameter \(\lambda\). Thus, we get
\begin{equation}
  g_{ab} (\lambda) \defn \Omega^2 \hat{g}_{ab} (\lambda) \eqsp \delta g_{ab} = \Omega^2 \delta \hat g_{ab}\,,
\end{equation}
where \(\delta g_{ab}\) is the perturbation of the unphysical metric.
Moreover, since the unphysical metric at $\scri$ is universal, we have that \(\delta g_{ab} \hateq 0\) and thus
\begin{equation}
  \delta g_{ab} \hateq \Omega \gamma_{ab},
\end{equation}
for some $\gamma_{ab}$ which is smooth at \(\scri\).

Since the conformal factor is chosen to satisfy the Bondi condition \cref{eq:Bondi-cond} in \emph{any} spacetime, varying the Bondi condition we get
\be
    \delta (\nabla_a n_b) \hateq 0 \implies \gamma_{ab}n^b \hateq 0\,.
\ee
Thus, since \(\gamma_{ab}\) is smooth at \(\scri\), there exists a smooth \(\gamma_a\) such that
\be
    \gamma_{ab}n^b = \Omega \gamma_a\,.
\ee
Thus, the perturbations \(\delta g_{ab}\) of the unphysical metric are given by the tensor fields \(\gamma_{ab}\) and \(\gamma_a\) with
\be\label{eq:pert-conds}
    \gamma_{ab} = \Omega^{-1} \delta g_{ab} \eqsp \gamma_a = \Omega^{-2} \delta g_{ab} n^b = \Omega^{-1} \gamma_{ab} n^b\,.
\ee\\
These tensor fields are constrained by the linearized vacuum Einstein equations.
A particularly important component of these equations is given by \cite{WZ}
\begin{equation}
  \nabla^b \gamma_{ab} - 3 \gamma_a - \nabla_a \gamma^b{}_b \hateq 0.
\end{equation}

Further, using the auxiliary foliation of \(\scri\) we can relate the unphysical metric perturbation characterized by \(\gamma_{ab}\) to the perturbation of the shear \(\sigma_{ab}\) as follows. Varying the definition of the shear (\cref{eq:shear-defn}), noting that the foliation is kept fixed and that \(\delta g_{ab} \hateq 0\), from \cref{eq:pert-conds} it is straightforward to compute that
\be \label{eqn:stf_gamma}
    \delta \sigma_{ab} = - \tfrac{1}{2} \STF \gamma_{ab}.
\ee
This equation shows that, although $\gamma_{ab}$ is a subleading quantity, its value is determined by the variation of an object that is constructed from quantities at $\scri$: it only depends on quantities off of $\scri$ through the dependence of $\sigma_{ab}$ on $\nabla_a$.

\section{Asymptotic symmetries at \(\scri\): the BMS Lie algebra}
\label{sec:bms}

Consider a smooth vector field $\xi^a$ in the physical spacetime \(\hat M\), and let \(\hat g_{ab}(\lambda)\) be the one-parameter family of physical metrics generated by diffeomorphims along \(\xi^a\). The physical metric perturbation corresponding to this family is given by
\begin{equation}
  \delta_\xi \hat{g}_{ab} = \Lie_\xi \hat{g}_{ab}.
\end{equation}
The corresponding perturbation of the unphysical metric is
\begin{equation} \label{eqn:delta_xi_g}
  \delta_\xi g_{ab} = \Omega^2 \Lie_\xi \hat{g}_{ab} = \Lie_\xi g_{ab} - 2 \Omega^{-1} \xi^c n_c g_{ab}.
\end{equation}
For \(\xi^a\) to be an asymptotic symmetry the infinitesimal diffeomorphism generated by \(\xi^a\) must preserve the universal structure discussed in \cref{sec:univ-str}. We now obtain the conditions on \(\xi^a\) for it to be an asymptotic symmetry.

Firstly, since the unphysical spacetime \(M\) is smooth up to and including \(\scri\), \(\xi^a\) must extend to a smooth vector field on \(M\) including at \(\scri\). Secondly, since the unphysical metric is smooth at \(\scri\), the perturbation \(\delta_\xi g_{ab}\) in \cref{eqn:delta_xi_g} is also smooth at \(\scri\). This condition implies that
\begin{equation}
  \xi^a n_a \hateq 0.
\end{equation}
That is, as expected, an asymptotic symmetry $\xi^a$ must be tangent to $\scri$ and thus preserves the asymptotic boundary. For convenience we define
\be\label{eq:alpha-xi-defn}
  \alpha_{(\xi)} \defn \Omega^{-1} \xi^a n_a\,,
\ee
which is smooth at \(\scri\).

Next, $\delta_\xi$ preserves the universal structure, which implies that (\cref{eq:pert-conds})
\begin{equation} \label{eqn:gamma_xi_def}
  \gamma_{ab}^{(\xi)} \defn \Omega^{-1} \delta_\xi g_{ab} \eqsp \gamma_a^{(\xi)} \defn \Omega^{-1} \gamma_{ab}^{(\xi)} n^b\,,
\end{equation}
must be smooth at $\scri$. Using \cref{eqn:delta_xi_g,eq:alpha-xi-defn}, the smoothness of $\gamma_{ab}^{(\xi)}$ implies
\begin{equation} \label{eqn:lie_xi_g}
  \Lie_\xi g_{ab} \hateq 2 \alpha_{(\xi)} g_{ab},
\end{equation}
while, the smoothness of $\gamma_a^{(\xi)}$ implies
\begin{equation} \label{eqn:lie_xi_n}
  \Lie_\xi n^a \hateq -\alpha_{(\xi)} n^a \eqsp \Lie_n \alpha_{(\xi)} \hateq 0.
\end{equation}
The pullback of \cref{eqn:lie_xi_g} gives
\begin{equation} \label{eqn:lie_xi_q}
  \Lie_\xi q_{ab} \hateq 2 \alpha_{(\xi)} q_{ab}.
\end{equation}
Intrinsically on \(\scri\), an asymptotic symmetry is, thus, given by a vector field \(\xi^a\) tangent to \(\scri\) which satisfies \cref{eqn:lie_xi_n,eqn:lie_xi_q} for some smooth function \(\alpha_{(\xi)}\). Such vector fields are the BMS symmetries on null infinity.

In any choice of foliation of \(\scri\) we can further characterize the BMS symmetries as follows. Since $\xi^a$ is tangent to \(\scri\), we can write
\begin{equation}\label{eq:xi-decomp}
  \xi^a \hateq \beta n^a + X^a,
\end{equation}
where
\begin{equation} \label{eqn:beta_X_def}
  \beta \defn -l_a \xi^a, \qquad X^a \defn Q^a{}_b \xi^b,
\end{equation}
We note their conformal weights:
\be\label{eq:beta_X_conf-wt}
    \beta : w = 1 \eqsp X^a : w = 0 \,.
\ee
That is, \(\beta\) is a smooth function of conformal weight \(1\) and \(X^a\) is a smooth vector field tangent to the cross-sections of the chosen foliation on \(\scri\).
The conditions on $\beta$ and $X^a$ follow from \cref{eqn:lie_xi_n,eqn:lie_xi_g} as we derive next. Note that \cref{eqn:lie_xi_n,eqn:lie_xi_g} only depend on the vector field \(\xi^a\) at \(\scri\) and are independent of how this vector field is extended away from \(\scri\).

The only non-trivial component of \cref{eqn:lie_xi_g} is given by its projection tangent to the cross-sections of the foliation which gives
\begin{equation}
  Q_a{}^c Q_b{}^d \Lie_\xi g_{cd} \hateq 2 \mathscr{D}_{(a} X_{b)} \hateq 2 \alpha_{(\xi)} Q_{ab},
\end{equation}
which shows both that
\begin{equation}
  \STF \mathscr{D}_a X_b \hateq 0,
\end{equation}
and that
\begin{equation}
  \alpha_{(\xi)} \hateq \tfrac{1}{2} \mathscr{D}_a X^a.
\end{equation}
Next, we consider \cref{eqn:lie_xi_n} on $\scri$ using \cref{eq:xi-decomp} to get
\begin{equation} \label{eqn:lie_xi_decomp_n}
  \Lie_\xi n^a = -\Lie_n \xi^a \hateq -n^a \Lie_n \beta - \Lie_n X^a \hateq -\alpha_{(\xi)} n^a.
\end{equation}
Projecting along the cross-sections we get
\begin{equation}
  Q^a{}_b \Lie_n X^b \hateq 0,
\end{equation}
whereas $X^a l_a \hateq 0$, the Bondi condition, and \cref{eqn:lie_n_l} imply that
\begin{equation}
  l_a \Lie_n X^a \hateq -X_a \Lie_n l^a \hateq -X_a \tau^a, \qquad n_a \Lie_n X^a \hateq 0,
\end{equation}
so that
\begin{equation} \label{eqn:lie_n_X_step}
  \Lie_n X^a \hateq n^a X_b \tau^b.
\end{equation}
Contracting \cref{eqn:lie_xi_decomp_n} with $l_a$, we find that
\begin{equation} \label{eqn:lie_n_beta_step}
  \Lie_n \beta + l_a \Lie_X n^a \hateq \Lie_n \beta - X_a \tau^a \hateq \alpha_{(\xi)},
\end{equation}
where the second equality follows from \cref{eqn:lie_n_X_step}.
As such, we find that \cref{eqn:lie_n_beta_step} becomes
\begin{equation}
  \Lie_n \beta \hateq \alpha_{(\xi)} - X_a \tau^a \hateq \tfrac{1}{2} (\mathscr{D}_a - 2 \tau_a) X^a.
\end{equation}
In summary, in a chosen foliation of \(\scri\) any BMS symmetry can be written as \(\xi^a \hateq \beta n^a + X^a\) where \(X^a\) is tangent to the cross-sections of the foliation and the following conditions are satisfied:
\begin{subequations} \label{eq:bms-decomp-conds}
  \begin{align}
    \Lie_n \beta &\hateq \tfrac{1}{2} (\mathscr{D}_a - 2 \tau_a) X^a, \label{eqn:lie_n_beta} \\
    \Lie_n X^a &\hateq n^a X_b \tau^b, \label{eqn:lie_n_X} \\
    \STF \mathscr{D}_a X_b &\hateq 0, \label{eqn:stf_dX} \\
    \alpha_{(\xi)} &\hateq \tfrac{1}{2} \mathscr{D}_a X^a. \label{eqn:div_X}
  \end{align}
\end{subequations}
Note that, using \cref{eq:beta_X_conf-wt,eqn:n_conf,eq:tau-tr,eq:alpha-xi-defn,eq:conf-freedom}, one can show that these equations are invariant under conformal transformations that preserve Bondi condition, that is, with $\Lie_n \omega \hateq 0$ (for the form of these equations when Bondi condition does not hold, see \cref{eq:bms-decomp-conds-Phi}).

Let \(\xi^a_1 \hateq \beta_1 n^a + X^a_1\) and \(\xi^a_2 \hateq \beta_2 n^a + X^a_2\) be two BMS symmetries (with \cref{eq:bms-decomp-conds} holding for each) then their Lie bracket can be computed to give
\be\label{eq:bracket-decomp}
    \xi^a &\hateq [\xi_1,\xi_2]^a \hateq \beta n^a + X^a \quad \text{with}\\[1.5ex]
    \beta &\hateq \Lie_{X_1} \beta_2 - \tfrac{1}{2} \beta_2 \ms D_a X^a_1 - (1 \leftrightarrow 2) \eqsp X^a \hateq \Lie_{X_1} X^a_2 \,.
\ee
It can be checked that \(\xi^a\) is also a BMS symmetry, i.e. the \(\beta\) and \(X^a\) in \cref{eq:bracket-decomp} also satisfy the conditions \cref{eq:bms-decomp-conds}. Thus, the BMS symmetries form a Lie algebra \(\mf b\).

The structure of the BMS algebra can be analyzed using \cref{eq:bracket-decomp}. Consider a BMS symmetry of the form \(\xi^a_1 \hateq f_1 n^a\) where \(f_1\) is a smooth function on \(\scri\) satisfying \(\Lie_n f_1 \hateq 0\) (from \cref{eqn:lie_n_beta}). Then, from \cref{eq:bracket-decomp} we see that the Lie bracket of \(\xi^a_1 \hateq f_1 n^a\) with any other BMS symmetry is also of the form \(\xi^a \hateq f n^a\) with \(\Lie_n f \hateq 0\), that is the set of such vector fields is invariant under the Lie bracket. Further, the Lie bracket of any two such symmetries vanishes. Thus, BMS symmetries of the form \(fn^a\) form a preferred infinite-dimensional abelian subalgebra \(\mf s\) which is a \emph{Lie ideal} of the BMS algebra \(\mf b\) consisting of \emph{supertranslations}. The quotient algebra \(\mf b/\mf s\) can then be parameterized by \(X^a\). From \cref{eqn:stf_dX} we see that this consists of conformal Killing fields on the cross-sections of \(\scri\). Since the cross-sections are diffeomorphic to \(\bb S^2\) we get that \(\mf b/\mf s\) is isomorphic to the Lorentz algebra \(\mf{so}(1,3)\). Note that the Lorentz algebra is only identified as a quotient algebra --- since the Lie bracket of a supertranslation and \(X^a\) is non-zero, there is no invariant choice of Lorentz subalgebra within the BMS algebra \(\mf b\). Thus the BMS algebra is the semi-direct sum
\be\label{eq:bms-semidirect}
    \mf b \cong \mf s \ltimes \mf{so}(1,3)\,,
\ee
of the Lie ideal \(\mf s\) of supertranslations with the Lorentz algebra; the \(\ltimes\) indicates the non-trivial Lie bracket between the two factors.

There is another finite-dimensional Lie ideal within the BMS algebra given by supertranslations \(fn^a \in \mf s\) which satisfy the additional condition
\be\label{eq:trans-cond}
    \STF (\ms D_a + \tau_a)(\ms D_b - \tau_b) f \hateq 0\,.
\ee
It can be checked that this equation is conformally-invariant. Further, the space of solutions to this equation is \(4\)-dimensional and is preserved under the Lie bracket of the BMS algebra (see \cref{rem:bms-harmonics} below). This \(4\)-dimensional Lie ideal \(\mf t\) can be viewed as the space of \emph{translations}.
In fact, if the physical spacetime $(\hat{M}, \hat{g}_{ab})$ possesses any Killing vectors $\xi^a$, then they can be extended to $\scri$; moreover, if the Killing field is of the form $\xi^a \hateq f n^a$, then $\xi^a$ must be a translation in the sense of $f$ obeying \cref{eq:trans-cond} (see \cite{AX}).
 
A special case of the translations are those that are given by a time-translation in some conformal frame.
A time-translation in a given conformal frame is given by $f = 1$, a definition that is motivated by the behavior of the time-translation vector field in Minkowski (and, similarly, the time-translation vector field in Kerr).
Since $f$ transforms by $f \mapsto \omega f$ under a conformal transformation, a conformally-invariant notion of a time-translation is given by $f > 0$, since $\omega > 0$.
\begin{remark}[Characterization of BMS symmetries through spherical harmonics]
\label{rem:bms-harmonics}
    Consider the Bondi frame --- where the conformal factor is chosen so that the metric on the cross-sections is the unit \(2\)-sphere metric and the foliation is chosen so that \(\tau_a \hateq 0\). A general supertranslation can then be expanded in spherical harmonics on \(\bb S^2\). The condition \cref{eq:trans-cond} implies that the translations are spanned by the \(\ell = 0,1\) harmonics which is indeed a \(4\)-dimensional space. The fact that the translations are preserved under Lie brackets can also be shown using some spherical harmonics technology (see the appendix in \cite{Flanagan:2019vbl,Bonga2020}). Note that since supertranslations have conformal weight \(1\), this spherical harmonic decomposition only holds in the Bondi frame. Similarly, the Lorentz vector fields satisfying
\be
    \STF \ms D_a X_b \hateq 0\,,
\ee
are spanned by vector spherical harmonics with \(\ell = 1\) (see, e.g. \cite{Flanagan:2015pxa}) which is the \(6\)-dimensional space of the Lorentz algebra \(\mf{so}(1,3)\).
\end{remark}

\begin{remark}[Supertranslation ``ambiguity'' in the Lorentz algebra]
  \label{rem:lorentz-amb}
  As noted above, while the supertranslations form a subalgebra of the BMS algebra, there is no preferred Lorentz subalgebra.
  Instead, the Lorentz algebra arises as the quotient algebra $\mf b/\mf s$: the set of equivalence classes $[\xi^a]$ of $\mf b$, where $\xi_1^a$ and $\xi_2^a$ are members of the same equivalence class if $\xi_1^a - \xi_2^a \hateq fn^a \in \mf{s}$.

  For a given cross-section $S$, one can consider the algebra $\mf l_S$ of BMS vector fields of the form $\xi^a|_S = X^a$ (that is, vector fields tangent to $S$).
  This is a subalgebra of $\mf b$, and in fact it is straightforward to show that $\mf l_S \cong \mf b/\mf s$.
  The issue, therefore, is not that there is no Lorentz subalgebra of $\mf b$ but that there is an uncountably infinite number of them, one per cross-section of $\scri$.

  The fact that the members of the quotient algebra $\mf b/\mf s$ are only defined up to supertranslations is known as the \emph{supertranslation ambiguity}.
  A similar situation occurs with the Poincar\'e algebra, which, while possessing a Lie ideal in the form of the algebra of translations, possesses no unique Lorentz subalgebra.
  For each origin in Minkowski space, there is a Lorentz subalgebra that consists of infinitesimal rotations and boosts that fix that point, which is analogous to the Lorentz subalgebra $\mf l_S$ associated with some cross-section $S$ of $\scri$.

  Much like how there is a Lorentz subalgebra associated with each cross-section of $\scri$, there is moreover a Poincar\'e subalgebra, given by considering the semidirect product of this Lorentz subalgebra with the translation subalgebra picked out by \cref{eq:trans-cond}.
  This Poincar\'e subalgebra can be shown to contain any \emph{exact} Killing vector field that might exist in the physical spacetime (see Theorem 1 of \cite{AX}).
  Similarly, in \cref{sec:twistor}, we review a construction (based on twistors) of a Poincar\'e subalgebra, once again depending explicitly on a choice of cross-section $S$.
  Finally, there is no invariant notion of a set of charges conjugate to Lorentz vector fields, which would form an invariant notion of angular momentum.
  Such a set of charges only exists for a given cross-section, so that (in particular) one cannot have an invariant notion of the flux of angular momentum between two cross-sections.
  This is similar to how, since the Poincar\'e algebra has no unique notion of a Lorentz subalgebra, with each origin determining a particular Lorentz subalgebra, the notion of angular momentum in flat space is origin-dependent.
\end{remark}

\begin{remark}[Extended BMS algebra]
There have been various recent proposals for extending the BMS algebra. These include proposals to extend the Lorentz quotient algebra, $\mf b/\mf s$, to the Virasoro algebra \cite{Kapec-Virasoro,Barnich-Virasoro} and to the algebra of all diffeomorphisms of $\mathbb{S}^2$ \cite{Campiglia-Laddha-2,Campiglia-gen-BMS,Compere-DiffS2}. The Virasoro vector fields are however singular at isolated points of $\mathbb{S}^2$ and hence do not preserve the smoothness of $\scri$. Similarly, as detailed in \cref{sec:univ-str}, it follows from the definition of asymptotic flatness that the conformal class of the induced metric on cross-sections of $\scri$ is always universal. As a result, arbitrary diffeomorphisms of $\mathbb{S}^2$  do not arise as symmetries in this context. It has also been shown that the extension to all diffeomorphisms of $\mathbb{S}^2$
cannot be implemented in a fully covariant manner \cite{Flanagan:2019vbl}. For these reasons, we do not work with these extended symmetries in this paper.
\end{remark}

\subsection{Extensions of BMS symmetries away from \(\scri\)}
\label{sec:bms-ext}

Up to this point, although we started this discussion with vector fields $\xi^a$ defined throughout the unphysical spacetime, we have worked mostly with the value of $\xi^a$ on $\scri$ itself. The extension of the BMS symmetries away from \(\scri\) into the spacetime is arbitrary --- one could choose some coordinate system or gauge conditions to obtain a particular extension, but then one needs to verify that any physical quantities (like charges and fluxes) associated with the BMS symmetries are independent of such choices. We collect some of the relevant results on the extensions of BMS symmetries away from \(\scri\) below.

First, we show that the extension of $\xi^a$ away from $\scri$ is determined up to $O(\Omega^2)$:
\begin{prop}[Equivalent representatives of a BMS symmetry]
\label{prop:bms-reps}
    If \(\xi^a\) and \({\xi'}^a\) are vector fields in \(M\) which represent the same BMS symmetry, i.e., \(\xi^a \hateq {\xi'}^a \in \mf b\) then \({\xi'}^a = \xi^a + O(\Omega^2)\).
\begin{proof}
    Since \({\xi'}^a \hateq \xi^a\), let \({\xi'}^a = \xi^a + \Omega Z^a\) from which we obtain
    \be
        \alpha_{(\xi')} - \alpha_{(\xi)} = \Omega^{-1} n_a ({\xi'}^a - \xi^a) = n_a Z^a\,.
    \ee
    Since \(\Lie_\xi g_{ab} - 2\alpha_{(\xi)} g_{ab} \hateq 0\) for any BMS vector field we have
    \be
    0 \hateq \lb( \Lie_{\xi'} g_{ab} - 2\alpha_{(\xi')} g_{ab} \rb) - \lb( \Lie_\xi g_{ab} - 2\alpha_{(\xi)} g_{ab} \rb) \hateq  2 n_{(a} Z_{b)}  - 2 n_c Z^c g_{ab}\,.
    \ee
    Taking the trace gives \(n_a Z^a \hateq 0\). Then we have \(n_{(a} Z_{b)} \hateq 0\) which implies \(Z^a \hateq 0\).    Thus \(Z^a = O(\Omega) \) and \({\xi'}^a = \xi^a + O(\Omega^2)\).
\end{proof}
\end{prop}

The perturbation \(\gamma_{ab}^{(\xi)}\) generated by a BMS symmetry \(\xi^a\vert_\scri\) will, in general, depend on the extension of the symmetry away from \(\scri\) and hence is \emph{not} well-defined for the BMS symmetries. However, using the above lemma it can be shown that $\STF \gamma_{ab}^{(\xi)}$ on \(\scri\) is, in fact, independent of the extension of the BMS symmetry $\xi^a$ away from $\scri$, and thus is well-defined for any BMS symmetry.
\begin{corollary} \label{cor:stf-gamma-bms}
    If \(\xi^a\) and \({\xi'}^a\) are any two extensions of a given BMS symmetry \(\xi^a\vert_\scri\) away from \(\scri\) then
\be
    \STF \gamma_{ab}^{(\xi)} \hateq \STF \gamma_{ab}^{(\xi')}.
\ee
\begin{proof}
    From \cref{prop:bms-reps}, we have \({\xi'}^a = \xi^a + \Omega^2 W^a\), from which (using \cref{eq:alpha-xi-defn}) we compute
\be
    \alpha_{(\xi')} = \Omega^{-1} n_a {\xi'}^a = \Omega^{-1} n_a \xi^a + \Omega n_a W^a = \alpha_{(\xi)} + \Omega n_a W^a \,.
\ee
Similarly
\be
    \Lie_{\xi'} g_{ab} = \Lie_\xi g_{ab} + 4 \Omega n_{(a} W_{b)} + O(\Omega^2)\,.
\ee
Thus, we find that
\be
    \Omega \gamma_{ab}^{(\xi')} &= \Lie_{\xi'} g_{ab} - 2 \alpha_{(\xi')} g_{ab} = \Lie_\xi g_{ab} - 2 \alpha_{(\xi)} g_{ab} + 4 \Omega \lb( n_{(a} W_{b)} - \tfrac{1}{2} n_c W^c g_{ab} \rb) + O(\Omega^2) \\
    \implies
    \gamma_{ab}^{(\xi')} &= \gamma_{ab}^{(\xi)} + 4 \lb( n_{(a} W_{b)} - \tfrac{1}{2} n_c W^c g_{ab} \rb) + O(\Omega)\,.
\ee
Evaluating the $\STF$ of both sides on \(\scri\) of the above equation we find the desired result.
\end{proof}
\end{corollary}

For later computations it will be useful to have an explicit expression for \(\STF \gamma_{ab}^{(\xi)}\) at \(\scri\) for any BMS symmetry \(\xi^a\) in terms of the fields defined on \(\scri\). We show below that
\begin{equation} \label{eqn:stf_gamma_xi}
  \STF \gamma_{ab}^{(\xi)} \hateq -2 \STF \left[\tfrac{1}{2} \beta N_{ab} + (\mathscr{D}_a + \tau_a) (\mathscr{D}_b - \tau_b) \beta + \Lie_X \sigma_{ab} - \tfrac{1}{2} (\mathscr{D}_c X^c) \sigma_{ab}\right].
\end{equation}
Comparing the above formula to \cref{eqn:stf_gamma} we find that under a BMS symmetry the shear transforms as
\be\label{eq:shear-trans}
    \delta_\xi \sigma_{ab} = \STF \left[\tfrac{1}{2} \beta N_{ab} + (\mathscr{D}_a + \tau_a) (\mathscr{D}_b - \tau_b) \beta + \Lie_X \sigma_{ab} - \tfrac{1}{2} (\mathscr{D}_c X^c) \sigma_{ab}\right] \, .
\ee
This is the infinitesimal version of the transformation of the shear found by Sachs \cite{Sachs1}. It is also the same as that given in, for example, Eq.~2.18b of \cite{Flanagan:2015pxa} for the transformation of the shear $C_{AB}$ in Bondi coordinates (which is related to $\sigma_{ab}$ by \cref{relations}).

In the remainder of this section we detail the computations which lead to \cref{eqn:stf_gamma_xi}. Let \(\xi^a\vert_\scri = \beta n^a + X^a\) be a BMS symmetry on \(\scri\). In the following, let \(l^a\) be \emph{any} vector field in a neighbourhood of \(\scri\) which coincides with the chosen auxiliary normal at \(\scri\) --- the result of the computation can be checked to be independent of how the auxiliary normal is extended away from \(\scri\). 

As shown above in \cref{cor:stf-gamma-bms}, \(\STF \gamma_{ab}^{(\xi)}\big\vert_\scri\) is independent of how the BMS symmetry \(\xi^a\vert_\scri\) is extended away from \(\scri\). Thus, we can choose to extend the BMS symmetry away from \(\scri\) as follows. We first extend \(\beta\) and \(X^a\) away from \(\scri\) to satisfy
\begin{equation} \label{eqn:lie_l_cond}
  \Lie_l \beta \hateq 0, \qquad \Lie_l X^a \hateq 0.
\end{equation}
With this choice any extension of the BMS symmetry \(\xi^a\vert_\scri\) can be written as
\begin{equation} \label{eqn:xi_decomp}
  \xi^a = \beta n^a + X^a + \Omega Z^a.
\end{equation}
where \(Z^a\) is some smooth vector field. Next from \cref{prop:bms-reps} we see that \(Z^a\vert_\scri\) can be determined in terms of \(\beta\) and \(X^a\). To do this, first note that, by \cref{eq:alpha-xi-defn},
\begin{equation}
  \alpha_{(\xi)} \hateq -\Lie_l (\xi^a n_a) \hateq -n_a \Lie_l \xi^a \hateq n_a Z^a.
\end{equation}
Next, we compute $l^b \Lie_\xi g_{ab}$, using \cref{eqn:lie_xi_g}, together with \cref{eqn:lie_l_cond}:
\begin{equation} \label{eqn:l_lie_xi_g}
  l^b \Lie_\xi g_{ab} \hateq -\nabla_a \beta + \Lie_X l_a - Z_a + n_a l^b Z_b \hateq 2 \alpha_{(\xi)} l_a.
\end{equation}
Contracting this equation with $l^a$, using \cref{eqn:lie_l_cond} and \(l^a \Lie_X l_a \hateq -l_a \Lie_X l^a \hateq 0\), we find that \(l_a Z^a \hateq 0\). Therefore, we can rearrange \cref{eqn:l_lie_xi_g} to solve for $Z_a$:
\begin{equation}
  Z_a \hateq -\nabla_a \beta - 2 \alpha_{(\xi)} l_a + \Lie_X l_a.
\end{equation}
Further, since \(l_a\) is the normal and \(X^a\) is tangent to the cross-sections of the chosen foliation, we have \(Q_a{}^b \Lie_X l_b \hateq 0\). So
\begin{equation}
  Z_a \hateq -\mathscr{D}_a \beta + l_a (\Lie_n \beta - 2 \alpha_{(\xi)} + l_b \Lie_X n^b) \hateq -\mathscr{D}_a \beta - \alpha_{(\xi)} l_a,
\end{equation}
where we have used \cref{eqn:lie_n_X,eqn:lie_n_beta}.
In summary, with the choice \cref{eqn:lie_l_cond}, we can write any extension of a given BMS symmetry \(\xi^a\vert_\scri\) as
\begin{equation} \label{eqn:xi_exp}
  \xi^a = \beta n^a + X^a - \Omega (\ms D^a \beta + \alpha_{(\xi)} l^a) + \Omega^2 W^a,
\end{equation}
for some smooth $W^a$.

With this choice of extension of the BMS symmetry, we compute $\STF \gamma_{ab}^{(\xi)}\big\vert_\scri$ in terms of $\beta$ and $X^a$.
Using the expansion in \cref{eqn:xi_exp} we have
\be
    \Omega \gamma_{ab}^{(\xi)} & = \Lie_\xi g_{ab} - 2 \alpha_{(\xi)} g_{ab} \\
    &= \beta \Lie_n g_{ab} + \Lie_X g_{ab} - 2 \alpha_{(\xi)} g_{ab} - 2 n_{(a} [ \ms D_{b)} \beta + \alpha_{(\xi)} l_{b)} ] \\
    &\hspace{1em}- \Omega [2 \nabla_{a} \nabla_{b} \beta + \alpha_{(\xi)} \Lie_l g_{ab} + 2 l_{(a} \nabla_{b)} \alpha_{(\xi)} - 4 n_{(a} W_{b)}] + \Omega^2 \Lie_W g_{ab} \\
\ee
We now want to solve this equation for $\STF \gamma_{ab}^{(\xi)}\big\vert_\scri$. This can be done by taking the \(\Lie_l\) of the above equation, evaluating on \(\scri\), and then taking the \(\STF\). Using \(\Lie_l \Omega \hateq l^a n_a = - 1\) and that \(\Lie_l n_a \propto n_a\) at \(\scri\), a long but straightforward computation gives
\begin{equation} \label{eqn:stf_gamma_lies}
  \STF \gamma_{ab}^{(\xi)} \hateq - \STF [\beta \Lie_l \Lie_n g_{ab} + \Lie_l \Lie_X g_{ab} + 2 \ms D_a \ms D_b \beta - \alpha_{(\xi)} \Lie_l g_{ab}],
\end{equation}
where we have used \cref{eqn:lie_l_cond}.
Note that by \cref{eqn:lie_l_cond}, we find that
\begin{equation}\label{eqn:lie_X_sigma}
  \STF \Lie_l \Lie_X g_{ab} \hateq \STF \Lie_X \Lie_l g_{ab} \hateq 2 \STF \Lie_X \sigma_{ab},
\end{equation}
where the final equality can be shown using \cref{eq:shear-defn,eqn:stf_dX}, together with the fact that $X^a n_a \hateq X^a l_a \hateq 0$.
Combining \cref{eqn:stf_gamma_lies,eqn:lie_X_sigma}, together with \cref{eq:shear-defn,eq:news_lie_ln_g,eqn:div_X}, we find that
\begin{equation} \label{eqn:stf_gamma_xi-2}
  \STF \gamma_{ab}^{(\xi)} = -2 \STF \left[\tfrac{1}{2} \beta N_{ab} + (\mathscr{D}_a + \tau_a) (\mathscr{D}_b - \tau_b) \beta + \Lie_X \sigma_{ab} - \tfrac{1}{2} (\mathscr{D}_c X^c) \sigma_{ab}\right] \, ,
\end{equation}
as claimed in \cref{eqn:stf_gamma_xi}.

\section{Asymptotic charges and fluxes: The Wald-Zoupas prescription}
\label{sec:WZ}

The prescription of Wald and Zoupas \cite{WZ} provides a method of determining charges and fluxes at null infinity, and can be applied to any local and covariant theory.
However, for simplicity, we will be specializing to the case of vacuum general relativity.
Note that the addition of matter, for example, does not significantly complicate this discussion (see for example, \cite{WZ,Bonga2019}).

We start with the Einstein-Hilbert Lagrangian \(4\)-form $\df L$:
\begin{equation}
  \df L = \frac{1}{16\pi} \hat{R}~ \hat{\df \varepsilon}_4.
\end{equation}
where \(\hat R\) is the Ricci scalar and \(\hat{\df \varepsilon}_4\) is the \(4\)-form volume element of the physical metric. The dynamical field is the physical metric \(\hat g_{ab}\), and varying this dynamical field we obtain
\begin{equation}\label{eqn:lagrangian_variation}
  \delta \df{L} = \df E^{ab} \delta \hat g_{ab} + d \df{\theta} (\delta \hat g),
\end{equation}
where the \(3\)-form $\df{\theta}$ is the \emph{symplectic potential}, and $\df E^{ab}$ is a tensor-valued \(4\)-form which gives the equations of motion in the form $\df E^{ab} = 0$. For general relativity, we have \cite{Burnett1990}
\begin{subequations}
  \begin{align}
    \df E^{ab} &= -\frac{1}{16\pi} \hat{\df \varepsilon}_4 \hat{G}^{ab}, \label{eq:eom-GR} \\
    \df \theta(\delta\hat g ) &\equiv - \frac{1}{8\pi} \hat{\varepsilon}_{abc}{}^{[d} \hat{g}^{e]f} \hat{\nabla}_e \delta \hat{g}_{df}, \label{eq:symp-potential-GR} 
  \end{align}
\end{subequations}

The \emph{symplectic current} is defined by taking a second, independent variation of the symplectic potential and antisymmetrizing in the perturbations:
\begin{equation} \label{eqn:omega}
  \df{\omega} (\delta_1 \hat g, \delta_2 \hat g) \defn \delta_1 \df{\theta} (\delta_2 \hat g) - \delta_2 \df{\theta} (\delta_1 \hat g).
\end{equation}
Computing $d \df \omega$ from this equation, using the fact that $d$ and $\delta$ commute, using the second variation of \cref{eqn:lagrangian_variation}, and using the fact that $\delta_1$ and $\delta_2$ commute, one finds that (for example, \cite{Burnett1990})
\begin{equation} \label{eqn:domega}
  d \df \omega (\delta_1 \hat g, \delta_2 \hat g) = \delta_2 \df{E}^{ab} \delta_1 \hat g_{ab} - \delta_1 \df{E}^{ab}  \delta_2 \hat g_{ab},
\end{equation}
which vanishes whenever the perturbations satisfy the linearized equations of motion $\delta\df{E}^{ab} = 0$.
When the dynamical fields $\hat g_{ab}$ satisfy the equations of motion, and $\delta \hat g_{ab}$ satisfy the linearized equations of motion, one can show that (see \cite{Lee1990, Iyer1994, Prabhu2015})
\begin{equation} \label{eq:omega_symmetry}
  \df \omega(\delta \hat g, \Lie_\xi \hat g) = d \left[\delta \df Q_\xi - \xi \cdot \df \theta(\delta \hat g)\right],
\end{equation}
for all infinitesimal diffeomorphisms generated by $\xi^a$, where the $2$-form $\df Q_\xi$ is the \emph{Noether charge} associated with $\xi^a$.
In general relativity, we have that \cite{Burnett1990,WZ}
\begin{equation} \label{eqn:omega_GR}
  \df \omega(\delta_1 \hat g, \delta_2 \hat g) \equiv \frac{1}{16\pi} \hat{\varepsilon}_{dabc} \hat{P}^{defghi} \left[\delta_2 \hat{g}_{ef} \hat{\nabla}_g \delta_1 \hat{g}_{hi} - (1 \leftrightarrow 2)\right],
\end{equation}
where
\begin{equation}
  \hat{P}^{abcdef} \defn \hat{g}^{ae} \hat{g}^{fb} \hat{g}^{cd} - \tfrac{1}{2} \hat{g}^{ad} \hat{g}^{be} \hat{g}^{fc} - \tfrac{1}{2} \hat{g}^{ab} \hat{g}^{cd} \hat{g}^{ef} - \tfrac{1}{2} \hat{g}^{bc} \hat{g}^{ae} \hat{g}^{fd} + \tfrac{1}{2} \hat{g}^{bc} \hat{g}^{ad} \hat{g}^{ef}.
\end{equation}
Moreover, the Noether charge is given by \cite{Iyer1994}
\begin{equation}\label{eq:noether-GR}
  \df Q_\xi \equiv -\frac{1}{16\pi} \hat{\varepsilon}_{cdab} \hat{\nabla}^c \xi^d.
\end{equation}
  
The symplectic current introduced above, when integrated over some hypersurface $\Sigma$, provides a symplectic product on phase space.
As discussed in \cite{Lee1990, WZ}, a perturbed Hamiltonian should be constructed by considering the symplectic product of an arbitrary perturbation $\delta g_{ab}$ with $\delta_\xi \hat g_{ab} = \Lie_\xi \hat g_{ab}$:
\begin{equation}\label{eq:Hamiltonian}
  \delta H_\xi = \int_\Sigma \df \omega(\delta \hat g, \Lie_\xi \hat g) = \int_\Sigma d \left[\delta \df Q_\xi - \xi \cdot \df \theta(\delta \hat g)\right] = \int_{\partial\Sigma} \left[\delta \df Q_\xi - \xi \cdot \df \theta(\delta \hat g)\right],
\end{equation}
where the second equality follows by \cref{eq:omega_symmetry} and \(\partial\Sigma\) is the boundary of \(\Sigma\). Thus, the Hamiltonian can then naturally be thought of as an integral over the boundary \(\partial\Sigma\).

Now consider the case where the hypersurface \(\Sigma\) extends as a smooth surface to \(\scri\) in the unphysical spacetime which intersects \(\scri\) at a cross-section \(S\). We take \cref{eq:Hamiltonian}, rewritten in terms of the unphysical fields which are smooth at $\scri$. In general relativity, using the behaviour of the unphysical metric perturbations detailed in \cref{sec:metric-pert}, the symplectic current has a finite limit to \(\scri\); as shown in \cite{WZ}. However, one should not conclude from \cref{eq:omega_symmetry} that \(\delta \df Q_\xi - \xi \cdot \df \theta(\delta \hat g)\) has a limit to \(\scri\); in general relativity it can be shown that \(\df\theta(\delta \hat g)\) again has a finite limit to \(\scri\) (see \cite{WZ}), but \(\delta \df Q_\xi\) diverges in the limit to \(\scri\). Note that any procedure to ``subtract out the diverging part'' is highly non-unique. Fortunately, there is no need to resort to any such ad hoc procedure as we can use some elementary differential geometry to proceed directly as follows.

\begin{lemma}\label{lem:charge-well-def}
Let \(S'\) be some sequence of \(2\)-spheres in unphysical spacetime which limits (continuously) to a chosen cross-section \(S\) of \(\scri\). Then, the limiting integral
\be
    \lim_{S' \to S} \int_{S'} \left[\delta \df Q_\xi - \xi \cdot \df \theta(\delta \hat g)\right]
\ee
defined by \emph{first} integrating \(\delta \df Q_\xi - \xi \cdot \df \theta(\delta \hat g)\) over the sequence of \(2\)-spheres \(S'\) and then taking the limit as the sequence tends to \(S \subset \scri\) exists and is independent of the chosen sequence of \(2\)-spheres used in the limit.
\begin{proof}
 Let \(S_0\) be some \(2\)-surface in the unphysical spacetime and let $\Sigma$ be a smooth \(3\)-surface which extends from \(S_0\) and intersects $\scri$ at a cross-section $S$. Note that this surface \(\Sigma\) is a \emph{compact} \(3\)-manifold in the unphysical spacetime. The integral of \(\df \omega(\delta \hat g, \Lie_\xi \hat g)\) over \(\Sigma\) is necessarily finite since \(\df \omega(\delta \hat g, \Lie_\xi \hat g)\) is a continuous \(3\)-form on \(\Sigma\) (including at the ``boundary'' at \(S\)). Thus integrating \cref{eq:omega_symmetry} over \(\Sigma\) we obtain%
\footnote{We note that on the left-hand side, we have taken $\Sigma$ to be future-oriented while on the right-hand side, the \(2\)-spheres \emph{do not} have the usual outward-facing orientation within $\Sigma$, but the opposite. This choice of orientation is more natural, since (as mentioned in \cref{rem:orientation}), the limit of the orientation as $S' \to S$ is the future-directed orientation of $S$ within $\scri$ as specified by $\df \varepsilon_2$. This choice of orientation is the opposite of the one that is used by Wald and Zoupas (see footnotes 2, 3, and 8 of \cite{WZ}), and so some of our equations have the opposite sign.}
\begin{equation} \label{eq:omega-dQ}
  \int_\Sigma \df \omega(\delta \hat g, \Lie_\xi \hat g) = \int_{S_0} \left[\delta \df Q_\xi - \xi \cdot \df \theta(\delta \hat g)\right] - \lim_{S' \to S} \int_{S'} \left[\delta \df Q_\xi - \xi \cdot \df \theta(\delta \hat g)\right],
\end{equation}
where the last expression on the right-hand side means ``integrate over some $2$-sphere $S' \subset \Sigma$, and then take the limit of this $2$-sphere to the boundary $S$''.
This limiting procedure is necessary because, although the integral on the left-hand side of \cref{eq:omega-dQ} is always finite (as argued above), the $2$-form integrand on the right-hand side need not have a finite limit to $\scri$ in general.
It follows that the limit in the last expression on the right-hand side of \cref{eq:omega-dQ} exists. The limit is also independent of the choice of sequence of $S'$ that is used in its definition due to \cref{eq:omega_symmetry} and the fact that $\Sigma$ is compact and \(\df \omega(\delta \hat g, \Lie_\xi \hat g)\) is continuous on \(\Sigma\). Further, by Stokes' theorem, this expression is the same for \emph{any} choice of the \(3\)-surface $\Sigma$ whose boundary is also $S$, since $d \df \omega(\delta \hat g, \delta_\xi \hat g) = 0$ by \cref{eqn:domega} and since \(\df \omega(\delta \hat g, \delta_\xi \hat g)\) extends continuously to \(\scri\).
\end{proof}
\end{lemma}

\begin{remark}[Necessity of \(\df\omega\) extending continuously to \(\scri\)]
We emphasize that the condition that \(\df \omega(\delta \hat g, \delta_\xi \hat g)\) extend continuously (as a \(3\)-form in \(M\)) to \(\scri\) is necessary in the above argument. For instance, if we instead only assume that \(\int_\Sigma \df \omega(\delta \hat g, \delta_\xi \hat g)\) is finite on \emph{every} surface \(\Sigma\) (or that the pullback of \(\df \omega(\delta \hat g, \delta_\xi \hat g)\) to \emph{every} \(\Sigma\) is continuous within \(\Sigma\)), then \(d\df\omega = 0\) does \emph{not} imply that the limit of \( \int_{S'} \left[\delta \df Q_\xi - \xi \cdot \df \theta(\delta \hat g)\right]\) is well-defined since it can depend on the choice of surface \(\Sigma\) used to define the limiting integral. As this point is often overlooked in the application of Stokes' theorem at \(\scri\) we provide a simple example on the Euclidean plane below.

On \(\bb R^2\), let \((x,y)\) be the usual Cartesian coordinates. Consider the \(1\)-form
\be
    \df\omega \equiv \frac{1}{(x^2+y^2)^{\nfrac{3}{2}}} \lb[ y^2~  dx - xy~ dy \rb]\,.
\ee
This \(1\)-form can be written as the exterior derivative of a \(0\)-form (i.e., a function) as\footnote{The choice of the constant in the function \(Q\) is irrelevant for our argument.}
\be
    \df\omega = d Q \eqsp Q = \frac{x}{(x^2+y^2)^\half} \,.
\ee
Note that \(\df\omega\) does not extend continuously to the origin \((x,y) = (0,0)\), but \(d\df\omega = 0\) everywhere else and extends continuously to the origin. Further, it can be checked that the pullback of this \(\df\omega\) to any smooth curve \(\Sigma\) through the origin is continuous at the origin \emph{within} this curve. However, we \emph{cannot} use Stokes' theorem to conclude that \(\int_\Sigma \df\omega\) is independent of the choice of curve \(\Sigma\) joining the origin to some other point, since \(\df\omega\) is not continuous at the origin as a \(1\)-form in \(\bb R^2\) . A direct computation shows that for any curve \(\Sigma\) from the origin to \((1,1)\), we have
\be
    \int_\Sigma \df\omega = \tfrac{1}{\sqrt{2}} - \cos\theta\,,
\ee
where \(\theta\) is the angle with the \(x\)-axis of the tangent of \(\Sigma\) at the origin. So the integral of \(\df\omega\) along any curve through the origin is finite but depends on the curve \(\Sigma\). Thus, we cannot define the value of \(Q\) at the origin by taking such integrals over curves \(\Sigma\) since it depends on the curve used in the limiting procedure. This can be explicitly checked from the expression of \(Q\) given above.
\end{remark}

Next we show that, in general relativity, the limiting integral of \(\left[\delta \df Q_\xi - \xi \cdot \df \theta(\delta \hat g)\right]\) is independent of the choice of extension of the BMS symmetry away from \(\scri\).

\begin{lemma}\label{lem:charge-well-def-2}
If \(\xi^a\) and \(\xi'^a\) are equivalent representatives of a BMS symmetry on \(\scri\) in general relativity then
\be
    \lim_{S' \to S} \int_{S'} \left[\delta \df Q_\xi - \xi \cdot \df \theta(\delta \hat g)\right] = \lim_{S' \to S} \int_{S'} \left[\delta \df Q_{\xi'} - \xi' \cdot \df \theta(\delta \hat g)\right],
\ee
for all background spacetimes, all perturbations \(\delta \hat g_{ab}\) and all cross-sections \(S\) of \(\scri\).
\begin{proof}
From \cref{prop:bms-reps}, two equivalent representatives of a BMS symmetry are related by \(\xi'^a = \xi^a + \Omega^2 W^a\) for some smooth \(W^a\). Hence to prove the desired result we only need to show that the above integral computed for the vector field \(\Omega^2 W^a\) vanishes. In general relativity it can be shown that the \(3\)-form \(\df\theta(\delta \hat g)\) is finite at \(\scri\) \cite{WZ} and thus \((\Omega^2 W) \cdot \df\theta(\delta \hat g) \hateq 0\). So we only need to the compute the Noether charge term which can be written as
\be
    \df Q_{\Omega^2 W} \equiv - \frac{1}{16\pi} \hat\varepsilon_{abcd} \hat\nabla^c (\Omega^2 W^d) = - \frac{1}{16\pi} \varepsilon_{abcd} \nabla^c W^d \,.
\ee
Note that this is manifestly finite at \(\scri\) and so we can dispense with the limiting procedure used in the integral and evaluate the variation of the above expression directly at \(\scri\). Using the asymptotic conditions on the metric perturbations (\cref{eq:pert-conds}), we get (we ignore the overall signs and factors)
\be
    \delta \df Q_{\Omega^2 W}\big\vert_\scri \propto \varepsilon_{abcd} n^c \gamma^d{}_e W^e\,.
\ee
Using the definition of the volume-forms (\cref{eqn:epsilon_32}) and \(n_a l^a \hateq -1\), the integral over the cross-section \(S\) is then (using \(\gamma_{ab}n^b \hateq \Omega \gamma_a\))
\be
    \int_S \delta \df Q_{\Omega^2 W} \propto \int_S \df\varepsilon_2~ n^a \gamma_{ab} W^b \hateq \int_S \df\varepsilon_2~ \Omega \gamma_{a} W^a \hateq 0 \,.
\ee
\end{proof}
\end{lemma}

The results of \cref{lem:charge-well-def,lem:charge-well-def-2} prove that the limiting integral of \(\left[\delta \df Q_\xi - \xi \cdot \df \theta(\delta \hat g)\right]\) is well-defined on BMS symmetries at \(\scri\). Thus, from \cref{eq:omega-dQ}, it would be natural to define a charge associated with the asymptotic symmetry $\xi^a$ at $S$ as a function $Q[\xi; S]$ in the phase space of the theory such that
\begin{equation} \label{eq:Q-naive}
  \delta Q[\xi; S] \defn \lim_{S' \to S} \int_{S'} \left[\delta \df Q_\xi - \xi \cdot \df \theta(\delta \hat g)\right],
\end{equation}
for all backgrounds \(\hat g_{ab}\), all perturbations $\delta \hat g_{ab}$, and all cross-sections \(S\). However, in general, no such function $Q[\xi; S]$ exists, since the right-hand side is not integrable in phase space; that is, it cannot be written as the variation of some quantity for \emph{all} perturbations.
To see this, suppose that the charge defined in \cref{eq:Q-naive} does exist.
Then, one must have $(\delta_1 \delta_2 - \delta_2 \delta_1) Q[\xi; S] = 0$ for \emph{all} backgrounds $\hat g_{ab}$ and \emph{all} perturbations $\delta_1 \hat g_{ab}$, and $\delta_2 \hat g_{ab}$ (satisfying the corresponding equations of motion).
However, it follows from \cref{eq:Q-naive,eqn:omega} and the commutativity of $\delta_1$ and $\delta_2$ that
\begin{equation} \label{eq:Q-integrablity}
  (\delta_1 \delta_2 - \delta_2 \delta_1) Q[\xi, S] = -\int_S \xi \cdot \df\omega(\delta_1 \hat g, \delta_2 \hat g).
\end{equation}
Thus, a charge defined by \cref{eq:Q-naive} will exist if the right-hand side of the above equation vanishes.
This is the case if $\xi^a$ is tangent to $S$.
However, in general, the right-hand side is non-vanishing, and so one cannot define any charge $Q[\xi; S]$ using \cref{eq:Q-naive}.

\begin{remark}[Subtracting the ``non-integrable part'']
\label{rem:non-integrable-subtraction}
    It might be tempting to simply stare at some explicit expression for the right-hand-side of \cref{eq:Q-naive} and then subtract off the ``non-integrable part'' of the expression to obtain an expression which is manifestly integrable to define the charge at \(S\). But this ``procedure'' is very ad hoc; for instance suppose one manages to write the right-hand side of \cref{eq:Q-naive} as \(\delta A + B\) where \(\delta A\) is a manifestly integrable expression and \(B\) is not. However, one can trivially write this in the alternative form \(\delta A + B = \delta (A +C) + (B - \delta C )\) where \(C\) is some tensorial expression in terms of the available fields. Obviously, \(\delta (A +C)\) is integrable while \((B - \delta C)\) is non-integrable for any choice of \(C\). Thus the procedure to ``subtract off the non-integrable part'' is highly ambiguous --- without any additional criteria one cannot know which ``non-integrable part'' \(B\) or \(B - \delta C\) should be ``subtracted off''.
\end{remark}

The obstruction to the non-integrability of \cref{eq:Q-naive} was resolved by the rather general prescription of Wald and Zoupas \cite{WZ}.
Their procedure for defining integrable charges associated with asymptotic symmetries can be summarized as follows: let $\df \Theta$ be a symplectic potential for the pullback of the symplectic current to $\scri$; that is,
\begin{equation} \label{eq:Theta-defn}
  \underleftarrow{\df \omega} (\delta_1 \hat g, \delta_2 \hat g) = \delta_1 \df \Theta(\delta_2 \hat g) - \delta_2 \df \Theta(\delta_1 \hat g),
\end{equation}
for \emph{all} backgrounds and \emph{all} perturbations, with appropriate asymptotic conditions and equations of motion imposed. Following \cite{WZ}, we require that the choice of $\df \Theta$ satisfies the following properties:
\begin{enumerate}
\item $\df \Theta$ must be locally and covariantly constructed out of the dynamical fields $g_{ab}$, the perturbations $\delta g_{ab}$, and finitely many of their derivatives, along with any fields in the ``universal background structure'' present at $\scri$ described in \cref{sec:univ-str};
\item $\df \Theta$ must be independent of any arbitrary choices made in specifying the background structure; that is, $\df \Theta$ is conformally invariant and independent of the choice of the auxiliary normal $l^a$; and
\item if $g_{ab}$ is a stationary background solution, then $\df \Theta(\hat g; \delta \hat g) = 0$, for \emph{all} (not necessarily stationary) perturbations $\delta g_{ab}$.
\end{enumerate}
The first of these criteria is motivated by the fact that, as laid out in the introduction, the prescription used to define charges should only require the tensor fields that exist on $\scri$, and should not depend on additional structure that can be used (for example, some choice of coordinates).
The second is required to ensure that the charges that are defined by this prescription are associated with the physical spacetime and do not depend that the particular choices that go into the conformal-completion (such as the conformal factor) or other additional choices like the foliation of \(\scri\).
The final criterion plays an important role in showing that the flux of these charges vanishes for stationary spacetimes as we shall see below.

If such a symplectic potential $\df \Theta$ can be found, define $\mc Q[\xi; S]$ to be a function on the phase space at $\scri$ by
\begin{equation} \label{eq:WZ-defn}
  \delta \mc Q[\xi; S] \defn \lim_{S' \to S} \int_{S'} \left[\delta \df Q_\xi - \xi \cdot \df \theta(\delta \hat g)\right] + \int_S \xi \cdot \df \Theta(\delta \hat g).
\end{equation}
It can easily be checked (using \cref{eq:Q-naive,eq:Q-integrablity,eq:Theta-defn}) that this expression is integrable in phase space; that is, $(\delta_1 \delta_2 - \delta_2  \delta_1) \mc Q[\xi; S] = 0$.
Together with some choice of reference solution $g_0$ on which $\mc Q[\xi; S] = 0$ for all asymptotic symmetries $\xi^a$ and all cross-sections $S$, \cref{eq:WZ-defn} can be integrated in phase space to define the \emph{Wald-Zoupas (WZ) charge} $\mc Q[\xi; S]$ associated with the asymptotic BMS symmetry $\xi^a$ at $S$. We note the following properties of the WZ charge defined by the above procedure:
\begin{enumerate}
    \item Since $\df \Theta$ is locally and covariantly constructed from the physical metric and the universal structure of $\scri$, the charge can similarly be written in terms of quantities that are defined at $\scri$. In particular the charge only depends on the BMS symmetry at \(\scri\) and not on any choice of its extension into the spacetime. 
      While this is clear for the term that is an integral of $\xi \cdot \df \Theta$, that this is also true for the remaining terms, which are defined by a limit in the unphysical spacetime, follows by \cref{lem:charge-well-def,lem:charge-well-def-2}.
    \item Moreover, since \(\df\Theta\) is chosen to be conformally invariant and independent of the choice of the auxiliary foliation of \(\scri\) and the choice of the reference solution is to be specified without any particular choice of its conformal completion or the choice of auxiliary foliation, the WZ charge is also conformally invariant and independent of the choice of the auxiliary foliation. Note, however that the WZ charge does depend on the chosen cross-section \(S\) on which it is evaluated and hence can depend on the auxiliary normal \(l_a\) at \(S\) (but does not depend on the choice of \(l_a\) away from \(S\)).
\end{enumerate}

The flux of the perturbed WZ charge, through a portion $\Delta \scri$ of $\scri$ whose boundary is given by two cross-sections $S_1$ and $S_2$, is given by (see Eqs.~28 and~29 of \cite{WZ})
\begin{equation} \label{eq:delta-F1}
  \delta \mc F[\xi; \Delta \scri] \defn \delta \mc Q[\xi; S_2] - \delta \mc Q[\xi; S_1] = \int_{\Delta \scri} \lb[ \underleftarrow{\df \omega} (\delta \hat g, \Lie_\xi \hat g) + d \{\xi \cdot \df\Theta(\delta \hat g)\} \rb].
\end{equation}
The last term of this equation can also be written as
\begin{equation}
    d [\xi \cdot \df \Theta(\delta \hat g)] = \Lie_\xi \df \Theta(\delta \hat g) = -\underleftarrow{\df \omega} (\delta \hat g, \Lie_\xi \hat g) + \delta \df\Theta(\Lie_\xi \hat g) \,,
\end{equation}
where in the first equality, we have used the fact that $\df \Theta$ is a \(3\)-form intrinsic to $\scri$, and so its exterior derivative is zero, and the second equality follows from the definition of $\df\Theta$ as a symplectic potential for $\underleftarrow{\df\omega}$ (\cref{eq:Theta-defn}).
The flux of the perturbed WZ charge is therefore simply given by
\begin{equation}
  \delta \mc F[\xi; \Delta\scri] = \int_{\Delta \scri} \delta \df \Theta(\Lie_\xi \hat g).
\end{equation}

To get the unperturbed charge and flux, we have to choose a reference solution $g_0$ on which the charges are required to vanish.
Since the symplectic potential $\df \Theta$ is required to vanish on stationary backgrounds, we choose the reference solution $g_0$ to also be stationary.
For our concrete case of general relativity, we will pick $g_0$ to be (any conformal completion of) Minkowski spacetime.
Then, the flux of the WZ charge is given by
\begin{equation} \label{eq:general-flux}
  \mc F[\xi; \Delta \scri] = \mc Q[\xi; S_2] - \mc Q[\xi; S_1] = \int_{\Delta \scri} \df \Theta(\Lie_\xi \hat g).
\end{equation}

There are two important properties that follow from this expression for the flux.
The first, which the flux inherits from $\df \Theta$, is that, for any stationary background, the flux will vanish.
This property captures the fact that, if there is no radiation, there should be no flux and the charges should be conserved quantities.
The second is that, if $\xi^a$ is an exact Killing vector field in the physical spacetime, then the flux will vanish as well.
This follows from the vanishing of $\Lie_\xi \hat g_{ab}$ for such vector fields.
This property is reminiscent of Noether's theorem: if there is an exact symmetry, then there should be a quantity related to that symmetry (in this case, the charge) that is conserved.
In the next two sections, we will find expressions for the WZ flux and the WZ charge in vacuum general relativity.
Since the charge and flux calculations will be performed in different ways, we will check these calculations by showing that they agree with \cref{eq:general-flux}.\\

Finally, we remark that the Wald-Zoupas prescription has certain ambiguities related to the choices of the symplectic potential \(\df\theta\) and the choice of \(\df\Theta\). However, it was argued in \cite{WZ} that these ambiguities do not affect the final result. As the proof is quite involved, we do not present it here, and turn our attention to computing explicit expressions for the WZ flux and charge.

\subsection{Wald-Zoupas flux in general relativity}
\label{sec:flux}

We first consider the flux of the Wald-Zoupas charge.
From \cref{eq:general-flux}, it is apparent that determining this flux requires finding $\df \Theta$.
First, a lengthy calculation starting with \cref{eqn:omega_GR}, then using \cref{eq:pert-conds} for the unphysical metric perturbation $\delta g_{ab}$, along with the variation of the vacuum Einstein equations (\cref{eq:S-ee}), shows that \cite{WZ}
\begin{equation}
  \pb{\df \omega} (\delta_1 \hat g, \delta_2 \hat g) \hateq -\frac{1}{32\pi} \left( \delta_1 S_{ab} \gamma_2^{ab} - \delta_2 S_{ab} \gamma_1^{ab} \right) \df \varepsilon_3.
\end{equation}
Since $\gamma^{ab} n_b \hateq 0$ by \cref{eq:pert-conds}, $\delta S_{ab}$ in this expression can be replaced with $\delta \pb{S}_{ab}$.
Moreover, $\rho_{ab}$ is universal, and so $\delta \rho_{ab} = 0$; as such, \cref{eqn:news_geroch_def} implies that we can replace $\delta \pb{S}_{ab}$ with $\delta N_{ab}$, yielding
\begin{equation} \label{eq:symp-current}
  \pb{\df \omega} (\delta_1 \hat g, \delta_2 \hat g) \hateq -\frac{1}{32\pi} \left( \delta_1 N_{ab} \gamma_2^{ab} - \delta_2 N_{ab} \gamma_1^{ab} \right) \df \varepsilon_3 \hateq \frac{1}{16\pi} \left( \delta_1 N^{ab} \delta_1 \sigma_{ab} - \delta_2 N^{ab} \delta_1 \sigma_{ab} \right) \df \varepsilon_3,
\end{equation}
where the final equality uses the fact that the News tensor is traceless and \cref{eqn:stf_gamma}.

A symplectic potential for $\pb{\df \omega}$ is therefore given by
\begin{equation} \label{eq:WZ-correction}
  \df \Theta(\delta \hat g) \hateq -\frac{1}{32 \pi} N^{ab} \gamma_{ab} \df \varepsilon_3 \hateq \frac{1}{16 \pi} N^{ab} \delta \sigma_{ab} \df \varepsilon_3 .
\end{equation}

One must check that this symplectic potential satisfies the requirements above.
First, it is constructed from $g_{ab}$, $\delta g_{ab}$, their derivatives (such as $S_{ab}$), and fields that are part of the universal structure at $\scri$ (such as $n^a$ and, less obviously, $\rho_{ab}$ \cite{WZ}).
Moreover, it does not depend on the arbitrary structure that we have provided at $\scri$, the choice of conformal factor $\Omega$ and auxiliary normal $l^a$.
To see this, first note that the only piece that depends on $l^a$ is $\df \varepsilon_3 = l \cdot \df \varepsilon_4$, but upon taking the pullback, this dependence drops out.
To see that it is conformally invariant, use the fact that $\gamma_{ab} = \Omega \delta \hat{g}_{ab}$, and so under a conformal transformation $\Omega \mapsto \omega \Omega$,
\be
  N^{ab} \mapsto \omega^{-4} N^{ab} \eqsp \gamma_{ab} \mapsto \omega \gamma_{ab} \eqsp \df \varepsilon_3 \mapsto \omega^3 \df \varepsilon_3,
\ee
and so we find that \(\df\Theta\) is conformally invariant.
Finally, note that this choice of $\df \Theta$ vanishes on stationary solutions $g_{ab}$, for \emph{any} perturbation $\delta g_{ab}$, by the argument on pp.~53--54 of \cite{Geroch-asymp} (which shows that $N_{ab}$ vanishes for stationary vacuum spacetimes).

Using a chosen foliation of \(\scri\), we now provide a more explicit expression for the WZ flux in terms of fields defined on \(\scri\). Note that, since the News tensor is traceless, we have that
\begin{equation}
  N^{ab} \gamma_{ab}^{(\xi)} \hateq N^{ab} \STF \gamma_{ab}^{(\xi)}.
\end{equation}
As such, from \cref{cor:stf-gamma-bms} we have that our flux is independent of the extension of $\xi^a$ off of $\scri$.
Plugging in our expression for $\STF \gamma_{ab}^{(\xi)}$ from \cref{eqn:stf_gamma_xi}, we therefore find that\footnote{The first form of the WZ flux in \cref{eq:flux-formula} was anticipated by Geroch and Winicour long before Wald and Zoupas (see Eq.~28 of \cite{GW})!}
\be\label{eq:flux-formula}
  \mathcal{F} [\xi; \Delta \scri]
  & \hateq -\frac{1}{16\pi} \int_{\Delta \scri} \df \varepsilon_3 N^{ab} \gamma_{ab}^{(\xi)} \\ 
  & \hateq -\frac{1}{16\pi} \int_{\Delta \scri} \df \varepsilon_3 N^{ab} [\tfrac{1}{2} \beta N_{ab}+(\mathscr{D}_a + \tau_a) (\mathscr{D}_b - \tau_b) \beta + \Lie_X \sigma_{ab} - \tfrac{1}{2} (\mathscr{D}_c X^c) \sigma_{ab}].
\ee
As mentioned in \cite{WZ} this flux formula is equal to the one obtained by Ashtekar and Streubel \cite{AS}; the explicit computation of this equivalence is given in \cref{sec:AS}.

Note that the first expression is manifestly independent of any choice of foliation of \(\scri\), a property it inherits from $\df \Theta$.
This is not obviously true for the second, which instead is more useful for explicit computations (such as comparing with expressions in Bondi coordinates).
Both, however, are clearly local and covariant, and can be easily shown to be independent of the conformal factor.
We reiterate that these properties of the flux are motivated by the desire that these expressions not be dependent on any arbitrary choices that we could make, such as a coordinate system, conformal factor etc. As mentioned below \cref{eq:general-flux}, the flux in \cref{eq:flux-formula} has both the property that it vanishes for stationary backgrounds, as well as when $\xi^a$ is an exact Killing vector field in the physical spacetime.
The first of these properties is evident both in the first and second expressions in \cref{eq:flux-formula}, as the integrands are both proportional to the News.
That the flux vanishes when $\xi^a$ is an exact Killing vector field follows immediately from the fact that $\gamma^{(\xi)}_{ab}$ vanishes for such vector fields.

In the case where $\xi^a$ is a translation, namely $\xi^a = f n^a$, where $f$ obeys \cref{eq:trans-cond}, it is the case that
\begin{equation}
  \mathcal{F} [f n; \Delta \scri] = -\frac{1}{32\pi} \int_{\Delta \scri} \df \varepsilon_3\; f N^{ab} N_{ab}.
\end{equation}
In the case where $f$ is everywhere positive, $\xi^a$ corresponds to a time-translation in some conformal frame, as mentioned below \cref{eq:trans-cond}.
In this case, the flux is negative, corresponding to the \emph{loss} of mass/energy during the emission of gravitational waves.

\begin{remark}[BMS fluxes and Hamiltonians]
\label{rem:flux-hamiltonian}
On any region \(\Delta\scri\) of null infinity we can consider the integral of the symplectic current \(\int_{\Delta \scri} \pb{\df \omega} (\delta_1 \hat g, \delta_2 \hat g)\) as defining a \emph{symplectic form} on the radiative phase on \(\Delta\scri\). Then, \cref{eq:delta-F1} implies that
  \begin{equation}
    \delta \mc F[\xi; \Delta \scri] = \int_{\Delta \scri} \pb{\df \omega} (\delta \hat g, \Lie_\xi \hat g) + \int_{S_2} \xi \cdot \df\Theta(\delta \hat g) - \int_{S_1} \xi \cdot \df\Theta(\delta \hat g),
  \end{equation}
  Note that if the boundary terms at $S_1$ and $S_2$ vanish for all perturbations $\delta g_{ab}$ and all background solutions $g_{ab}$ then the WZ flux \(\mc F[\xi; \Delta \scri]\) would define a \emph{Hamiltonian generator} corresponding to the BMS symmetry \(\xi^a\), as defined by \cite{WZ, Lee1990}. But these boundary terms do not vanish in general and the WZ flux does not define a Hamiltonian. However, in the case where we consider all of null infinity, instead of some finite portion, and use appropriate boundary conditions at timelike and spatial infinity, then $\mc F[\xi; \scri]$ will be a Hamiltonian generator on the full radiative phase space on $\scri$. To see this, let \(u\) be a parameter along the null generators of \(\scri\) such that \(n^a \nabla_a u \hateq 1\), and impose, as $u \to \pm \infty$, the following boundary conditions:
  \begin{equation} \label{eqn:uinf_conds}
    N_{ab} = O(1/|u|^{1 + \epsilon}), \quad \gamma_{ab} = O(1),
  \end{equation}
  for some $\epsilon > 0$.
  These conditions ensure that the integral of the symplectic current over all of $\scri$, as given in \cref{eq:symp-current}, is finite. Further, \cref{eqn:lie_n_beta} implies that \(\xi^a = O(|u|)\),
  and so it follows from \cref{eqn:uinf_conds,eq:WZ-correction} that
  \begin{equation} \label{eqn:hamiltonian_cond}
    \lim_{u \to \pm \infty} \xi \cdot \df\Theta(\delta \hat g) = 0.
  \end{equation}
Note that the conditions \cref{eqn:uinf_conds} are preserved by all BMS symmetries, and thus the total flux \(\mc F[\xi; \scri]\) defines a Hamiltonian generator for the BMS symmetry \(\xi^a\) on the radiative phase space on \(\scri\).
\end{remark}

\subsection{Wald-Zoupas charge in general relativity}
\label{sec:charge}

Having obtained the WZ flux we now wish to find an expression for the WZ charge in terms of fields on \(\scri\). From \cref{eq:WZ-defn}, the WZ charge \(\mc Q[\xi;S]\) is determined by
\be\label{eq:pert-Q2}
    \delta \mc Q[\xi;S] = \lim_{S' \to S} \int_{S'} \lb[ \delta \df Q_\xi - \xi \cdot \df\theta(\delta \hat g) \rb] + \int_S \xi \cdot \df\Theta(\delta \hat g)\,,
\ee
along with the requirement that \(\mc Q[\xi;S]\) vanish on Minkowski spacetime for all BMS symmetries \(\xi^a\) and all cross-sections \(S\). The main difficulty in carrying out this computation directly is that the \(2\)-form \(\delta \df Q_\xi\) does not have a limit to \(\scri\); as mentioned earlier, it can be shown that \(\df\theta(\delta \hat g)\) does have a limit to \(\scri\) in general relativity. To compute the right-hand-side of \cref{eq:pert-Q2}, one must first choose some family of \(2\)-spheres inside the spacetime, evaluate the integral and then take the limit as these \(2\)-spheres tend to the chosen cross-section \(S\) of \(\scri\).

To compute this term, we note that by the general arguments in \cref{lem:charge-well-def,lem:charge-well-def-2},
\be\label{eq:Q-xi-theta}
    \lim_{S' \to S} \int_{S'} \lb[ \delta \df Q_\xi - \xi \cdot \df\theta(\delta \hat g) \rb]\,,
\ee
is guaranteed to exist, is independent of the family of \(2\)-spheres chosen to take the limit and is independent of how the BMS symmetry is extended into the spacetime away from \(\scri\). Moreover, it is manifestly independent of the choice of the conformal factor and the choice of the foliation of \(\scri\). Thus, we can compute \cref{eq:Q-xi-theta} in the choices given by the conformal Bondi-Sachs coordinates in a neighbourhood of \(\scri\). The detailed construction of these coordinates and the form of the unphysical and physical metrics in these coordinates is given in \cref{sec:BS}. The conformal Bondi-Sachs coordinates give us a family of null surfaces \(\mc N_u\) labeled by a coordinate \(u\) and a family of \(2\)-spheres \(S'\) labeled by the coordinates \(\Omega\) and \(u\) along each null surface \(\mc N_u\) such that as \(\Omega \to 0\) the \(2\)-spheres limit to a cross-section \(S\) of \(\scri\). We use this family of \(2\)-spheres to evaluate \cref{eq:Q-xi-theta} and then take the limit \(\Omega \to 0\) along this family.

 With this setup, we take the form of the unphysical metric in Bondi-Sachs coordinates (given by \cref{eq:unphys-BS,eq:bondi-expansion}), the corresponding expressions for the unphysical metric perturbations along with the expression for $\xi^a$ derived in \cref{eq:BMS-Bondi-vec}. We use this to evaluate \cref{eq:Q-xi-theta} using \cref{eq:symp-potential-GR,eq:noether-GR}, after converting the physical metric and physical metric perturbations to their unphysical counterparts. The resulting expression for the \(2\)-form $\delta \df Q_\xi$ has a term that diverges as $\Omega \to 0$, which is given by 
\be \label{eq:div-term}
    \df\varepsilon_{2}~ \Omega^{-1} \lb( 2 X^{A} \delta U^{(2)}_{A} \rb) = - \df\varepsilon_{2}~ \Omega^{-1} X^{A} \ms{D}^{B} \delta C_{AB}\,.
\ee
The indices \(A,B\) are abstract indices for tensor fields on the chosen family of \(2\)-spheres, and \(\ms D_A\) is the covariant derivative on these \(2\)-spheres. The equality \cref{eq:div-term} follows from \cref{eq:order-2} which is a consequence of the (linearized) Einstein equation in the Bondi-Sachs coordinates. Now, since $X^{A}$ is a Lorentz vector field (satisfying \cref{eq:CKV} in the conformal Bondi-Sachs coordinates) and \(\delta C_{AB}\) is traceless, this term vanishes when integrated over the \(2\)-spheres. As a result, the limit of $\int_{S'} \delta \df Q_\xi$ as \(S' \to S\), i.e. \(\Omega \to 0\), is finite. Then, computing the remaining terms in \cref{eq:Q-xi-theta} (which are manifestly finite as \(\Omega \to 0\)) we obtain
\be\label{eq:charge-bondi-2}
    & \lim_{S' \to S} \int_{S'} \lb[\delta \df Q_\xi - \xi \cdot \df\theta(\delta \hat g) \rb] \\
     & = - \frac{1}{8\pi} \delta \int_{S} \df\varepsilon_{2} \big[ \beta\, ( \mc{P} + \tfrac{1}{2} \sigma_{ab} N^{ab} )+ X^a \mc{J}_{a}+ X^{a} \sigma_{ab} \ms{D}_{c} \sigma^{bc} - \tfrac{1}{4} \sigma_{ab} \sigma^{ab} \ms{D}_{c} X^{c} \big] \\
    &\,\quad - \frac{1}{16\pi} \int_{S} \df\varepsilon_{2}~ \beta\, N^{ab} \delta\sigma_{ab} \, .
\ee
where we have used \cref{relations} to convert the metric components in the conformal Bondi-Sachs coordinates to covariant quantities defined on \(\scri\). We can immediately see this expression is non-integrable due to the last term.\footnote{As emphasized in \cref{rem:non-integrable-subtraction} above, this \emph{should not} be taken to mean that the last line above is the ``non-integrable part'' of the expression which should be ``simply subtracted away'' without specifying any additional criteria.}

Using \cref{eq:WZ-correction,eq:BMS-Bondi-vec,eqn:epsilon_32}, we have
\be
    \int_S \xi \cdot \df\Theta(\delta \hat g) = \tfrac{1}{16\pi}\int_S\df\varepsilon_2~ \beta N^{ab} \delta\sigma_{ab}\,.
\ee
Thus we have the perturbed WZ charge
\be
    \delta \mc Q[\xi;S] & = \lim_{S' \to S} \int_{S'} \lb[ \delta \df Q_\xi - \xi \cdot \df\theta(\delta \hat g) \rb] + \int_S \xi \cdot \df\Theta(\delta \hat g) \\
    & = - \frac{1}{8\pi} \delta \int_{S} \df\varepsilon_{2} \big[ \beta\, ( \mc{P} + \tfrac{1}{2} \sigma_{ab} N^{ab} )+ X^a \mc{J}_{a}+ X^{a} \sigma_{ab} \ms{D}_{c} \sigma^{bc} - \tfrac{1}{4} \sigma_{ab} \sigma^{ab} \ms{D}_{c} X^{c} \big]\,.
\ee
Therefore, the expression for the WZ charge is given by
\be\label{eq:Q-GR-defn}
    \mc Q[\xi; S] \hateq - \frac{1}{8\pi} \int_S \df\varepsilon_2 \left[ \beta ( \mc P + \tfrac{1}{2}\sigma^{ab} N_{ab} ) + X^a \mc J_a + X^a \sigma_{ab} (\ms D_c - \tau_c) \sigma^{bc} - \tfrac{1}{4} \sigma_{ab} \sigma^{ab} (\ms D_c - 2 \tau_c) X^c \right] \,.
\ee

Note that in \cref{eq:Q-GR-defn} we have added terms which depend on \(\tau_a\); these terms cancel amongst each other using the identity \cref{eq:the-Grant-equation}. These additional terms make each term in the integrand of conformal weight \(-2\) (which can be verified using the conformal weights given in \cref{eq:beta_X_conf-wt,eq:c-weights,eq:shear-conf-wt,eq:tau-tr}) which, along with the conformal weight \(+2\) of \(\df\varepsilon_2\), makes the charge conformally-invariant. Further, since these \(\tau_a\) terms cancel the charge is independent of the choice of the foliation and only depends on the chosen cross-section \(S\). We will verify below that the flux of this charge across any region \(\Delta\scri\) is given by \cref{eq:flux-formula}. Finally, since the flux vanishes in Minkowski spacetime we can compute the charge on any cross-section of \(\scri\). Using a shear-free cross-section we see that the charge also vanishes on \emph{any} cross-section of \(\scri\) in Minkowski spacetime. Thus, the charge formula in \cref{eq:Q-GR-defn} satisfies all the properties required by the WZ charge.

For a supertranslation symmetry with \(\beta \hateq f\) satisfying \(\Lie_n f \hateq 0\) it is straightforward to verify that \cref{eq:Q-GR-defn} reproduces Geroch's supermomentum \cite{Geroch-asymp,WZ}. Further, if at some chosen cross-section \(S\) we pick \(\beta\vert_S \hateq 0\) then the resulting formula is equal to the linkage charge defined by Geroch and Winicour \cite{GW,WZ}; this was proven in \cite{WZ} and we show this more explicitly in \cref{sec:linkage}. Now, while on any fixed cross-section \(S\) we can decompose a general BMS symmetry into a supertranslation part and a part tangent to \(S\), this decomposition is not preserved along \(\scri\) (see also \cref{rem:lorentz-amb}). Thus, in general the WZ charge is \emph{not} the sum of Geroch's supermomentum with the Geroch-Winicour charge. For an exact Killing vector field $\xi^a$ in the physical spacetime, the WZ charge agrees with the charge that is given by the Komar formula. We discuss in more detail \cref{sec:linkage}. To see this, note that, for exact Killing vector fields, the linkage charge agrees with the Komar formula for Killing vector fields which asymptotically become Lorentz vector fields at $S$, and agrees up to a factor of two for Killing vector fields that become translations at $S$. The WZ charge, similarly, equals the linkage charge for Lorentz vector fields at $S$, and up to a factor of two for translations; as such, the WZ charge and the Komar formula agree for exact Killing vector fields.
Note that the Komar formula is (up to constant factors) the integral of the Noether charge in \cref{eq:noether-GR}.\\

By the general arguments following \cref{eq:delta-F1}, the change in the charge \cref{eq:Q-GR-defn} between two cross-sections is given by the flux formula \cref{eq:flux-formula}. However, showing this explicitly is a non-trivial computation, which we detail in the remainder of this section.

Let \(S_2\) and \(S_1\) be \emph{any} two cross-sections of \(\scri\), with \(S_2\) to the future of \(S_1\), and let \(\Delta\scri\) be the portion of \(\scri\) bounded by these cross-sections. Then, the change in the charge is given by
\be\label{eq:flux1}
    & \mc F[\xi;\Delta\scri] \hateq \mc Q[\xi; S_2] - \mc Q[\xi; S_1] \\
    & \hateq - \frac{1}{8\pi} \int_{\Delta\scri} \df\varepsilon_3 \Lie_n \left[ \beta ( \mc P + \tfrac{1}{2}\sigma^{ab} N_{ab} ) + X^a \mc J_a + X^a \sigma_{ab} \ms D_c \sigma^{bc} - \tfrac{1}{4} \sigma_{ab} \sigma^{ab} \ms D_c X^c \right] \,.
\ee
Note that we have dropped the \(\tau_a\) terms from the expression since they do not contribute to the charge as explained above.

Now we simplify \cref{eq:flux1} term-by-term starting with the first and second terms. Using $\mc J_{a} n^{a} \hateq 0$, as well as \cref{eq:lie-P,eq:lie-J,eq:weyl-News,eqn:lie_n_beta,eqn:lie_n_X} and the integration-by-parts formula \cref{eq:IBP-D3}, we see that the first two terms in \cref{eq:flux1} contribute 
\be \label{contrib3}
    & - \frac{1}{8\pi} \int_{\Delta \scri}\df\varepsilon_{3}\bigg[ N^{ab} \bigg( \tfrac{1}{4} \beta N_{ab} + \tfrac{1}{2}(\ms{D}_{a} + \tau_{a})(\ms{D}_{b}-\tau_{b})\beta + (\ms D_{a} + \tau_{a}) (X^{c} \sigma_{bc}) + \tfrac{1}{4} \sigma_{ab}(\ms D_{c} -\tau_{c} )X^{c}\bigg) \\
   &\qquad\qquad  - \tfrac{1}{2}\varepsilon_{a}{}^{b} X^{a}\big(\ms D_{b} + 3\,  \tau_{b}) \mc P^{\ast}\bigg]\,, 
\ee
to the flux. Now consider the contribution of the third term in \cref{eq:flux1}:
\be\label{eq:temp1}
    \Lie_{n} (X^{a} \sigma_{ab} \ms{D}_{c}\sigma^{bc}) \hateq \tfrac{1}{2} X^{a} N_{ab} \ms{D}_{c}\sigma^{bc} + X^{a} \sigma_{ab} \Lie_n \ms{D}_{c}\sigma^{bc}
\ee
where we have used \cref{eqn:lie_n_X} and \cref{eqn:news_l_def}. For the last term above we have
\be \label{first}
    \Lie_n \ms D_{c} \sigma^{bc} & \hateq \Lie_n (Q^{a}{}_{c} Q^{b}{}_{d}) \nabla_{a} \sigma^{cd} + Q^{a}{}_{c} Q^{b}{}_{d} \Lie_n \nabla_{a} \sigma^{cd} \\
    & \hateq (Q^{a}{}_{c} n^{b}\tau_{d} + Q^{b}{}_{d} n^{a} \tau_{c}) \nabla_{a} \sigma^{cd} + \tfrac{1}{2} \ms D_{a} N^{ab} \\
    & \hateq (Q^{a}{}_{c} n^{b}\tau_{d} ) \nabla_{a} \sigma^{cd} + \tfrac{1}{2} (\ms D_{a} + \tau_a) N^{ab}\,,
\ee
where we have made liberal use of the Bondi condition (\cref{eq:Bondi-cond}) along with \cref{eq:Q-defn,eqn:lie_n_l}, and commuted the \(\Lie_n\) past the \(\nabla_a\) using $\sigma^{ab} n_{a} \hateq Q_{ab} n^{a} \hateq 0$. Using the above expression in \cref{eq:temp1} the first term vanishes by \(\sigma_{ab} n^b \hateq 0\) and we get
\be \label{contrib1}
    \Lie_{n} (X^{a} \sigma_{ab} \ms{D}_{c}\sigma^{bc}) \hateq \tfrac{1}{2} X^{a} N_{ab} \ms D_{c} \sigma^{bc} + \tfrac{1}{2} (\ms D_{c} + \tau_c )(X^{a}\sigma_{ab} N^{bc}) - \tfrac{1}{2} N_{bc} \ms D^{b} (X_{a} \sigma^{ac})\,.
\ee
Note that the second term on the right-hand side above drops out of the flux formula using \cref{eq:IBP-D3}. 

Consider now the fourth term in \cref{eq:flux1}. Since \(\alpha_{(\xi)} \hateq \tfrac{1}{2} \ms D_a X^a\) and \(\Lie_n \alpha_{(\xi)} \hateq 0\) (\cref{eqn:lie_xi_n}), using \cref{eqn:news_l_def} we get
\be \label{contrib2}
-\tfrac{1}{4} \Lie_{n} (\sigma_{ab} \sigma^{ab} \ms{D}_{c} X^{c}) \hateq -\tfrac{1}{4} \sigma^{ab} N_{ab}\ms D_{c} X^{c}\,.
\ee
Putting together \cref{contrib3,contrib1,contrib2}, we see that the flux is given by
\be \label{flux}
    \mc F[\xi;\Delta\scri] &\hateq - \frac{1}{8\pi}\int_{\Delta \scri} \df\varepsilon_{3} \bigg[ N^{ab} \bigg( \tfrac{1}{4} \beta N_{ab} + \tfrac{1}{2}(\ms{D}_{a} + \tau_{a})(\ms{D}_{b}-\tau_{b})\beta + \tfrac{1}{2}  (\ms D_{a}+2 \tau_{a}) (X^{c} \sigma_{bc}) \\
    &\qquad - \tfrac{1}{2} \sigma_{ab}\tau_{c} X^{c}
+ \tfrac{1}{2} X^{a} \ms D_{c} \sigma^{bc}\bigg) - \tfrac{1}{2} \varepsilon_{a}{}^{b} X^{a}( \ms D_{b} + 3\tau_{b})\mc P^*\bigg]\,. 
\ee
Next we simplify the last term using the identity \cref{eq:Pstar-identity} for \(\mc P^*\). The term arising from this identity which involves derivatives of the shear reads $\varepsilon_{3}\,\varepsilon^{de}\varepsilon_{a}{}^{b} X^{a} (\ms{D}_{b} + 3 \tau_{b}) [\ms D_d (\ms D_c - \tau_c) \sigma_e{}^c] $. This term can be shown to vanish upon integrating over the cross-sections as follows. Note that this term is conformally-invariant  (which can be seen using \cref{eq:Pstar-identity} and the conformal weights given in \cref{eq:fol-conf-wts,eq:beta_X_conf-wt,eq:shear-conf-wt,eq:c-weights}). Therefore, we can evaluate this term in the Bondi frame where the metric on the cross-sections is chosen to be the unit round metric and $\tau_{a}\hateq 0$. This allows us to make use of spherical harmonics --- $\sigma_{ab}$ is a symmetric and trace-free tensor and thus is supported on $\ell \geq 2$ tensor spherical harmonics while $X^{a}$ is supported on $\ell=1$ vector spherical harmonics (see \cref{rem:bms-harmonics}). Using the orthogonality of the spherical harmonics, the integral of this term over the cross-sections vanishes. The remaining term arising from \cref{eq:Pstar-identity} contains the News tensor which, after an integration-by-parts using \cref{eq:IBP-D3}, becomes
\be
-\tfrac{1}{4}\varepsilon^{ab}\varepsilon_e{}^{d}  N_{ac} \sigma_b{}^c (\ms{D}_{d}-2 \tau_{d}) X^{e} \hateq \tfrac{1}{2} N_{ab} \sigma^b{}_{c} \ms D^{[a} X^{c]} - N_{ab} \sigma^b{}_{c} X^{[c} \tau^{a]}\,.
\ee
Replacing the above for the last term in \cref{flux} we get
\be\label{eq:intermediate}
\mc F[\xi;\Delta\scri] & \hateq - \frac{1}{8\pi} \int_{\Delta \scri} \df\varepsilon_{3}N^{ab} \bigg[\tfrac{1}{4} \beta N_{ab} + \tfrac{1}{2}(\ms{D}_{a} + \tau_{a})(\ms{D}_{b}-\tau_{b})\beta + \tfrac{1}{2} (\ms D_{a} +2 \tau_{a}) (X^{c} \sigma_{bc}) \\ 
    &\qquad\qquad -\tfrac{1}{2} \sigma_{ab}\tau_{c} X^{c}+ \tfrac{1}{2} X_{a} \ms D^{c} \sigma_{bc} +\tfrac{1}{2}\sigma_b{}^c \ms D_{[a} X_{c]} - \sigma_b{}^{c} X_{[c} \tau_{a]}\bigg] \,.
\ee
The last term on the first line and first term on the second line in the expression above can together be written as 
\be\label{intermediate-eq}
    N^{ab}(X^{c} \sigma_{bc} \tau_{a}- \tfrac{1}{2} \sigma_{ab}\tau_{c} X^{c}) \hateq N^{ab} \sigma_b{}^c (X_c \tau_{a} - \tfrac{1}{2} Q_{ca} X_{d} \tau^{d}) \,. 
\ee
Moreover, it follows from \cref{eq:the-Grant-equation} that $N_{ab} \sigma^{b}{}_{c} \hateq \text{pure trace term}+ N_{b[a} \sigma^b{}_{c]}$  and so \cref{intermediate-eq} becomes $N^{ab} \sigma_b{}^{c} X_{[c} \tau_{a]}$. This exactly cancels the last term in \cref{eq:intermediate}, and therefore the flux formula simplifies to
\be\label{eq:final-flux}
\mc F[\xi;\Delta\scri] \hateq - \frac{1}{8\pi}\int_{\Delta \scri} \df\varepsilon_{3} N^{ab} \bigg[& \tfrac{1}{4} \beta N_{ab} + \tfrac{1}{2} (\ms{D}_{a} + \tau_{a}) (\ms{D}_{b} - \tau_{b})\beta  \\
    & + \tfrac{1}{2} \ms D_{a} (X^{c} \sigma_{bc}) + \tfrac{1}{2} X_{a} \ms D^{c} \sigma_{bc} +\tfrac{1}{2}\sigma_b{}^c \ms D_{[a} X_{c]}\bigg] \,.
\ee
The second line above can be simplified using \cref{eq:diff-grant} to get
\be\label{eq:final-flux-2}
  \mc F [\xi; \Delta \scri] \hateq -\frac{1}{16\pi} \int_{\Delta \scri} \df \varepsilon_3 N^{ab} [\tfrac{1}{2} \beta N_{ab}+(\mathscr{D}_a + \tau_a) (\mathscr{D}_b - \tau_b) \beta + \Lie_X \sigma_{ab} - \tfrac{1}{2} (\mathscr{D}_c X^c) \sigma_{ab}].
\ee
which matches the WZ flux derived in \cref{eq:flux-formula}.

\section{Expressions in some coordinate systems}
\label{sec:coord}

Our entire preceding analysis was completely covariant, without referring to any particular coordinate  systems. In this section, we consider two examples of coordinate systems for the unphysical spacetime in a neighbourhood of null infinity. These coordinates can be used to also obtain asymptotic coordinates for the physical spacetime. We will find the asymptotic form of the metric, both physical and unphysical, and derive the expressions for the BMS symmetries and their charges in these coordinates.

As described in \cref{sec:univ-str}, one can construct a geometrically defined coordinate system \((\Omega,u,x^A)\)  at \(\scri\) where \(x^A\) are coordinates on the cross-sections of \(\scri\), \(u\) satisfies \(n^a \nabla_a u \hateq 1\) and the conformal factor \(\Omega\) is chosen so that the Bondi condition is satisfied and so that the induced metric on the cross-sections is the unit round metric \(s_{AB}\) on \(\bb S^2\). Then for any asymptotically flat spacetime the line element of the unphysical metric at \(\scri\) is
\be\label{eq:g-at-scri}
	ds^2 \hateq 2 d\Omega du + s_{AB} dx^A dx^B \,.
\ee
These coordinates can be extended away from \(\scri\) in different ways, and these give rise to the different coordinates that are often used in the analysis of asymptotic symmetries and their associated charges. Two commonly used coordinates are the Bondi-Sachs coordinates and conformal Gau{\ss}ian null coordinates, and we will focus on these in the remainder of this section. We emphasize that the form of the (physical or unphysical) metric in these coordinates follows directly from the construction of the coordinate systems and the covariant definition of asymptotic flatness without any additional assumptions. We also show that in any such coordinate system one can obtain the BMS symmetries as coordinate transformations which preserve the asymptotic form of the metric.

\subsection{Bondi-Sachs coordinates}
\label{sec:BS}

One way to extend the coordinates described above away from \(\scri\) is as follows. Let \(S_u\) be cross-sections of \(\scri\) with \(u = \text{constant}\), and consider a family of null surfaces \(\mc N_u\) which intersect \(\scri\) transversely in the cross-sections \(S_u\). These surfaces \(\mc N_u\) foliate a neighbourhood of \(\scri\) by null surfaces. We first extend the coordinate \(u\) away from \(\scri\) so that it is constant along each null surface \(\mc N_u\). Then \(l_a \defn - \nabla_a u\) is the null normal to each \(\mc N_u\) with \(l^a l_a = 0\) in addition to \(l^a n_a \hateq -1\) as above. Then we extend the angular coordinates \(x^A\) on each cross-section \(S_u\) by parallel transport i.e. \(l^a \nabla_a x^A = 0\).

Fixing the induced metric on cross sections of $\scri$ to be the unit round sphere metric fixes the conformal factor, $\Omega$, on $\scri$. To extend \(\Omega\) away from \(\scri\), we use the freedom in the conformal
factor off of \(\scri\) to demand that the \(2\)-spheres at
constant \(u\) and \(\Omega\) have the same area element as the unit sphere,
that is, if \(h_{AB}\) is the \(2\)-metric on the surfaces of constant \(u\) and \(\Omega\) then we demand that \(\det h = \det s\) in the \(x^A\)-coordinates. This fixes \(\Omega\) uniquely away from \(\scri\). Thus we have set up the conformal Bondi-Sachs coordinate system \((\Omega, u, \theta^A) \) in a neighborhood of \(\scri\).

The most general form of the unphysical metric in conformal Bondi-Sachs coordinates is given by\footnote{Note that the function we denote by \(B\) is conventionally denoted by \(\beta\), but we use a different symbol to avoid conflict with the \(n^a\)-component of a BMS vector field.}
\be\label{eq:unphys-BS}
	ds^2 \equiv - W e^{2 B} du^2 + 2 e^{2 B} d\Omega du + h_{AB} (dx^A - U^A du )(dx^B - U^B du )\,,
\ee
where the metric components \(g_{\Omega \Omega}\) and \(g_{\Omega A}\) vanish everywhere due to the conditions \(l^a l_a = l^a \nabla_a x^A = 0\), and \(W\), \(B\), \(h_{AB}\), and \(U^A\) are smooth functions of the coordinates \((\Omega, u, x^A)\). Since the metric at \(\scri\) is given by \cref{eq:g-at-scri} we also have
\be 
 W = O(\Omega) \eqsp B = O(\Omega) \eqsp U^A = O(\Omega) \eqsp h_{AB} = s_{AB} + O(\Omega)\,.
\ee
Further, evaluating the Bondi condition \(\nabla_a n_b \hateq 0\)  (\cref{eq:Bondi-cond}) gives
\be
     W = O(\Omega^2) \eqsp B = O(\Omega^2) \eqsp U^A = O(\Omega^2) \,.
\ee
We therefore consider the following expansion of the metric components
\be\label{eq:bondi-expansion}
W = \Omega^2 W^{(2)} - 2\Omega^3 M + O(\Omega^4) \eqsp& U^A = \Omega^2 U^{(2)}{}^A + 2 \Omega^3 L^A + O(\Omega^4)\,, \\
B = \Omega^2 B^{(2)} + O(\Omega^3) \eqsp& h_{AB} = s_{AB} + \Omega C_{AB} + \Omega^2 d_{AB} + O(\Omega^3)\,.
    \ee
Next, imposing \(\det h = \det s\), we get
\be \label{eq:detheq}
  s^{AB} C_{AB}  = 0 \eqsp s^{AB} d_{AB} = \tfrac{1}{2} C^{AB} C_{AB}\,.
 \ee
  We then impose the Einstein equation, \cref{eq:S-ee}, order by order in $\Omega$. At $O(\Omega^{0})$, it gives
 \be\label{eq:order-2}
        W^{(2)} = 1 \eqsp B^{(2)} = - \tfrac{1}{32} C^{AB} C_{AB} \eqsp U^{(2)}_A = - \tfrac{1}{2} \ms D^B C_{AB}\,,
    \ee
while at $O(\Omega)$ we get
\be
    \partial_{u} \STF d_{AB} = \ms{D}^{A} \STF d_{AB} = 0\,.
\ee
 Since \(d_{AB}\) is a smooth tensor on a \(2\)-sphere this implies that \(\STF d_{AB} = 0\) and thus (from \cref{eq:detheq})
\be\label{eq:d-form}
    d_{AB} =\tfrac{1}{4} C^{CD} C_{CD} s_{AB}\,.
\ee

The metric component \(C_{AB}\) is related to the shear and the News tensor while \(M\) and \(L_A\) are related to the Weyl tensor components (\cref{eq:weyl-defn}) through
\be\label{relations}  
    \sigma_{AB} = -\tfrac{1}{2} C_{AB} &\eqsp N_{AB} = - \partial_u C_{AB}\,, \\
     \mc{P} =-2 M + \tfrac{1}{4} C_{AB} N^{AB} &\eqsp \mc{J}_{A} = 3 L_{A} -\tfrac{3}{32} \ms{D}_{A} (C_{BC} C^{BC}) - \tfrac{3}{4} C_{A}{}^{B} \ms{D}^{C} C_{BC}\,.
\ee

Imposing the Einstein equation to higher order in \(\Omega\) either relates the higher order metric components to the lower order ones or gives evolution equations along \(u\) ---  for instance, one gets equations for \(\partial_u M\) and \(\partial_u L_A\) which are equivalent to \cref{eq:weyl-evol} using \cref{relations}. We will not need the explicit form of these higher order equations in our analysis.\\

The conformal Bondi-Sachs coordinates defined above can be used to define the \emph{physical Bondi-Sachs coordinates} \((r,u,x^A)\) which are often used in the asymptotic analysis near \(\scri\) \footnote{We present these equations in the two-sphere covariant form appearing in~\cite{TW}, and in a more modern form in~\cite{Jezierski1997, Flanagan:2015pxa}, instead of the original notation of~\cite{BBM}.}. Define the physical ``radial coordinate'' \(r \defn \Omega^{-1}\), and using \((r,u,x^A)\) as coordinates, the physical metric \(\hat g_{ab} = \Omega^{-2} g_{ab} = r^2 g_{ab}\) has the line element
\be\label{eq:phys-BS}
    d\hat s^2 \equiv - U e^{2B} du^2 - 2 e^{2B} du dr + r^2 h_{AB} (d x^A - U^A du ) (d x^B - U^B du )\,,
\ee
where, from \cref{eq:unphys-BS,eq:bondi-expansion,eq:order-2,eq:d-form}, we have the asymptotic expansions
\be\label{eq:phys-BS-exp}
    U = r^2 W & = 1-\tfrac{2}{r}M + O(1/r^2)\,, \\
    B & = - \tfrac{1}{32 r^{2}} C^{AB}C_{AB} + O(1/r^3)\,, \\
    U^A & = -\tfrac{1}{2 r^{2}} \ms{D}_{B} C^{AB} + \tfrac{2}{r^{3}} L^{A} + O(1/r^4)\,, \\
    h_{AB} & = s_{AB} + \tfrac{1}{r} C_{AB} +\tfrac{1}{4 r^{2}} s_{AB} C^{CD} C_{CD} + O(1/r^3)\,.
\ee
Note that since \(\Omega\), and hence \(r\), is chosen so that \(\det h = \det s\), the area (in the physical metric) of the \(2\)-spheres of constant \(u\) and \(r\) is precisely \(4 \pi r^2\). Thus \(r\) is a ``radial coordinate'' along the outgoing null surfaces \(\mc N_u\) as constructed by Bondi and van der Burg \cite{BBM}.  We emphasize that the Bondi-Sachs asymptotic form of the physical metric (\cref{eq:phys-BS,eq:phys-BS-exp}) directly follows from the covariant definition of asymptotic flatness in a particular choice of coordinate system and \emph{does not} involve any additional assumptions.\\

Having chosen the conformal Bondi-Sachs coordinate system, the asymptotic BMS symmetries are coordinate transformations which preserve the Bondi-Sachs form of the unphysical metric. We shall only consider the infinitesimal coordinate transformations, i.e., we take \((\Omega, u, x^A)\) and \((\Omega', u', {x'}^A)\) to be two conformal Bondi-Sachs coordinates, as constructed above, related by an infinitesimal coordinate transformation parametrized by a vector field \(\xi^a\). We next obtain an expression for this vector field in the coordinate system \((\Omega,u,x^A)\).

Since in both coordinates \((\Omega, u, x^A)\) and \((\Omega', u', {x'}^A)\), null infinity \(\scri\) lies at \(\Omega = \Omega' = 0\), the component \(\xi^\Omega\) vanishes at \(\Omega = 0\), i.e., \(\xi^a\) must be tangent to \(\scri\). Thus, \(\xi^a\) can be written in the coordinate system \((\Omega, u, x^A)\) as
\be
    \xi^a  \equiv \beta \partial_{u} + X^{A} \partial_{A} + \Omega Z^{a} \partial_{a} + \Omega^{2} W^{a} \partial_a + O(\Omega^{3})\,,
\ee
where each of $\beta, X^{A}, Z^{a}$ and $W^{a}$ are functions of $(u,x^{A})$.

Next we note that, since we are using the conformal factor itself as a coordinate, an infinitesimal change in the coordinate system is accompanied by an infinitesimal change of the conformal factor parametrized by the component \(\xi^\Omega\). Thus, when changing the coordinate system, the unphysical metric changes infinitesimally by \(\Lie_{\xi} g_{ab} - 2 \Omega^{-1} \xi^\Omega g_{ab}\); note that this is finite at \(\scri\) since \(\xi^\Omega\) vanishes there. We now require that this change preserve the Bondi-Sachs form of the metric obtained in \cref{eq:unphys-BS} along with \cref{eq:bondi-expansion} and the equations below.

As discussed above, the metric on $\scri$, given by \cref{eq:g-at-scri}, is universal and so we require that at \(O(\Omega^0)\), $\Lie_{\xi} g_{ab} - 2 \Omega^{-1} \xi^\Omega g_{ab} \hateq 0$. This gives us the following conditions~\cite{Sachs1962symm}: $X^{A}$ is constant along \(u\) and satisfies the conformal Killing equation on the cross-sections, that is,
\be \label{eq:CKV}
    \ms{D}_{(A} X_{B)}= \tfrac{1}{2} q_{AB} \ms{D}_{C}X^{C} \eqsp \partial_{u} X^{A}=0  \,,
\ee
while the components of \(Z^a\) satisfy
\be \label{eq:bondivecfield-cond1}
    Z^{\Omega} = \partial_{u} \beta = \tfrac{1}{2} \ms{D}_{A} X^{A}\eqsp Z^{u}=0 \eqsp Z_{A} = -\ms{D}_{A} \beta\,.
\ee
The first condition in \cref{eq:bondivecfield-cond1} allows us to solve for the \(u\)-dependence of \(\beta\) to get
\be \label{beta-bondi}
\beta = f + \tfrac{1}{2} (u-u_0) \ms{D}_{A} X^{A}\,,
\ee
where $f$ is an arbitrary function of $x^{A}$ which denotes the value of \(\beta\) at some choice of cross-section with \(u = u_0\).

At $O(\Omega)$, requiring that the form of the unphysical metric be preserved, we further obtain
\be
 W^{u} =0 \,, \quad W_{A} = \tfrac{1}{2} C_{AB} \ms{D}^{B}\, \beta\,,
\ee
and the requirement that the metric component $C_{AB}$ remain trace-free with respect to the unit round sphere metric gives us
\be
W^{\Omega} = \tfrac{1}{2} \ms{D}_{A} Z^{A} = -\tfrac{1}{2} \ms{D}^{2} \beta\,.
\ee
At this order one also obtains the transformation of \(C_{AB}\) under a BMS symmetry which, upon using \cref{relations}, coincides with \cref{eq:shear-trans} (see also Eq.~2.18b of \cite{Flanagan:2015pxa}).

Putting all of this together, we obtain
\be \label{eq:BMS-Bondi-vec}
    \xi^a \equiv \beta \partial_{u} + X^{A} \partial_{A} + \Omega (- \ms{D}^{A} \beta\, \partial_{A} + \tfrac{1}{2} \ms{D}_{A} X^{A}\partial_{\Omega} ) + \tfrac{1}{2} \Omega^{2} (- \ms{D}^{2} \beta\, \partial_{\Omega} + C^{AB} \ms{D}_{B} \beta\, \partial_{A})  + O(\Omega^{3})\,.
 \ee
Up to $O(\Omega)$, this expression agrees with the covariant expression derived in \cref{sec:bms}. The form of the \(O(\Omega^2)\) terms here is fixed by the choice of the conformal Bondi-Sachs coordinates and that also matches Eq.~2.16 of \cite{Flanagan:2015pxa}. One could continue this computation to higher orders in \(\Omega\), which gives expressions for higher order terms of the components of \(\xi^a\) (which appear in Eqs. III.5-7 of~\cite{Sachs1962symm}) and also the transformation laws for the various metric components, but we will not require these expressions.

Using the relations \cref{relations} in \cref{eq:Q-GR-defn} the WZ charge on \emph{any} cross-section \(S\) of \(\scri\) can be written in terms of the metric components in the Bondi-Sachs form of the metric to get 
\be \label{eq:charge-bondi}
    \mc{Q}[\xi,S] &\hateq - \frac{1}{8\pi} \int_{S} \df\varepsilon_{2} \bigg[-2 M \beta + X^{A} \big(3 L_{A} - \tfrac{1}{32} \ms{D}_{A} (C_{BC} C^{BC}) - \tfrac{1}{2} C_{A}{}^{B} \ms{D}^{C} C_{BC} \big) \bigg] \,.
\ee 
The above charge expression matches the charge expression given by Flanagan and Nichols in Eq.~3.5 of \cite{Flanagan:2015pxa}, even though their expression was calculated on a cross section of $\scri$ where the News vanishes. A similar expression was also obtained by Barnich and Troessaert \cite{BT} with non-vanishing News but they do not find the integrable charge. 

From the above expression we also see that the function \(M\) determines the Bondi mass at any cross-section of \(\scri\) and can be called the \emph{mass aspect}. In the Bondi-Sachs coordinates it coincides with the (constant) mass parameter of Kerr spacetimes. Similarly, in the usual choice of Bondi-Sachs coordinates in Kerr spacetime the angular momentum parameter \(a\) appears in the metric component \(L_A\) \cite{CJK}. Note however, on cross-sections with shear, there are other terms containing the \(C_{AB}\) in the charge formulae. In any case, since there is no preferred Lorentz subalgebra in the BMS symmetries (see \cref{rem:lorentz-amb}), there is no preferred notion of ``angular momentum aspect'' --- so identifying any particular metric component with ``the angular momentum'' is a moot point. 

\subsection{Conformal Gau{\ss}ian null coordinates}
\label{sec:CGNC}

Instead of extending the coordinates $(\Omega, u,x^{A})$ away from \(\scri\) along null hypersurfaces, we can extend them into the (unphysical) spacetime along affine null geodesics, transverse to $\scri$. We recall this construction below which leads to the \emph{conformal Gau\ss ian null coordinates} in a neighborhood of $\scri$.

We fix the conformal factor to leading order away from \(\scri\) as follows. Consider the \emph{expansion} of \(l^a\) at \(\scri\) defined by
\be
    \vartheta \defn Q^{ab}\nabla_a l_b \hateq \tfrac{1}{2} Q^{ab} \Lie_l g_{ab}\,.
\ee
We can set this expansion to vanish by suitably choosing the conformal factor away from \(\scri\) as follows. Note that the conformal factor on \(\scri\) has already been fixed so that the induced metric  on cross-sections of \(\scri\) is the unit-sphere metric. So consider a new conformal factor \(\tilde\Omega = \omega \Omega\) (with \(\omega \hateq 1\)) so that \(\tilde g_{ab} = \omega^2 g_{ab}\). Then, the expansion of the new auxiliary normal \(\tilde l^a \hateq l^a\) can be computed to be (the behaviour of the auxiliary normal away from \(\scri\) is not relevant here)
\be
    \tilde\vartheta \hateq \tfrac{1}{2} \tilde Q^{ab} \Lie_{\tilde l} \tilde g_{ab} \hateq \vartheta + 2 \Lie_l \omega\,.
\ee
Then, choosing \(\omega\) to be any solution of \(\vartheta + 2 \Lie_l \omega = 0\) with \(\omega\hateq 1\) we can set \(\tilde\vartheta \hateq 0\). In the rest of this section we work the choice of conformal factor where the auxiliary normal is expansion-free at \(\scri\) --- we drop the ``tilde'' from the notation.

Having made this choice, we then extend the vector field $l^{a} \hateq -\partial/\partial\Omega$ away from \(\scri\) such that it is the generator of affine null geodesics so that \(l^b \nabla_b l^a  = 0\). We can further use the remaining freedom in the conformal factor $\Omega$ so that \(\Omega\) is the affine parameter along these null geodesics generated by \(l^a\). To summarize, we can always choose the conformal factor \(\Omega\) and extend the auxiliary normal away from \(\scri\) so that in a neighbourhood of \(\scri\) we have
\be
    l^a = - \frac{\partial}{\partial\Omega} \eqsp \vartheta \hateq 0 \eqsp l^a l_a = 0 \eqsp l^b \nabla_b l^a = 0 .
\ee
 Finally, we extend \((u,x^A)\) into the spacetime by parallel-transport along \(l^{a}\), that is, we require
\be
l^{a} \nabla_{a} u=l^{a}\nabla_{a}x^{A}=0\,.
\ee
This construction gives us the conformal Gau\ss ian null coordinates in a neighbourhood of $\scri$.

The most general form of the unphysical metric in these coordinates is given by
\be\label{eq:unphys-CGN-metric}
    ds^2 & = 2 du (d\Omega - \alpha du - \beta_A dx^A) + h_{AB} dx^A dx^B\,,
\ee
To see why this is the most general form, note that \(g_{\Omega\Omega} = 0\) by \(l^al_a = 0\) and \(g_{\Omega A} = 0\) by \(l^a\nabla_a x^A = 0\). Then, \(l^b \nabla_b l^a = 0\) gives \(\pd[g_{u\Omega}]{\Omega} = 0\) which implies \(g_{u\Omega} = 1\) by the condition \(l^a n_a \hateq -1\). Further,
\be
    \alpha = O(\Omega) \eqsp \beta_A = O(\Omega) \eqsp h_{AB} = s_{AB} + O(\Omega)\,,
\ee
since the metric at \(\scri\) is given by \cref{eq:g-at-scri}.
Imposing the Bondi condition \(\nabla_a n_b \hateq 0\) leads to the following conditions
\be
    \alpha = O(\Omega^2) \eqsp \beta_A = O(\Omega^2)\,.
\ee
We therefore consider the asymptotic expansions
\be\label{eq:CGN-expansions}
    &\alpha = \Omega^2 \alpha^{(2)} + \Omega^3 \alpha^{(3)} + O(\Omega^4) \eqsp \beta_A = \Omega^2 \beta^{(2)}_A + \Omega^3 \beta_A^{(3)} + O(\Omega^4) \\
    & h_{AB} = s_{AB} + \Omega C_{AB} + \Omega^2 h^{(2)}_{AB} + O(\Omega^3)\,.
\ee
The condition that the expansion of $l^{a}$  vanishes on $\scri$ gives us
\be
    s^{AB} C_{AB} = 0\,.
\ee
We then impose the Einstein equation \cref{eq:S-ee}. At \(O(\Omega^0)\), this gives the conditions
\be\label{eq:GNC-order-2}
    \alpha^{(2)} = \tfrac{1}{2} \eqsp \beta^{(2)}_A = - \tfrac{1}{2} \ms D^B C_{AB} \eqsp s^{AB} h^{(2)}_{AB} = \tfrac{1}{4} C^{AB} C_{AB}\,,
\ee
while at \(O(\Omega)\), it implies 
\be
 \ms D^{B} h^{(2)}_{AB} = \tfrac{1}{8} \ms{D}_{A}(C_{BC} C^{BC}) \eqsp \partial_{u} h^{(2)}_{AB} = \tfrac{1}{8} s_{AB} \partial_{u}(C_{CD}C^{CD})\,.
\ee
As in the Bondi-Sachs case, this implies that \(h^{(2)}_{AB}\) is pure trace.

The metric coefficients \(C_{AB}\), \(\alpha^{(3)}\) and \(\beta_A^{(3)}\) are directly related to the shear and the Weyl tensor components as follows
\be\label{eq:GNC-weyl}
    \sigma_{AB} = -\tfrac{1}{2} C_{AB} \eqsp \mc P = 2 \alpha^{(3)} \eqsp \mc J_A = \tfrac{3}{2} \beta_A^{(3)} \,.
\ee
One could continue computing the Einstein equations to higher orders but we will not need to do so.\\

To write an expression for the physical metric define \(\lambda \defn \Omega^{-1}\) so that in the coordinates \((\lambda, u, x^A)\) the physical metric \(\hat g_{ab} = \Omega^{-2} g_{ab} = \lambda^2 g_{ab}\) has the components
\be\label{eq:phys-metric-GNC}
    \hat{g}_{\lambda\lambda} & =0 \eqsp
    \hat{g}_{A\lambda}=0 \eqsp 
    \hat{g}_{u\lambda}=-1 \eqsp \\
    \hat{g}_{uu} & =-1 - \frac{1}{\lambda} \mc{P}+ O(1/\lambda^2) \eqsp \\
    \hat{g}_{uA} & =\tfrac{1}{2} \ms{D}^{B}C_{AB}- \tfrac{2}{3 \lambda}  \mc{J}_{A} + O(1/\lambda^2) \eqsp \\
    \hat{g}_{AB} & = \lambda^{2} s_{AB} + \lambda C_{AB}+ \tfrac{1}{8} s_{AB} C_{CD} C^{CD} + O(1/\lambda) \,. 
\ee
Note that the vector field
\be
    \hat l^a \equiv \pd{\lambda} = - \Omega^2 \pd{\Omega} = \Omega^2 l^a\,,
\ee
generates outgoing null geodesics which are affinely parametrized with respect to the physical metric \(\hat g_{ab}\) with the affine parameter being \(\lambda\). \cref{eq:phys-metric-GNC} is consistent with Eqs.~5 and 58 of \cite{HIW} put together (with the additional condition that our \(C_{AB}\) is tracefree since we picked \(l^a\) to be expansion-free; see \cref{rem:CGNC-comparison}). This also gives the physical metric in the coordinates used in the \emph{affine-null} form used in \cite{an-Sachs,an-W}, as well as the Newman-Unti coordinates \cite{NU}.

We can also derive the form of the BMS vector fields by considering infinitesimal coordinate transformations between two conformal Gau{\ss}ian null coordinate systems and demanding that the conditions on the unphysical metric derived above be preserved. Since the computation proceeds exactly as in the case of Bondi-Sachs coordinates detailed above, we skip the details. The end result is that the BMS vector field in conformal Gaussian null coordinates takes the same form as \cref{eq:BMS-Bondi-vec} above --- the difference in the form of the BMS vector fields written in conformal Gau{\ss}ian null coordinates and conformal Bondi-Sachs coordinates only appears at \(O(\Omega^3)\) and higher. Note that this form is different than the one obtained by \cite{HIW} since our coordinates differ slightly from theirs as explained in \cref{rem:CGNC-comparison} below. The WZ charge (\cref{eq:Q-GR-defn}) can be straightforwardly written in these coordinates using \cref{eq:GNC-weyl}.

\begin{remark}[Comparison of different conformal Gaussian null coordinates]
\label{rem:CGNC-comparison}
    The conformal Gau{\ss}ian null coordinates constructed in this section are closely related to, but not the same as, the ones used in \cite{Thorne,Hol-Th,HIW}. Note that these references use the freedom in the conformal factor to set the metric coefficient \(\alpha^{(2)} = \half\) (as in \cref{eq:GNC-order-2}) and then the Einstein equations imply that the expansion \(\vartheta\) of the auxiliary normal \(l^a\) is constant along \(n^a\) on \(\scri\), i.e., \(\Lie_n \vartheta \hateq 0\). In contrast, we used the conformal freedom to set \(\vartheta \hateq 0\) and then \(\alpha^{(2)} = \half\) follows from the Einstein equation.
\end{remark}

\section{Discussion}
\label{sec:disc}

The aim of this paper was to obtain, using the Wald-Zoupas prescription, manifestly covariant expressions for the charges and fluxes corresponding to the Bondi-Metzner-Sachs (BMS) symmetries at null infinity in asymptotically flat spacetimes in vacuum general relativity. While (special cases of) these expressions have appeared in various places in the literature, they are usually written in a way that obscures their covariant and conformally-invariant nature, for example, by restricting to specific coordinate frames near null infinity. The expressions we obtained are manifestly covariant, conformally-invariant and do not rely on a preferred choice of foliation of null infinity or on any arbitrary extension of the BMS symmetries away from null infinity. We also recast our charge expression in two specific choices of coordinate frames near null infinity and showed, for example, that in Bondi-Sachs coordinates, it reduces to the expression obtained by Flanagan and Nichols, even on cross-sections of $\scri$ where the News tensor is non-vanishing. In \cref{sec:AS,sec:twistor,sec:linkage}, we compare the Wald-Zoupas charge and flux formulae to some other expressions that have appeared in the literature, including the Ashtekar-Streubel flux formula, the Komar and linkage formulae and Penrose's twistor charge formula. In particular, we explicitly show that the flux of the Wald-Zoupas charge matches the flux expression of Ashtekar and Streubel.\\

While our analysis was limited to null infinity in asymptotically flat spacetimes in vacuum general relativity, the Wald-Zoupas prescription is, in fact, much more general and can be used to obtain local and covariant charges for arbitrary diffeomorphism covariant Lagrangian theories of gravity including gravity coupled to electromagnetism \cite{Bonga2019} and Brans-Dicke theory \cite{Hou2020}. The Wald-Zoupas prescription has also been applied to the context of symmetries and charges associated with finite null surfaces and horizons \cite{CFP,Chandrasekaran:2019ewn} and spatial infinity \cite{PS} in asymptotically flat spacetimes in vacuum general relativity. We hope that the explicit computations presented here will be useful for similar analyses in other contexts, for instance, in spacetimes with compact extra dimensions \cite{Ferko:2021bym}.

\section*{Acknowledgements}
We thank \'Eanna \'E. Flanagan for useful discussions. A.M.G. thanks David Nichols for helpful conversations, as well as for sharing preliminary results from work with Arwa Elhashash.
This work is supported in part by the NSF grant PHY-1707800 to Cornell University. K.P. is supported in part by the NSF grant PHY-1801805.

\appendix

\section{General choices of conformal factor}
\label{sec:non-Bondi}

In the body of the paper, we worked in a conformal frame where 
\be\label{eq:Phi-defn-2}
    \Phi \defn \tfrac{1}{4} \nabla_a n^a \hateq 0 \,,
\ee
which, as a result of the Einstein equation (\cref{eq:S-ee}), implies the conditions \cref{eq:Bondi-cond,eq:nn-cond}. This choice, however, was made purely for convenience and is not essential to the results of this paper. In this appendix, we state some of our main results in general conformal frames where $\Phi \neq 0$ and therefore the Bondi condition does not hold. In this context, one is allowed more general conformal transformations of the form 
\be \label{eq:gen-conf}
\Omega \mapsto \omega \Omega\,, \quad \mbox{where} \quad \Lie_{n} \omega \not\hateq 0\,.
\ee

Using the fact that \(S_{ab}\) is smooth at \(\scri\), \cref{eq:S-ee} implies that in general
\be\label{eq:n-Phi}
    \nabla_a n_b \hateq \Phi g_{ab} \eqsp \lim_{\to \scri} \Omega^{-1} n_a n^a = 2 \Phi \,,
\ee
which generalize \cref{eq:Bondi-cond,eq:nn-cond}.
 In these conformal frames, the pullback of the unphysical metric to $\scri$, $q_{ab}$, satisfies
\be 
\Lie_{n} q_{ab} \hateq 2 \Phi q_{ab}\,.
\ee

It is important to note that $\Phi$ is universal in the sense of \cref{sec:univ-str}, that is, it is independent of the physical spacetime under consideration and can, without any loss of generality, be picked to be the same for the conformal completion of any asymptotically-flat physical spacetime. Hence, $\delta \Phi$ =0 on phase space. \\

In these general conformal frames, evolution equations for components of the Weyl tensor on $\scri$ given in \cref{eq:weyl-evol} also get generalized and may be found in Eq.~3.3 of \cite{KP-GR-match} (see also \cref{eq:weyl-evol-GHP} for their expressions in the GHP formalism). Moreover, the definition of the News tensor is generalized to 
\begin{equation} \label{eqn:news_l_def-2}
  N_{ab} \defn 2 Q_a{}^c Q_b{}^d (\Lie_n-\Phi) \sigma_{cd} \,,
\end{equation}
(given by \cref{eq:News-GHP} in the GHP formalism) where $\sigma_{ab}$ is (still) defined by \cref{eq:shear-defn} and has conformal weight 1. In addition,  \cref{eq:news-Schouten} is generalized to 
\be\label{eq:news-Schouten-Phi}
N_{a b} \hateq \STF \left[S_{ab}-2 \Phi \sigma_{ab}+2\left(\mathscr{D}_{a} +\tau_{a}\right) \tau_{b}\right]\,,
\ee
(which corresponds to \cref{eq:News-Ric-GHP} in the GHP formalism).
 Note that these expressions for the News tensor are defined in any conformal frame and are conformally invariant under all conformal transformations, including those of the form \cref{eq:gen-conf}, whereas, for example, the expression due to Geroch, discussed in \cref{eqn:news_geroch_def}, is only defined in frames where the Bondi condition holds.\\

 In general conformal frames where $\Phi \neq 0$, a BMS symmetry is given by $\xi^{a} \hateq \beta n^{a} + X^{a}$ where $X^{a} l_{a} \hateq 0$. Here, $\beta \,, X^{a}$ satisfy the relations (which are derived using the same method as for \cref{eq:bms-decomp-conds})
\be\label{eq:bms-decomp-conds-Phi} 
    (\Lie_n-\Phi) \beta \hateq \tfrac{1}{2} (\mathscr{D}_a - 2 \tau_a) X^a \eqsp \Lie_n X^a \hateq n^a X_b \tau^b \eqsp \STF \mathscr{D}_a X_b \hateq 0 \eqsp \alpha_{(\xi)} \hateq\Phi \beta+ \tfrac{1}{2} \mathscr{D}_a X^a \,,
\ee
which correspond to \cref{eq:bms-decomp-conds-GHP} in the GHP formalism. It is worth emphasizing that this is \emph{not} a different symmetry algebra than the one considered in the body of the paper. The observation that, for example, in these conformal frames a pure supertranslation, $\xi^{a} \hateq f n^{a}$, is no longer constrained to satisfy $\Lie_{n} f \hateq 0$ and that $f$ can have a non-trivial functional dependence on the normal direction along $\scri$, should not be taken to mean that symmetry algebra has become bigger. The point is simply that supertranslations are associated with conformally-weighted functions which have different-looking functional forms in different conformal frames.\\

With these considerations all of our computations can be carried out in a similar manner; one only has to keep track of the chosen \(\Phi\) through the calculations. For any BMS symmetry with $\beta$ and $X^{a}$ subject to \cref{eq:bms-decomp-conds-Phi} and $N_{ab}$ given by \cref{eqn:news_l_def-2}, the expressions for the WZ flux (\cref{eq:final-flux-2}) and charge (\cref{eq:Q-GR-defn}), remain unchanged in these conformal frames and are given by
\be\label{eq:final-flux-3}
  \mc F [\xi; \Delta \scri] \hateq -\frac{1}{16\pi} \int_{\Delta \scri} \df \varepsilon_3 N^{ab} [\tfrac{1}{2} \beta N_{ab}+(\mathscr{D}_a + \tau_a) (\mathscr{D}_b - \tau_b) \beta + \Lie_X \sigma_{ab} - \tfrac{1}{2} (\mathscr{D}_c X^c) \sigma_{ab}]\,,
\ee
and
\be\label{eq:Q-GR-defn-2}
    \mc Q[\xi; S] \hateq - \frac{1}{8\pi} \int_S \df\varepsilon_2 \left[ \beta ( \mc P + \tfrac{1}{2}\sigma^{ab} N_{ab} ) + X^a \mc J_a + X^a \sigma_{ab} (\ms D_c - \tau_c) \sigma^{bc} - \tfrac{1}{4} \sigma_{ab} \sigma^{ab} (\ms D_c - 2 \tau_c) X^c \right] \,,
\ee
respectively. These expressions are invariant under all conformal transformations including those of the form \cref{eq:gen-conf}.

\section{The GHP formalism at \(\scri\)}
\label{sec:GHP}
In this appendix, we include a brief review of the Geroch-Held-Penrose (GHP) formalism (see \cite{NP,GHP,KP-GR-match} for details) applied at null infinity.
We also present some of the expressions written in the body of the paper, including those for the WZ charge and flux, in this formalism.
\\

In order to be consistent with our conventions for the signature and Riemann curvature of the metric, we will use the sign conventions of \cite{SK}.\footnote{Note however that \(l^a\) in the notation of \cite{SK} corresponds to our \(n^a\) while their \(k^a\) corresponds to our \(l^a\).\label{fn:SK}} We pick a null tetrad \((n^a, l^a, m^a, \bar m^a)\) which is normalised such that \( n_a l^a = - m_a \bar m^a = - 1 \), with all other inner products vanishing. The metric then takes the form \(g_{ab} = - 2 n_{(a} l_{b)} + 2 m_{(a} \bar m_{b)}\). It is easy to see that the form of the metric is invariant under the following transformations,
\be\label{eq:GHP-transform}
	l^a \mapsto \lambda \bar\lambda l^a \eqsp n^a \mapsto (\lambda \bar\lambda)^{-1} n^a \eqsp m^a \mapsto \lambda (\bar\lambda)^{-1} m^a\,,
\ee
for any complex scalar \(\lambda\). These are called \emph{GHP transformations}. A scalar field \(\xi\), associated with a choice of null tetrad, is said to have \emph{GHP weight} \((p,q)\) if, under the GHP transformations in \cref{eq:GHP-transform}, \(\xi\) transforms as
\be\label{eq:GHP-wt-defn}
    \xi \mapsto \lambda^p \bar\lambda{}^q \xi \,.
\ee
We will indicate the GHP weight of such fields by writing \(\xi \wt (p,q)\).\\

We now adapt the choice of the null tetrad to \(\scri\). Note that the conformal factor \(\Omega\), and therefore \(\nabla^a\Omega\), are independent of the choice of null tetrad. To use the GHP formalism consistently with the transformations \cref{eq:GHP-transform}, we therefore need to define (on \(\scri\)) a scalar field $A$ such that
\be\label{eq:A-defn}
    An^a \hateq \nabla^a\Omega \quad\text{where } A \wt(1,1)\,.
\ee
When converting to tensor expressions used in the body of the paper, we set \(A=1\) at the end of the calculation. As in \cref{sec:aux}, we choose \(l^a\) to be the auxiliary null normal to a foliation of \(\scri\). The tetrads \((m^a, \bar m^a)\) are then a \emph{complex} orthonormal basis for the cross-sections of \(\scri\) in the chosen foliation, and are related to $Q_{ab}$ and $\varepsilon_{ab}$ by
\be
    Q_{ab} \hateq 2 m_{(a} \bar m_{b)} \eqsp \varepsilon_{ab} \hateq 2 i m_{[a} \bar m_{b]}\,.
\ee
The definition of the GHP spin coefficients in our conventions can be found in Eq.7.2~ of \cite{SK}; see also \cref{fn:SK} above. This definition differs by a sign from the ones used in \cite{NP,GHP}. At \(\scri\), using \(\nabla_a \nabla_b \Omega\vert_\scri = \Phi g_{ab} \) (from \cref{eq:n-Phi}) and \cref{eq:A-defn} we obtain
\be\label{eq:spin-scri}\begin{aligned}
    \kappa' &\hateq \sigma' \hateq \tau' \hateq 0 \eqsp & \rho' & \hateq - A^{-1}\Phi \,,\\[1.5ex]
    \epsilon &\hateq - \tfrac{1}{2} (n^a l^b \nabla_b l_a - \bar m^a l^b \nabla_b m_a) \eqsp & \beta & \hateq \tfrac{1}{2}( m^a \nabla_a \ln A + \bar m^a m^b \nabla_b m_a )\,, \\
    \epsilon' &\hateq \tfrac{1}{2} (- \Lie_n \ln A + A^{-1}\Phi + m^a n^b \nabla_b \bar m_a) \eqsp & \beta' & \hateq \tfrac{1}{2}( - \bar m^a \nabla_a \ln A + m^a \bar m^b \nabla_b \bar m_a )\,,
\end{aligned}\ee
while the spin coefficients \(\kappa, \rho, \sigma, \tau\) are arbitrary. Note that when the conformal factor is chosen to satisfy the Bondi condition we have \(\Phi \hateq 0\) and thus \(\rho' \hateq 0\). Also, \cref{eq:spin-scri} implies that
\be\label{eq:A-relations}
    (\thorn'+\rho')A \hateq \eth A \hateq 0 \,,
\ee
which is Eq.~9.8.26 of \cite{PR2}.\\

Further, since \(l_a\) is chosen to be normal to the cross-sections of \(\scri\), we have \(\rho = \bar \rho\) (i.e., the twist of \(l_a\) vanishes), and the expansion of \(l_a\) is given by
\be\label{eq:exp-GHP}
    \vartheta \hateq -2 \Re \rho \hateq - 2 \rho \,.
\ee
 The spin coefficient $\kappa$ depends on the choice of extension of \(l_a\) off of \(\scri\), while \(\sigma\) and \(\tau\) are related to the tensors \(\sigma_{ab}\) and \(\tau_a\) defined in \cref{eq:tau-defn,eq:shear-defn} by
\be\label{eq:tau-sigma-GHP}
     \sigma_{ab} & \hateq - \sigma \bar m_a \bar m_b - \bar\sigma m_a m_b \eqsp& \sigma & \hateq - \sigma_{ab} m^a m^b\,, \\
     \tau_a & \hateq - \tau \bar m_a - \bar \tau m_a  \eqsp& \tau & \hateq - \tau_a m^a\,.
\ee

Note that the spin coefficients \(\epsilon, \beta, \epsilon', \beta'\) are not GHP-weighted quantities, and in fact their GHP transformations contain derivatives of \(\lambda\). These are used to define the \emph{GHP derivative operators} \((\thorn, \thorn',\eth,\eth')\) (see Eq.~7.41 of \cite{SK}) which map GHP-weighted scalars to (other) GHP-weighted scalars.

Recall that, under a change of conformal factor \(\Omega \mapsto \omega \Omega\), the transformations of the metric and null tetrad on \(\scri\) are given by
\be\label{eq:conf-g-tetrad}
    g_{ab} \mapsto \omega^2 g_{ab} \eqsp (n^a, l^a, m^a, \bar m^a)  \mapsto \omega^{-1} (n^a, l^a, m^a, \bar m^a)\,,
\ee
where $\omega$ has GHP-weight \((0,0)\). Under this transformation, a scalar field \(\xi\) associated with the choice of null tetrad will have an additional conformal weight \(w\) if it transforms as \(\xi \mapsto \omega^w \xi\). We will indicate the combined \emph{conformal-GHP weight} of such scalars by writing \(\xi \wt (p,q; w)\).

Under the conformal transformation in \cref{eq:conf-g-tetrad}, the GHP derivatives of a conformally-weighted GHP scalar \(\xi\)  will, in general, pick up derivatives of \(\omega\). However, using the fact that the spin coefficients \(\rho'\) and \(\tau\) are not conformally-weighted and transform as (see also \cref{eq:Phi-conf-trans,eq:tau-tr})
\be
    \rho' \mapsto \omega^{-1} (\rho' - \thorn' \ln\omega) \eqsp \tau \mapsto \omega^{-1} (\tau - \eth \ln\omega)\,,
\ee
we can ``correct'' the conformal behaviour of the GHP derivatives by adding suitable combinations of \(\rho'\) and \(\tau\) to define the \emph{conformal-GHP derivatives}. For our purposes we only need the conformal-GHP derivatives tangential to \(\scri\), which are given by (see Eq.~5.6.36 of \cite{PR1}, where we set \(w_0 = w_1 = -\half\) as in Eq.~5.6.26 (iii) of \cite{PR1} for compatibility with the conformal transformations of the null tetrad given in \cref{eq:conf-g-tetrad})
\be\label{eq:conf-GHP-d}\begin{split}
    \lb[ \thorn' + (w + \tfrac{1}{2}(p+q))\rho' \rb]\xi &\wt (p-1,q-1; w-1)\,, \\
    \lb[ \eth + (w - \tfrac{1}{2}(p-q))\tau\rb]\xi &\wt (p+1,q-1; w-1)\,, \\
    \lb[ \eth' + (w + \tfrac{1}{2}(p-q))\bar\tau \rb]\xi &\wt (p-1,q+1; w-1)\,,
\end{split}\ee
where \(\xi \wt (p,q; w)\) is any conformally-weighted GHP scalar. We note that the spin coefficient \(\tau'\) is, in our formalism, conformally-invariant on \(\scri\), contrary to its transformation given in Eq.~5.6.27~(iii) of \cite{PR1}. This difference arises because, under \(\Omega \mapsto \omega \Omega\), the normal \(n_a\), in our definition, transforms away from \(\scri\) as \(n_a \mapsto \omega n_a + \Omega A^{-1} \nabla_a \omega\). For this reason, we have defined the conformal-GHP derivative corresponding to \(\eth'\) with \(\bar\tau\) instead of \(\tau'\). This definition corresponds to the conformal operators \(\tfrac{1}{2}(\eth_{\ms C}+\bar\eth'_{\ms C})\) and \(\tfrac{1}{2}(\bar\eth_{\ms C} + \eth'_{\ms C})\) that were introduced in Eq.~5.6.36 of \cite{PR1}. The conformal-GHP derivatives are useful for translating expressions from a Bondi frame to arbitrary choices of conformal factor and foliation, as follows: in the Bondi frame, where \(\rho' = \tau = 0\), the conformal-GHP derivatives \cref{eq:conf-GHP-d} are equivalent to the usual GHP derivatives. Thus, to generalise any expression in the Bondi frame to arbitrary conformal frames, we replace the usual GHP derivatives in that expression with the conformal-GHP derivatives, taking into account the appropriate weights. We will summarize the essential relations below; see Appendix~A of \cite{KP-GR-match} for details of this procedure.

By the peeling theorem (see Theorem~9.6.41 of \cite{PR2}) the usual GHP components \(\Psi_i\) of the Weyl tensor vanish at \(\scri\) and the components of \(\Omega^{-1} C_{abcd}\) are smooth at \(\scri\). We denote the components of \(\Omega^{-1} C_{abcd}\) in the chosen tetrad at \(\scri\), which we call the (unphysical) \emph{Weyl scalars}, by
\be\label{eq:weyl-GHP}\begin{aligned}
    \psi_4 & \defn \mc R_{ab}\bar m^a \bar m^b &&\wt (-4,0;-3)\,,\\
    \psi_3 & \defn \mc S_a \bar m^a &&\wt (-2,0;-3)\,, \\
    \psi_2 & \defn \tfrac{1}{2} (\mc P + i \mc P^*) &&\wt (0,0;-3)\,, \\
    \psi_1 & \defn \mc J_a m^a &&\wt (2,0;-3)\,, \\
    \psi_0 & \defn \mc I_{ab} m^a m^b &&\wt (4,0;-3)\,.
\end{aligned}\ee
The relations between the spin coefficients, $\Psi_i = \Omega \psi_i$, and components of the Ricci tensor in the GHP formalism can be found in \cite{GHP}.\\

\cref{eq:curl-weyl} implies that the Weyl scalars \(\psi_i\) in the unphysical spacetime satisfy the same Bianchi identity at \(\scri\) as the Weyl scalars \(\hat\Psi_i\) defined by the physical Weyl tensor, with all the physical Ricci tensor components set to zero by the vacuum Einstein equations (see Eqs.~9.10.1 and 9.10.2 of \cite{PR2}). Taking the prime transform (defined by Eq.~2.7 of \cite{GHP}) of the Bianchi identities Eqs.~2.33--2.36 of \cite{GHP}, with their \(\Psi_i\) replaced by \(\psi_i\) and the Ricci tensor terms set to zero, we have
\begin{subequations}\label{eq:weyl-evol-GHP}\begin{align}
    (\thorn' - 4\rho') \psi_3 & \hateq (\eth - \tau) \psi_4\,, \\
    (\thorn' - 3\rho') \psi_2 & \hateq (\eth - 2\tau) \psi_3 + \sigma \psi_4\,, \\
    (\thorn' - 2\rho') \psi_1 & \hateq (\eth - 3 \tau) \psi_2 + 2 \sigma \psi_3 \,,\\ 
    (\thorn' - \rho') \psi_0 & \hateq (\eth - 4 \tau) \psi_1 + 3 \sigma \psi_2 \,.
\end{align}\end{subequations}
If the conformal factor satisfies the Bondi condition (i.e., \(\rho' = 0\)), these reduce to Eqs.~9.10.4--9.10.7 of \cite{PR2}. Using \cref{eq:weyl-GHP,eq:spin-scri} and setting \(A=1\), we see that \cref{eq:weyl-evol-GHP} is equivalent to \cref{eq:weyl-evol}.\\

The spin coefficient \(\sigma\) has conformal-GHP weight \((3,-1; -1)\). The \emph{complex News function} \(N\) is defined by
\be\label{eq:News-sigma-conf}
    \bar N \defn - \thorn' \sigma \wt (2,-2; -2)\,.
\ee
The \emph{conformally-invariant} News tensor \(N_{ab}\) is given by
\be\label{eq:News-GHP}
    N_{ab} \hateq 2 N m_a m_b + 2 \bar N \bar m_a \bar m_b \eqsp 2 \bar N \hateq N_{ab} m^a m^b \,.
\ee
Using \cref{eq:tau-sigma-GHP,eq:News-Ric-GHP,eq:News-sigma-conf,eq:News-GHP}, we reproduce the definition of the News tensor in \cref{eqn:news_l_def}. Note that \cref{eq:News-sigma-conf} differs from the definition of the News function in Eq.~9.8.73 of \cite{PR2} by terms involving the spin coefficient \(\tau\), which vanishes in the Bondi frame. In arbitrary conformal frames we choose \cref{eq:News-sigma-conf} as the definition of the News function, since the corresponding News tensor in \cref{eq:News-GHP} is conformally-invariant.

The News function is related to the Weyl scalars by
\be\label{eq:weyl-News-conf}
   A \psi_4 \hateq (\thorn' - 2\rho') N \eqsp A \psi_3 \hateq \eth N\,.
\ee
Rewriting the above in terms of tensors using \cref{eq:weyl-GHP,eq:News-GHP} and setting \(\rho' \hateq 0\) and \(A = 1\), we get \cref{eq:weyl-News}. Moreover, using Eq.~2.25 of \cite{GHP}, along with \cref{eq:spin-scri,eq:News-sigma-conf}, we have
\be\label{eq:News-Ric-GHP}
    \bar N \hateq \Phi_{02} - \rho' \sigma - (\eth - \tau) \tau \,,
\ee
where, in our conventions, \(\Phi_{02} \defn \tfrac{1}{2} S_{ab} m^a m^b\) \cite{SK}; this corresponds to \cref{eq:news-Schouten} when \(\rho' \hateq 0\), or \cref{eq:news-Schouten-Phi} when the Bondi condition is not imposed.

Using \(\Psi_i = \Omega \psi_i\) with \cref{eq:News-Ric-GHP,eq:spin-scri} in Eq.~2.34 of \cite{GHP}, we also get
\be\label{eq:Pstar-identity-GHP}
    A\Im~ \psi_2 \hateq \Im \lb[(\eth' - \bar\tau) (\eth'+ \bar\tau) \sigma + \sigma N \rb]\,,
\ee
which corresponds to \cref{eq:Pstar-identity}. Similarly, from Eq.~2.21 of \cite{GHP} we have
\be\label{eq:exp-Ric-GHP}
    \Phi_{01} \hateq \eth \rho - \eth'\sigma\,,
\ee 
where \(\Phi_{01} = \tfrac{1}{2} S_{ab} l^a m^b\); using \cref{eq:exp-GHP}, this corresponds to \cref{eq:exp-Ric}.\\

For any vector \(\xi^a\) tangent to \(\scri\), we have the tetrad components
\be
    \xi^a \hateq \beta \nabla^a \Omega + X^a \hateq (A\beta) n^a + X m^a + \bar X \bar m^a\,,
\ee
where the conformal-GHP weights of $\beta$ and $X$ are:
\be
    A\beta \wt (1,1;1) \eqsp X \wt (-1,1;1)\,.
\ee
The conditions for \(\xi^a\) to be a BMS symmetry are given by
\be\label{eq:bms-decomp-conds-GHP}
    (\thorn' + 2\rho') (A\beta) \hateq \Re [(\eth + 2\tau) X ]\,, \\
    (\thorn' + 2\rho') X \hateq 0 \eqsp \eth' X \hateq 0 \,.
\ee
For a supertranslation \((\beta = f, X = 0)\) to be a translation we have the additional condition
\be\label{eq:trans-cond-GHP}
    (\eth - \tau)(\eth + \tau) (Af) \hateq 0\,.
\ee
Using \cref{eq:spin-scri} and setting \(A=1\), we reproduce \cref{eq:bms-decomp-conds-Phi,eq:trans-cond}; when \(\rho' = 0\), the former reduces to \cref{eq:bms-decomp-conds}.\\

Finally, the WZ charge in \cref{eq:Q-GR-defn}, in GHP notation, takes the form
\be\label{eq:Q-GHP}
    \mc Q[\xi; S] \hateq -\tfrac{1}{4\pi} \Re \int_S \df\varepsilon_2~\bigg[ \beta ( \psi_2 - A^{-1} \sigma N ) + A^{-1} X \psi_1 + A^{-2} X \sigma (\eth + \tau) \bar\sigma - \tfrac{1}{2} A^{-2} \sigma \bar\sigma (\eth+2\tau) X \bigg] ,
\ee
while the WZ flux is given by
\begin{equation} \label{eqn:flux_GHP}
  \begin{split}
    \mc F [\xi; \Delta \scri] \hateq -\tfrac{1}{4\pi} \Re \int_{\Delta \scri} \df\varepsilon_3~ N \Big[ &\tfrac{1}{2} \beta \bar{N} + (\eth - \tau) (\eth + \tau) \beta \\
    &- 2 A^{-1} (\Re[X \eth] \sigma + \sigma \eth X) + A^{-1}\sigma \Re[\eth X]\Big].
  \end{split}
\end{equation}

\section{Comparison to other charge formulae}
\label{sec:other-charges}

In this appendix we compare the Wald-Zoupas prescription to other formulations of the charge and fluxes associated with the BMS symmetries. In particular, we consider the Hamiltonian formulation of Ashtekar and Streubel, the Noether/Komar charge formula and their linkage versions and also the charges defined by considering twistors at null infinity.

\subsection{Ashtekar-Streubel flux and charge}
\label{sec:AS}

In this appendix, we compare the expression for the WZ flux given in \cref{eq:final-flux} to the flux given by Ashtekar and Streubel \cite{AS}. Rewriting the expression for the flux given in Eq.~4.14 of \cite{AS} in our  conventions, we obtain
\begin{align} \label{eq:AS-flux}
\mc F^{\rm (AS)}[\xi;\Delta\scri] & \hateq - \frac{1}{16\pi} \int_{\Delta\scri}\df\varepsilon_{3} N^{ab} \left(\Lie_{\xi} D_{a}-D_{a} \Lie_{\xi}\right) l_{b} \,,
\end{align}
where $D_a$ is the derivative operator induced on \(\scri\) by the unphysical derivative operator $\nabla_a$ as defined on pp.~46 of \cite{Geroch-asymp}. The derivative operator \(D_a\) satisfies
\be 
    D_{a} n^{b} \hateq 0 \eqsp D_{a} q_{bc} \hateq 0 \eqsp 
D_{a} \pb{v_b} \hateq (\delta_a{}^{c}+ n_{a} l^{c}) (\delta_b{}^{d} + n_{d} l^{b})\nabla_{c} v_{d} \,,
\ee
where \(v_a\) is any covector field in in the unphysical spacetime and \(\pb{v_a}\)is its pullback to \(\scri\).
\\

To compare the flux \cref{eq:AS-flux} to the WZ flux we need to compute the quantity \(\STF [\left(\Lie_{\xi} D_{a}-D_{a} \Lie_{\xi}\right) l_{b}]\) for a BMS symmetry \(\xi^a\). First, note that \(\left(\Lie_{\xi} D_{a}-D_{a} \Lie_{\xi}\right) l_{b}\) can be expressed in terms of the Riemann tensor of \(D_a\) which can, in turn, be written in terms of \(S_{ab}\) using Eq.~3.5 of \cite{AS}. This gives
\be
 \left(\Lie_{\xi} D_{a}-D_{a} \Lie_{\xi}\right) l_{b} \hateq \xi^{c}\left(q_{b[c} S_{a]}{}^{d}+S_{b[c} \delta_{a]}{}^{d}\right) l_{d}-l_{c} D_{a} D_{b} \xi^{c}\,.
 \ee
For a BMS symmetry $\xi^{a} \hateq \beta n^{a} + X^{a}$, we get
\be\label{eq:AS-goal}
    \STF [\left(\Lie_{\xi} D_{a}-D_{a} \Lie_{\xi}\right) l_{b}] \hateq \STF \lb[ \tfrac{1}{2} X_{b} S_{a}{}^{c} l_{c} + \tfrac{1}{2} S_{ab} \beta  + D_{a} D_{b} \beta - l_{c} D_{a} D_{b} X^{c} \rb] \,.
\ee
Next, we use the following relation (which follows from \cref{eq:exp-Ric-GHP} in the GHP notation)
\be \label{eq:exp-Ric}
    \tfrac{1}{2} Q_{b}{}^{c} S_{ac} l^{a} \hateq \ms D^{a} \sigma_{ab} - \tfrac{1}{2} \ms D_{b} \vartheta\,,
\ee
where $\vartheta \hateq Q^{ab} \nabla_a l_b$ is the expansion of $l^{a}$. Therefore, the first term in \cref{eq:AS-goal} gives
\be \label{eq:term1}
    \STF [\tfrac{1}{2} X_{b} S_{a}{}^{c} l_{c}] \hateq \STF \lb [X_{b} \ms{D}^{c} \sigma_{ac}-\tfrac{1}{2} X_{b} \ms{D}_{a} \vartheta \rb]\,.
\ee
Using \cref{eq:news-Schouten}, the second term in \cref{eq:AS-goal} is
 \be\label{eq:term2}
 \STF [\tfrac{1}{2 } S_{ab} \beta] \hateq \beta \STF [\tfrac{1}{2}N_{ab} - (\ms{D}_{a} \tau_{b} + \tau_{a}\tau_{b})]\,.
 \ee
The third term in \cref{eq:AS-goal} can be expressed as
\be\label{eq:term3}
 \STF D_{a} D_{b} \beta & \hateq \STF( \ms{D}_{a} \ms{D}_{b} \beta - \Lie_n \beta \nabla_{a} l_{b} ) = \STF \lb[ \ms{D}_{a}\ms{D}_{b} \beta - \tfrac{1}{2} (\ms{D}_{c} - 2\tau_c )X^{c} \sigma_{ab} \rb]\,, 
\ee
where we have used $\Lie_{n} \beta \hateq \frac{1}{2} (\ms{D}_a- 2\tau_a)X^{a}$ (\cref{eqn:lie_n_beta}). For the fourth term in \cref{eq:AS-goal}, using \(l_a X^a \hateq 0\) we have \(- l_{c} D_{a} D_{b} X^{c} = D_{a} (X^{c} D_{b} l_{c}) + D_{b} X^{c} D_{a} l_{c} \). We substitute $D_{a}l_{b} \hateq \sigma_{ab} + \frac{1}{2}Q_{ab} \vartheta - l_{a} \tau_{b}$ and take the \(\STF\) of the resulting expression. Using \(\STF \ms D_a X_b \hateq 0\) (\cref{eqn:stf_dX}),  we get
\be\label{eq:term4}
    \STF [-l_{c} D_{a} D_{b} X^{c} ] \hateq \STF [\ms{D}_{a} (X^{c} \sigma_{bc})+ \tfrac{1}{2}X_{b} \ms{D}_{a} \vartheta - X^{c} \tau_{c} \sigma_{ab} + \sigma_{a}{}^{c} \ms{D}_{b} X_{c} ]\,.
\ee
Putting together \cref{eq:term1,eq:term2,eq:term3,eq:term4} we get
\be
    \STF [ \left(\Lie_{\xi} D_{a}-D_{a} \Lie_{\xi}\right) l_{b} ] &\hateq \STF \bigg[ \tfrac{1}{2} \beta N_{ab} + (\ms{D}_{a} +\tau_a) (\ms{D}_{b} - \tau_b) \beta \\
     & \qquad + X_{b} \ms{D}^{c} \sigma_{ac} - \tfrac{1}{2} \ms{D}_{c} X^{c} \sigma_{ab} + \ms{D}_{a} (X^{c} \sigma_{bc}) + \sigma_{a}{}^{c} \ms{D}_{b} X_{c}  \bigg]\,.
\ee
Finally, we simplify the second line above using \cref{prop:Grant}. As a result, we find
  \be \label{eq:AS-goal-final}
  \mc F^{\rm (AS)}[\xi;\Delta\scri] \hateq - \frac{1}{16\pi} \int_{\Delta \scri} \df\varepsilon_{3}~ N^{ab}\left[\tfrac{1}{2} \beta N_{ab} + (\ms{D}_{a} +\tau_a) (\ms{D}_{b} - \tau_b) \beta + \Lie_X \sigma_{ab} - \tfrac{1}{2} \sigma_{ab} \ms D_c X^c  \right] \,.
  \ee
This is the same as the WZ flux formula given in \cref{eq:final-flux-2}. A charge expression whose flux is given by the Ashtekar-Streubel formula, for the case where $\tau_{a} =\vartheta=0$, was also considered in \cite{AK}. Our charge expression, when adapted to $\tau_{a} = \vartheta=0$, differs from the one in \cite{AK} which we believe to be a result of sign errors in some of the terms given in \cite{AK}.

\subsection{Komar formulae and linkage charges}
\label{sec:linkage}

In this section we review the Komar formulae for conserved charges associated with Killing vector fields and the linkage charges which generalize these to asymptotic BMS symmetries. We show that for translations and Lorentz symmetries these reproduce the WZ charge formula (up to proportionality constants), however for supertranslations the linkage charge does not coincide with the WZ charge.

For physical spacetimes with a Killing vector field a prescription for a corresponding conserved quantity was given by Komar \cite{Komar}. For stationary spacetimes (with a timelike Killing field \(\hat t^a\)) and axisymmetric spacetimes (with an axial Killing field \(\hat \phi^a\)) this gives a prescription for the Komar mass \(\mc M^{\rm (K)}\) and Komar angular momentum \(\mc J^{\rm (K)}\), respectively, defined as
\be\label{eq:Komar}
    \mc M^{\rm (K)} \defn - \frac{1}{8\pi}\int_S \hat\varepsilon_{abcd} \hat\nabla^c \hat t^d \eqsp \mc J^{\rm (K)} \defn \frac{1}{16\pi} \int_S \hat\varepsilon_{abcd} \hat\nabla^c \hat \phi^d \,.
\ee
Using the vacuum Einstein equation and the Killing equation it can be checked that these expressions are identically conserved and so can be evaluated on any \(2\)-sphere \(S\) in the spacetime. The relative difference in the sign arises because \(\hat t^a\) is timelike while \(\hat\phi^a\) is spacelike with the choice dictated by standard conventions for fields in Minkowski spacetime. The relative ``factor of \(2\)'' between the two formulae is essentially put in ``by hand'' to match the standard notions of mass and angular momentum in the case of the Kerr spacetime.

Note that the integrands in \cref{eq:Komar} are proportional to the Noether charge \(2\)-form (\cref{eq:noether-GR}) evaluated for the respective Killing fields. Thus, for asymptotic BMS symmetries we could define the asymptotic charges using the Noether charge to get
\be\label{eq:Q-noether}
    \mc Q^{\rm (N)}[\xi;S] = - \frac{1}{16\pi} \lim_{S' \to S} \int_{S'} \hat\varepsilon_{abcd} \hat\nabla^c \xi^d = - \frac{1}{16\pi} \lim_{S' \to S} \int_{S'} \varepsilon_{abcd} \nabla^c ( \Omega^{-2} \xi^d )\,.
\ee
It can be shown that the limit of the integrals defined above is finite and is independent of the manner in which \(S'\) limits to \(S\) (this is similar to the proof of \cref{lem:charge-well-def}; see \cite{GW}). However, a major drawback of using the Noether charge as the charge associated with a BMS symmetry is that \cref{eq:Q-noether} depends on how the BMS symmetry is extended away from \(\scri\): if \({\xi'}^a\) and \(\xi^a\) are equivalent representatives of a BMS symmetry, so that \({\xi'}^a = \xi^a + \Omega^2 W^a\) (from \cref{prop:bms-reps}) then we have
\be
    \mc Q^{\rm (N)}[\xi';S] - \mc Q^{\rm (N)}[\xi;S] = - \frac{1}{16\pi} \int_S \varepsilon_{abcd} \nabla^c W^d\,.
\ee
It is clear that one can obtain \emph{any} value for the Noether charge on \(S\) associated with the BMS symmetry simply by using different extensions of the symmetry into the spacetime. Note, if \(\xi^a\) is an exact Killing vector field of the physical spacetime then \(\xi^a\) is determined everywhere by its value on \(\scri\) by the Killing equation in the physical spacetime, or equivalently the conformal Killing equation in the unphysical spacetime and this ambiguity does not arise. Thus, for a Killing vector field the Noether charge (i.e., the Komar formula) at \(S\) is unambiguously defined. But for asymptotic BMS symmetries which need not be exact Killing fields the Noether charge cannot be unambiguously defined.\\

A way of avoiding this dependence on the extension of the BMS vector fields would be to suitably modify the Noether charge formula (so that it reduces to the Noether charge for a Killing field) and/or prescribe a choice of extension for the BMS vector field. Such a prescription is given by the \emph{linkage charge} which we summarize next.

\begin{enumerate}
    \item One formulation of the linkage charge is given as follows \cite{TW,Winicour}: Consider a \emph{fixed} cross-section \(S\) of \(\scri\), and let \(\mc N\) be an outgoing null hypersurface which intersects \(\scri\) in \(S\). Let \(l^a\) be a null geodesic vector field (defined on \(\mc N\)) which is tangent to \(\mc N\) and coincides with the auxiliary normal at \(S\). Given a BMS vector field \(\xi^a\) on \(S\) we extend it along the null surface \(\mc N\) by the condition
\be\label{eq:linkage-gauge}
    \lb(\gamma_{ab}^{(\xi)} - \tfrac{1}{2} \gamma^{(\xi)} g_{ab} \rb) l^b \big\vert_{\mc N} = 0\,,
\ee
where \(\gamma_{ab}^{(\xi)}\) is the unphysical metric perturbation generated by \(\xi^a\) (see \cref{eqn:delta_xi_g,eqn:gamma_xi_def}) and \(\gamma^{(\xi)}\) is its trace. Then the linkage charge at \(S\) is defined by
\be\label{eq:Q-linkage1}
    \mc Q^{\rm (L)}[\xi;S] \defn - \frac{1}{16\pi} \lim_{S' \to S} \int_{S'} \lb[\varepsilon_{abcd} \nabla^c ( \Omega^{-2} \xi^d ) + \Omega^{-1} \gamma^{(\xi)} \varepsilon_{ab} \rb] \, ,
\ee
where the integral is evaluated on \(2\)-sphere cross-sections \(S'\) of the null surface \(\mc N\), with area-element \(\varepsilon_{ab} \defn \varepsilon_{abcd}n^c l^d (l^e n_e )^{-1}\), and then the limit is taken to \(S\) along \(\mc N\). 
    \item Another formulation of the linkage charge, which does not require fixing a cross-section and a null surface was given by Geroch and Winicour \cite{GW}. In this formulation the BMS vector field \(\xi^a\) on \(\scri\) is extended away from \(\scri\) by the condition
\be\label{eq:GW-gauge}
    \gamma^{(\xi)} = 0\,,
\ee
and the linkage charge is defined by
\be\label{eq:Q-linkage2}
    \mc Q^{\rm (L)}[\xi;S] \defn - \frac{1}{16\pi} \lim_{S' \to S} \int_{S'} \varepsilon_{abcd} \nabla^c ( \Omega^{-2} \xi^d )\,,
\ee
where now the integral is evaluated on \emph{any} family of \(2\)-spheres \(S'\) in \(M\) and then the limit is taken as \(S'\) limits to \(S\).
\end{enumerate}
 It was shown in \cite{GW} that both formulations of the linkage charge give the same charge on \emph{any} choice of cross-section \(S\) (when the respective conditions in each formulation are satisfied). The advantages and disadvantages of both formulations are also discussed in \cite{GW}. An explicit expression for the linkage charge in terms of fields on \(\scri\) was computed by Winicour \cite{Winicour}. After integration-by-parts on the cross-section \(S\), the expression for the linkage charge given in Eq.~5.6 of \cite{Winicour} (in the GHP notation) can be written as\footnote{Note that the function \(\bar N\) defined in Eq.~3.30 of \cite{Winicour} is in fact \(\bar{\Phi_{02}}\) in our notation and we have used \cref{eq:News-Ric-GHP} to convert to our notation. Similarly, the quantity \(P\) defined in Eq.~3.31 of \cite{Winicour} vanishes at \(\scri\) due to the Bondi condition. We have also restored factors of \(A\) (defined in \cref{eq:A-defn}) to make the expression invariant under GHP transformations (\cref{eq:GHP-transform}).}
\be\label{eq:Q-linkage}
    \mc Q^{\rm (L)}[\xi; S] \hateq \tfrac{1}{4\pi} \Re & \int_S \df\varepsilon_2~ \bigg[ \tfrac{1}{2} \beta \big( \psi_2 - A^{-1}\sigma N \big) - \tfrac{1}{2} \beta A^{-1} (\eth - \tau)(\eth+\tau) \bar\sigma \\
    & \qquad\qquad + A^{-1} X \psi_1 + A^{-2} X \sigma (\eth + \tau) \bar\sigma - \tfrac{1}{2} A^{-2} \sigma \bar\sigma (\eth+2\tau) X \bigg] \,.
\ee

The linkage charge expression is conformally-invariant and thus only depends on the physical spacetime and \emph{not} on the choice of the conformal-completion. If we pick \(\beta\vert_S = 0\), i.e. \(\xi^a\) is a Lorentz vector field at \(S\), the linkage charge coincides with the WZ charge \cref{eq:Q-GHP}; this was argued by Wald and Zoupas \cite{WZ}. For a BMS translation (with \(\beta = f \) satisfying \cref{eq:trans-cond-GHP} and \(X = 0\)) note that the third term in the first line of \cref{eq:Q-linkage} vanishes upon integration-by-parts (also using \cref{eq:A-relations}) and we get \emph{half} of the WZ charge. Thus, the linkage charges also have a ``factor of \(2\)'' discrepancy between the mass and angular momentum similar to the Komar formulae \cref{eq:Komar}. Note that the WZ charge gives the correct relative factor between mass and angular momentum without any need to fix these factors in an ad hoc manner.
Since exact Killing vector fields in the physical spacetime can be shown to be part of some Poincar\'e algebra at $\scri$ (see \cref{rem:lorentz-amb}), we conclude that the WZ charge of an exact Killing vector field agrees with the Komar formula.

 However for supertranslations which are not translations the third term in \cref{eq:Q-linkage} does not vanish when the shear \(\sigma\) on the cross-section \(S\) is non-zero, and the linkage charge \emph{does not} coincide with the WZ charge or the Geroch supermomentum \cite{Geroch-asymp}. Thus, the linkage charge for supertranslations is non-zero even in Minkowski spacetime when the cross-section is not shear-free. This also implies that the flux of the linkage charge between arbitrary cross-sections does not vanish in Minkowski spacetime; though the flux does vanish when computed between two shear-free cross-sections. A detailed comparison of the linkage flux and the Ashtekar-Streubel flux (which equals the WZ flux, see \cref{sec:AS}), with similar conclusions, was given in \cite{AW-linkage}.
 
\subsection{Twistor charge}
\label{sec:twistor}

Another expression for a charge associated with the BMS symmetries was given by Penrose \cite{Penrose-charges} motivated by twistor theory (see also \cite{DS,Shaw,PR2}). To compare to these references we will work with the GHP formalism, set \(\rho' \hateq 0\) (i.e. the Bondi condition is satisfied) and \(\tau \hateq 0\) (equivalent to a choice of foliation where \(\tau_a \hateq 0\)).

With these choices being made, we choose a cross-section \(S\) of \(\scri\) and define \(2\)-surface twistors \((\omega^0, \omega^1)\) on \(S\) satisfying\footnote{The pair \((\omega^0,\omega^1)\) are the components of a \(2\)-spinor defined at \(S\) in a spin frame compatible with the GHP tetrad frame at \(S\); see \cite{Shaw}.}

\be\label{eq:2-twistors}
    \eth' \omega^0 \hateq 0 \eqsp \eth \omega^1 \hateq \sigma \omega^0 \,, \\
    \omega^0 \wt (-1,0; \half) \eqsp \omega^1 \wt (1,0;\half)\,,
\ee
Given two solutions \((\omega^0, \omega^1)\) and \((\tilde\omega^0, \tilde\omega^1)\) of \cref{eq:2-twistors} we can form the (complex) vector field
\be\label{eq:BMS-twistor}
    \xi^a_{\bb C} = (A \beta) n^a + X m^a \eqsp 
    \beta = - i (\omega^0 \tilde\omega^1 + \omega^1 \tilde\omega^0) \eqsp A^{-1} X = -2 i \omega^0 \tilde\omega^0\,,
\ee
where \(A\) is as defined in \cref{eq:A-defn}. Then \cref{eq:2-twistors} implies
\begin{subequations}\label{eq:twistor-poin}\begin{align}
    \eth' X & \hateq 0\,, \\ 
    \eth^2 (A \beta) & \hateq \tfrac{1}{2} \sigma \eth X + \eth (\sigma X) \label{eq:twistor-trans}\,.
\end{align}\end{subequations}
The first equation says that \(X m^a\) generates a (complex) Lorentz transformation on \(S\), while the second is a restriction to a (complex) Poincar\'e subalgebra at \(S\). The ``complexification'' is necessary since \cref{eq:twistor-trans} in general does not have any real solutions for \(\beta\). A fixed choice of cross-section \(S\) is necessary since \cref{eq:twistor-trans} is \emph{not} preserved along the null generators of \(\scri\) unless \(\beta = 0\) or \(\bar N = 0\), i.e. the News vanishes. For this choice of subalgebra the (complex) twistor charge at \(S\) proposed by Penrose \cite{Penrose-charges} is
\be
    \mc Q^{\rm (T)}[\xi_{\bb C}; S] \hateq - \tfrac{i}{4\pi} \int_S \df\varepsilon_2 \lb[ (\omega^0 \tilde\omega^1 + \omega^1 \tilde\omega^0) \psi_2 + \omega^0 \tilde\omega^0 \psi_1 + \omega^1 \tilde\omega^1 \psi_3 \rb]\,.
\ee
Using \(A \psi_3 = \eth N\) (\cref{eq:weyl-News-conf}), and integrating by parts on \(S\) we can write this charge in terms of the vector field \(\xi^a_{\bb C}\), defined in \cref{eq:BMS-twistor}, as
\be\label{eq:Q-twistor}
    \mc Q^{\rm (T)}[\xi_{\bb C}; S] \hateq \tfrac{1}{8\pi} \int_S \df\varepsilon_2 \lb[ 2 \beta (\psi_2 - A^{-1}\sigma N)  + A^{-1} X \psi_1 \rb]\,.
\ee
Using \cref{eq:twistor-poin} and integrating-by-parts on \(S\), \cref{eq:Q-twistor} can be written as (see also Eq.~11 of \cite{DS}) 
\be
    \mc Q^{\rm (T)}[\xi_{\bb C}; S] & \hateq \tfrac{1}{8\pi} \int_S \df\varepsilon_2~ \big[ 2 \beta (\psi_2 - A^{-1}\sigma N - i A^{-1} \Im \eth'\eth' \sigma ) \\
    &\qquad\qquad + A^{-1}X \psi_1 + A^{-2} X \sigma \eth \bar\sigma - \tfrac{1}{2} A^{-2} \sigma \bar\sigma \eth X \big]\,.
\ee
To compare this expression to the WZ charge we can use \cref{eq:Pstar-identity-GHP} (with \(\tau \hateq 0\)) to get
\be
    \mc Q^{\rm (T)}[\xi_{\bb C}; S] \hateq \tfrac{1}{8\pi} \int_S \df\varepsilon_2 \lb[ 2 \beta \Re (\psi_2 - A^{-1} \sigma N ) ++ A^{-1}X \psi_1 + A^{-2} X \sigma \eth \bar\sigma - \tfrac{1}{2} A^{-2} \sigma \bar\sigma \eth X \rb]\,.
\ee
Comparing to \cref{eq:Q-GHP} we see that the twistor charge is the same as the WZ charge for the complex vector field \(\xi^a_{\bb C} = \beta n^a + X m^a\) in the Poincar\'e subalgebra at \(S\) chosen by \cref{eq:twistor-poin}. The flux of the twistor charge was shown to be equivalent to the (complex) Ashtekar-Streubel flux in \cite{Shaw}.

\section{Symmetric and tracefree tensors in two dimensions}
\label{sec:2d-identities}

This appendix collects some useful identities for symmetric, tracefree tensors in two dimensions. In the main body of the paper these are applied to tensors on a cross-section \(S\) of \(\scri\), but they hold for any \(2\)-dimensional orientable manifold \(S\) with a Riemannian metric \(Q_{ab}\); as such, for simplicity, we drop the hats on all equalities.

The results of this appendix primarily rely on the following relation between the \(2\)-dimensional Riemannian metric \(Q_{ab}\) and the corresponding area-element $\varepsilon_{ab}$, (see Eq.~B.2.12 of Wald \cite{Wald-book}):
\begin{equation} \label{eqn:epsilon_2_square}
  \varepsilon^{ab} \varepsilon_{cd} = 2 Q^{[a}{}_c Q^{b]}{}_d.
\end{equation}

For a tensor field \(A_{ab}\), let \(\STF A_{ab}\) be the symmetric tracefree part as defined in \cref{eq:STF-defn}. Then, we can decompose \(A_{ab}\) into an antisymmetric part, a pure trace part, and a symmetric, tracefree part:
\begin{equation}
  A_{ab} = A_{[ab]} + \tfrac{1}{2} A^c{}_c Q_{ab} + \STF A_{ab}.
\end{equation}\\

Most of the tensors on \(S\) which arise in our main analysis are symmetric and tracefree, so we collect the identities satisfied by such tensors in the following proposition: 

\begin{prop}\label{prop:Grant}
    Let $A_{ab}$ and $B_{ab}$ be symmetric and tracefree tensors on \(S\); then the following identities hold:
    \be\label{eq:the-Grant-equation}
    \STF \lb( A_{ac} B^c{}_b \rb) = 0 \,,
    \ee
    \be\label{eqn:stf_curl_div}
    \mathscr{D}_{[a} A_{b]c} = Q_{c[a} \mathscr{D}^d A_{b]d} \,,
    \ee
    where \(\ms D\) is the covariant derivative compatible with \(Q_{ab}\), and
    \be\label{eq:diff-grant}
  \STF \left(\Lie_X A_{ab} - \tfrac{1}{2} A_{ab} \mathscr{D}_c X^c\right) = \STF\left[X_a \mathscr{D}^c A_{bc} + \mathscr{D}_a (X^c A_{bc}) + A_a{}^c \mathscr{D}_{[b} X_{c]}\right] \,,
    \ee
    for any vector field \(X^a\) on \(S\).
\begin{proof}
    For any tensor \(A_{ab}\) on \(S\), define a notion of left and right dual, respectively, by
    \begin{equation}\label{eq:dual-defn}
      (^* A)_{ab} \defn \varepsilon_{ca} A^c{}_b \eqsp (A^*)_{ab} \defn A_a{}^c \varepsilon_{cb}.
    \end{equation}
    It follows from \cref{eqn:epsilon_2_square} that
    \begin{equation} \label{eqn:2d_double_dual}
      (^{**} A)_{ab} = -A_{ab} \eqsp (A^{**})_{ab} = -A_{ab} \eqsp (^* A^*)_{ab} = A_c{}^c Q_{ab} - A_{ba}.
    \end{equation}
    Using \cref{eqn:2d_double_dual,eqn:2d_double_dual}, we therefore can characterize $A_{ab}$ by how its left and right duals are related:
    \begin{equation} \label{eqn:dual_characterization}
        (A^*)_{ab} = \begin{cases}
          ({}^* A)_{ab} & \iff A_{ab} = \STF A_{ab} \\
          -({}^* A)_{ab} & \iff \STF A_{ab} = 0
        \end{cases}
    \end{equation}
    Now, let \(A_{ab}\) and \(B_{ab}\) be two symmetric, tracefree tensors, and consider the left and right duals of $U_{ab} = A_{ac} B^c{}_b$. Using \cref{eq:dual-defn,eqn:dual_characterization} we obtain
    \begin{equation}
      (U^*)_{ab} = A_{ac} (B^*)^c{}_b = A_{ac} (^* B)^c{}_b  = A_{ac} \varepsilon^{dc} B_{db} = - (A^*)_{ac} B^c{}_b = - (^* A)_{ac} B^c{}_b = - (^* U)_{ab} \, .
    \end{equation}
    Then, using \cref{eqn:dual_characterization} we get that \(\STF(A_{ac}B^c{}_b) = 0\).

To show \cref{eqn:stf_curl_div} we start with the fact that
\begin{equation}
  \varepsilon^{cd} \varepsilon^{ab} \mathscr{D}_a A_{bc} = 2 \mathscr{D}_a (Q^{a[c} Q^{d]b} A_{bc}) = \mathscr{D}_c A^{dc},
\end{equation}
where the first equality uses \cref{eqn:epsilon_2_square} and the compatibility of $\mathscr{D}_a$ and $\varepsilon^{ab}$, and the second follows from the fact that $Q^{ab} A_{ab} = 0$. As such, the trace of \cref{eqn:epsilon_2_square} implies that
\begin{equation}
  \varepsilon^{ab} \mathscr{D}_a A_{bc} = \varepsilon_{cd} \mathscr{D}_e A^{de}.
\end{equation}
The result \cref{eqn:stf_curl_div} then follows by \cref{eqn:epsilon_2_square}.

Finally, consider the Lie derivative of $A_{ab}$ with respect to some vector field $X^a$ on $S$.
We have that
\be
    \STF \left(\Lie_X A_{ab} - \tfrac{1}{2} A_{ab} \mathscr{D}_c X^c\right) &= \STF \left(X^c \mathscr{D}_c A_{ab} + 2 A_a{}^c \mathscr{D}_b X_c - \tfrac{1}{2} A_{ab} \mathscr{D}_c X^c\right) \\
    &= \STF \big[2 X^c \mathscr{D}_{[c} A_{a]b} + \mathscr{D}_a (X^c A_{bc}) + A_a{}^c \mathscr{D}_b X_c - \tfrac{1}{2} A_{ab} \mathscr{D}_c X^c\big],
\ee
where the first equation follows by the definition of the Lie derivative and the second from the Leibniz rule. Using \cref{eqn:stf_curl_div} for the first term in this expression and rewriting the last two terms as
\be
    \STF \lb[ A_a{}^c \ms D_{[b} X_{c]} + A_a{}^c \STF \ms D_b X_c \rb] = \STF \lb[ A_a{}^c \ms D_{[b} X_{c]}\rb]\,,
\ee
and using \cref{eq:the-Grant-equation}, we find \cref{eq:diff-grant}.
  \end{proof}
\end{prop}

Finally, we show that divergence-free, symmetric, and trace-free tensors $A_{ab}$ on $S$ must vanish.
Unlike the above proposition, which holds for any two-dimensional manifold $S$, this is a consequence of the fact that any cross-section $S$ of $\scri$ has the topology of $\bb S^2$.
Our proof is based on one given in Lemma~5 of \cite{Geroch-asymp} (for similar discussion, see Sec.~A.4 of \cite{AK} or Appendix~C of \cite{Ashtekar-Magnon1984}):

\begin{prop} \label{prop:div_STF}
  Let $A_{ab}$ be a symmetric, trace-free tensor on $S$ that satisfies \(\ms D_a A^{ab} = 0\). Then $A_{ab}$ vanishes.
  \begin{proof}
    Consider any conformal Killing vector $\xi^a$ on $S$, and define
    \begin{equation}
      v_a \defn A_{ab} \xi^b.
    \end{equation}
    First note that, since \(A_{ab}\) is symmetric and trace-free and \(\xi^a\) is a conformal Killing field on \(S\),  $v_a$ is divergence-free:
    \begin{equation}
      \ms D_a v^a = A^{ab} \ms D_a \xi_b = 0.
    \end{equation}
    Next, we show that $v_a$ is curl-free.
    To do so, we use~\cref{eqn:epsilon_2_square} to show that
    \begin{equation}
      \varepsilon_{ab} \varepsilon^{ac} A_c{}^b = Q_{ab} A^{ab} = 0, \quad Q_{ab} \varepsilon^{ac} A_c{}^b = \varepsilon_{ab} A^{ab} = 0,
    \end{equation}
    so that $\varepsilon^{ac} A_c{}^b$ is also a symmetric, trace-free tensor.
    Then, \cref{prop:Grant} implies that
    \begin{equation}
      \ms D_a (\varepsilon^{ac} A_c{}^b) = \varepsilon^{ac} Q^b{}_a \ms D^d A_{cd} = 0,
    \end{equation}
    and so
    \begin{equation}
      \varepsilon^{ab} \ms D_a v_b = \ms D_a (\varepsilon^{ac} A_c{}^b \xi_b) = \varepsilon^{ac} A_c{}^b \ms D_a \xi_b = 0.
    \end{equation}
    As such, $v_a$ is both divergence- and curl-free, and so it must vanish (as $S$ has the topology of $\bb S^2$): that is,
    \begin{equation}
      A_{ab} \xi^b = 0,
    \end{equation}
    for \emph{any} conformal Killing vector $\xi^a$ on $S$.
    However, at each point, the set of conformal Killing vectors on $S$ spans the tangent space, and so we find that $A_{ab} = 0$.
  \end{proof}
\end{prop}


\bibliographystyle{JHEP}
\bibliography{WZ-explicit}
\end{document}